\documentclass[aps,prc,twocolumn,letterpaper,superscriptaddress,nofootinbib,showpacs,floatfix,10pt]{revtex4-2}
\usepackage[utf8]{inputenc}
\usepackage{graphicx}
\usepackage{comment}
\usepackage{setspace}
\usepackage{amsmath}
\usepackage{amssymb}
\usepackage[textwidth=16cm,textheight=22cm]{geometry}
\usepackage{color}
\usepackage{indentfirst}
\usepackage{xspace}
\usepackage{hyperref}
\usepackage{verbatim}
\usepackage{epstopdf}
\usepackage{graphicx}
\usepackage{longtable}
\usepackage{lineno}
\usepackage{subfigure}
\usepackage{parskip}
\usepackage{xcolor}
\usepackage{float}
\newcommand{\pt}   {\ensuremath{p_\mathrm{T}}\xspace}

\newcommand{\nch}   {\ensuremath{N_\text{ch}}\xspace}
\newcommand{\pythia} {\textsc{pythia8}\xspace}
\newcommand{\epos} {\textsc{epos-lhc}\xspace}

\begin{document}


\title{Two-particle number and transverse momentum balance function with event topology in pp collisions at $\sqrt{s}=13$ TeV}

\author{Subash Chandra Behera}
\email{subash.chandra.behera@cern.ch}
\affiliation{
INFN--Sezione di Roma, Piazzale Aldo Moro, 2 - 00185 Roma RM, Italy}
\author{Arvind Khuntia}
\email{arvind.khuntia@cern.ch}

\affiliation{INFN--Sezione di Bologna, via Irnerio 46, 40126 Bologna BO, Italy}

\begin{abstract}
 The first study of charge-dependent two-particle differential number ($B$) and momentum balance functions ($P_{2}^\mathrm{CD}$) with respect to an event shape variable, transverse spherocity, is reported. Results are presented from \textsc{pythia8} and \epos model calculations in proton-proton (pp) collisions at $\sqrt{s} = 13$ TeV. To distinguish between back-to-back jetlike topologies and isotropic events, low and high transverse spherocity values are chosen. The correlation functions are measured as a function of charged-particle multiplicity classes in relative pseudorapidity ($\Delta\eta$) and relative azimuthal angle ($\Delta\phi$) with $|\eta| < 2.4$ and $0.2 < \pt < 2.0$ GeV. A narrowing of the balance function width is observed in $\Delta\eta$ and $\Delta\phi$ from low- to high-multiplicity collisions. Wider balance functions are found in isotropic events as compared to jet-like events. However, for the momentum correlations, a nearly flat dependence is observed with charged particle multiplicity. This study investigates charge conservation mechanisms and their correlations for events classified with jetlike and isotropic topologies. To isolate medium-driven effects, we compare \epos with its hydrodynamic core enabled and disabled and observed narrowing patterns in $B$ and $P_{2}^{\mathrm{CD}}$ as a quantitative handle on radial-flow–induced localization of charge-balancing pairs.
\end{abstract}


\maketitle

\section{Introduction}
Measurements in ultra-relativistic heavy-ion collisions can provide information about the deconfined state of matter, known as the quark-gluon plasma~\cite{cmswhite, ALICE:2022wpn, PHENIX:2004vcz, starqgp, qgpmed2, QGPmedium1, ALICE:pruneau,cmsbf, jet1, jet2, jet3, Shuryak}. Interestingly, recent results from high-multiplicity proton-proton (pp) and proton-lead (p\text{--}Pb) collisions have revealed phenomena that closely resemble those seen in heavy-ion collisions~\cite{ALICE:2023ulm, ALICE:2016fzo, Shuryak, Baty:2021ugw, cmsppflow, CMS:2023iam}. One of the important observations is collectivity in high-multiplicity pp collisions. These observations raise important questions about the applicability of hydrodynamic models to small systems and whether collective behavior can emerge from mechanisms such as color reconnection, initial momentum correlations, or hadronization effects even in the absence of a large medium~\cite{ALICE:2023ulm, cmsppflow}. An open question is whether such collective phenomena can be probed using charge-sensitive observables. In this context, the balance function, $B$, serves as a powerful observable to investigate the underlying physics mechanisms in small collision systems, particularly as a function of charged particle multiplicity \cite{cmsbf, Manea:2024qgd, scottpratt, bfscott, alicebfpbpbpb, alicep2r2pp, Parida, STARBF1, STARBF2}.
The presence of collective radial flow modifies the balance function by introducing competing spatial and momentum-space effects. As the system expands, the velocity gradients impart a collective momentum boost to all particles, including charge-balancing pairs. This flow-induced boost enhances momentum correlations between opposite-sign partners, resulting in a narrowing of the $B$ in both relative pseudorapidity ($\Delta\eta$) and relative azimuthal angle ($\Delta\phi$). However, this narrowing effect competes with the spatial separation of balancing charges that occurs during earlier stages of the collision. Earlier quark production allows more time for charge separation, which would otherwise broaden the balance function. The observed balance function width, therefore, reflects an interplay between the flow magnitude (which tightens momentum correlations) and the production time of charges (which increases spatial decorrelation). This duality necessitates careful interpretation of balance function measurements, as a narrow width could indicate either strong radial flow development or late-stage charge production, while broadening may signal either weak flow or very early charge separation. The narrowing of the balance function in pseudorapidity is largely driven by radial flow, with the width showing an inverse dependence on the transverse mass, $m_\mathrm{T} = \sqrt{m^{2} + p_\mathrm{T}^{2}}$ ~\cite{icpaqgp, cmsbf}. On the other hand, Refs.~\cite{Bialas:1, Bialas:2} interpret the narrowing observed in high multiplicity collisions as a consequence of enhanced short-range correlations at freeze-out time. The sensitivity of the charge balance functions to event topology variables can provide crucial insights into hadronization dynamics, distinguishing between scenarios dominated by back-to-back jet-like structures and those with isotropic particle distributions. 

In this study, we present the first differential analysis of the charge balance function with respect to transverse spherocity, an event-shape variable that discriminates between soft (bulk-dominated) and hard (jet-dominated) QCD dynamics in small collision systems~\cite{originalsp, arvind_sp}. In high-multiplicity collisions, jet-like and isotropic  topologies likely represent events in which different underlying physics processes have occurred. For example, isotropic events likely emerge from multiple soft scatterings, while jetlike events arise from high-momentum parton fragmentation processes~\cite{ALICE_SP1, ALICE_SP2}. By examining $B$ as a function of both transverse spherocity and charged-particle multiplicity, we aim to elucidate how collective effects, parton showering, and event topology interplay in shaping charge-dependent correlations. Variations in the width and shape of $B$ with multiplicity and transverse spherocity thus provide sensitive signatures of medium effects, collective flow, and the underlying hadronization mechanisms. For this analysis, we select tracks within $|\eta| < 2.4$ and $0.2 < p_{\rm T} < 2.0$ GeV to facilitate comparison with prior CMS results~\cite{cmsbf}. We refer to transverse spherocity simply as ``spherocity" throughout this paper for the sake of convenience.

The charge-dependent correlation studies are particularly interesting using \textsc{pythia8} and \epos calculations~\cite{pythia_ref, eposLHC, eposCC}. \pythia hadronizes partons through the string fragmentation mechanism, which conserves charge locally and provides a baseline where correlations arise mainly from fragmentation and resonance decays, with no hydrodynamic evolution~\cite{Pruneau:2019baa, icpaqgp}. It helps to isolate noncollective effects such as jets or string breaking in charge correlations, which are crucial to understanding features of collective phenomena. On the other hand, \epos employs a core-corona model with statistical hadronization. The hydrodynamic evolution of the core generates radial flow, which reduces the width of the balance function. The corona, consisting of nonthermalized partons, hadronizes via string fragmentation. The interplay of the core and corona varies with charged-particle multiplicity, and thus it is very important to study the evolution of the balance function from low- to high-multiplicity events~\cite{eposLHC}. To quantify how radial flow influences $B$, we analyze results from \epos with the hydrodynamic core turned on and off.

This paper is organized as follows. Section~\ref{anaproedure} discusses the procedure used for the analysis. Section~\ref{modeldes} describes the \textsc{pythia8} and \textsc{epos} model simulations. Section~\ref{results} presents the results of the charge-dependent correlation as a function of $\Delta\eta$ and $\Delta\phi$, and the width in different multiplicity and spherocity classes~\cite{mult_def}. Section~\ref{summary} summarizes the findings of this work. 
\begin{figure*}[!ht]  
    \centering
    \subfigure[]{
        \includegraphics[width=0.45\textwidth]{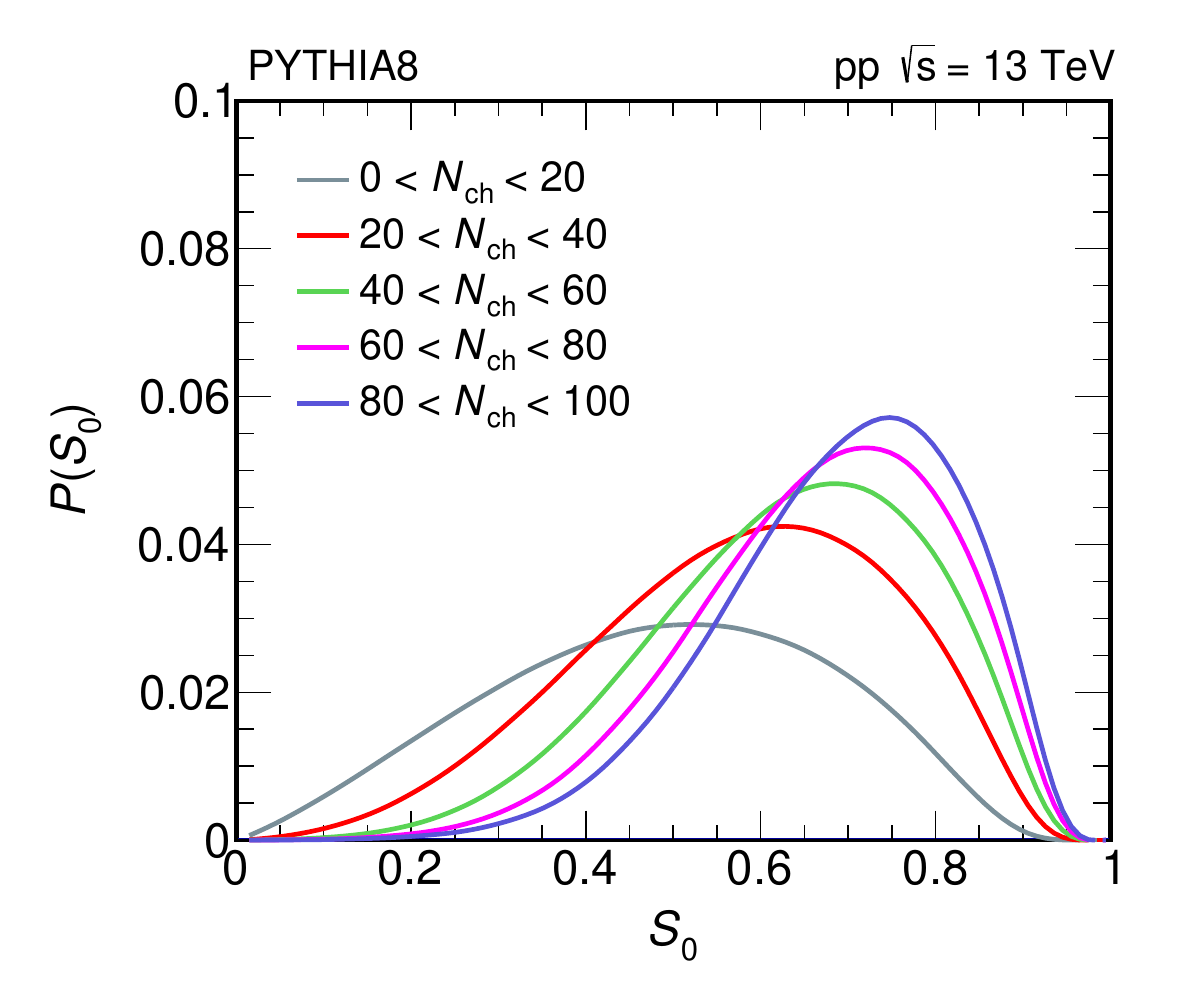}
    }
    \subfigure[]{
        \includegraphics[width=0.45\textwidth]{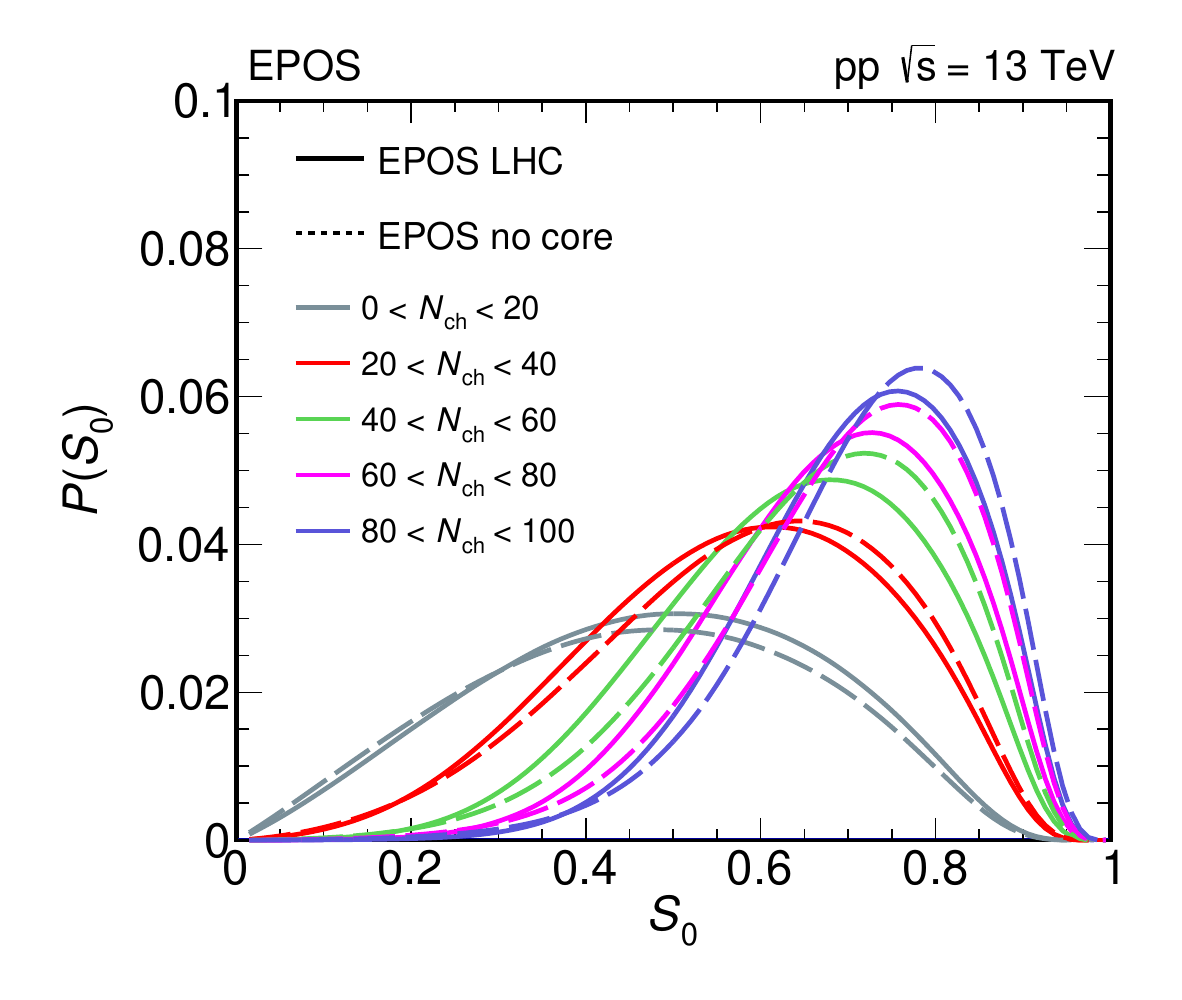}
    }
 \caption{ Colored lines show the spherocity distributions~\cite{arvind_sp, ALICE_SP1, ALICE_SP2} for different multiplicity classes~\cite{mult_def} in pp collisions at $\sqrt{s} = 13$ TeV using the
\pythia (left) and \textsc{epos} (right) event generators.}
    \label{fig:sp_epos_pythia}
\end{figure*}
\section{Analysis Methodology}
\label{anaproedure}
In high-energy collisions, particle production is subject to local charge conservation. This implies that to balance the charge, if a positively charged particle is created, a negatively charged particle must likewise be produced. The charge balance function aims to quantify the correlation between these oppositely charged pairs in phase space ($p_\mathrm{T}$,  $\eta$, or $\phi$)~\cite{cmsbf, STAR:bf1, alicebfpbpbpb}. The $B$ is constructed as a function of 
$\Delta \eta$ and $\Delta\phi$. Mathematically, this can be written as
\begin{equation} \label{eqn_balfun}
B =\frac{1}{2}[C_{2}^{(+, -)}  + C_{2}^{(-, +)} - C_{2}^{(-, -)} - C_{2}^{(+, +)}],
\end{equation}
where $C_{2}$ represents the two-particle correlations of the positively and negatively charged pairs. These correlation functions are constructed using the normalized signal and mixed event distributions~\cite{Baty:2021ugw, CMS:2023iam, cmsbf, cmsppridge, cmspbpbflow, hin18008, CMSPP}.  The signal distribution ($S$) is calculated by pairing the particles in the same event,
\begin{equation} \label{eqn_sig}
S(\Delta \eta,\Delta \phi) = \frac{1}{N_\text{trig}}\frac{d^{2}N^\text{same}}{{d\Delta \eta} \ {d\Delta \phi}},
\end{equation}
where $N_\text{trig}$ is the number of trigger particles within a given $p_\mathrm{T}$ interval and $N^\text{same}$ represents the total number of trigger-associated pairs in $\Delta \eta$ and $\Delta\phi$. The mixed event distribution, $M(\Delta \eta, \Delta\phi)$, is constructed using the mixed event technique. In this method, trigger particles from each event are paired with associated particles selected from ten different randomly chosen events within the same multiplicity classes in $|\eta|<2.4$ and $0.2 < p_\mathrm{T} < 2.0$ GeV.
\begin{equation} \label{eqn_mix}
M(\Delta \eta,\Delta \phi) = \frac{1}{N_\text{trig}}\frac{d^{2}N^\text{mix}}{{d\Delta \eta} \ {d\Delta \phi}},
\end{equation}
where $N_\text{mix}$ represents the number of mixed event pair for a given $\Delta\eta$ and $\Delta\phi$ bin. The background distribution corrects the acceptance effects caused by the finite $\eta$ range of the detector. The two-dimensional angular correlation function is calculated as 
\begin{align} \label{eqn_2pc}
\frac{1}{N_\text{trig}}\frac{d^{2}N^\text{pair}}{{d\Delta \eta} \ {d\Delta \phi}}& =C_{2}(\Delta\eta, \Delta\phi) \\
&= M(0,0) \frac{S(\Delta\eta, \Delta\phi)}{M(\Delta\eta, \Delta\phi)}.
\end{align}
The ratio $M(0,0)/M(\Delta\eta, \Delta\phi)$ mainly accounts for the effects of pair acceptance, where $M(0,0)$ denotes the mixed-event yield when both particles in the pair are emitted in nearly the same direction, leading to the highest possible pair detection efficiency~\cite{cmspbpbflow, cmsbf, cmsppridge}. 
\begin{figure*}[!htb]  
    \centering
        \includegraphics[width=0.9\textwidth]{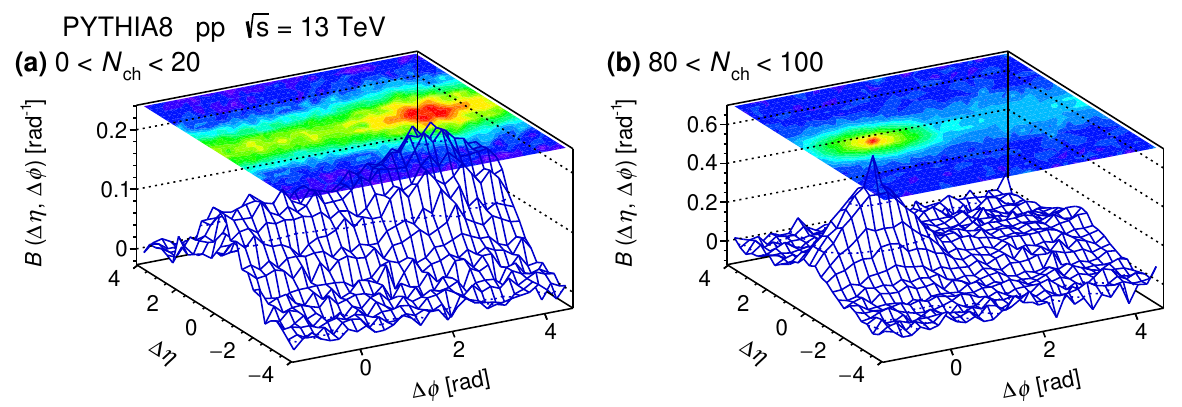}
        \includegraphics[width=0.9\textwidth]{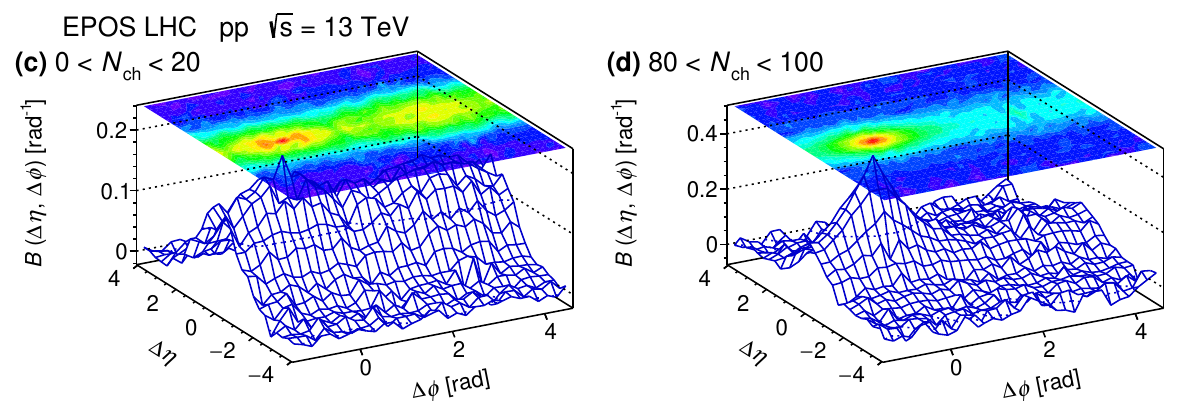}
 \caption{Two-dimensional number balance function ($B$) for \pythia and \epos model simulation in pp collisions at $\sqrt{s}$ = 13 TeV. The plots are shown for the integrated spherocity class of the lowest multiplicity $0 < N_\mathrm{ch} < 20$ (left column) and highest multiplicity $80 < N_\mathrm{ch} < 100$ (right column) collisions.}
    \label{fig:cbf_epos_pythia}
\end{figure*}
\begin{figure*}[!htb]  
    \centering
        \includegraphics[width=0.9\textwidth]{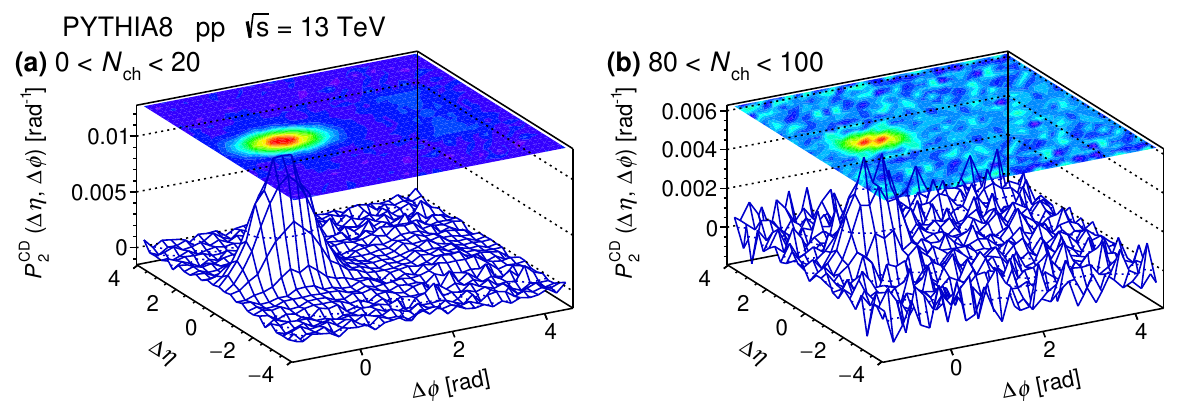}
        \includegraphics[width=0.9\textwidth]{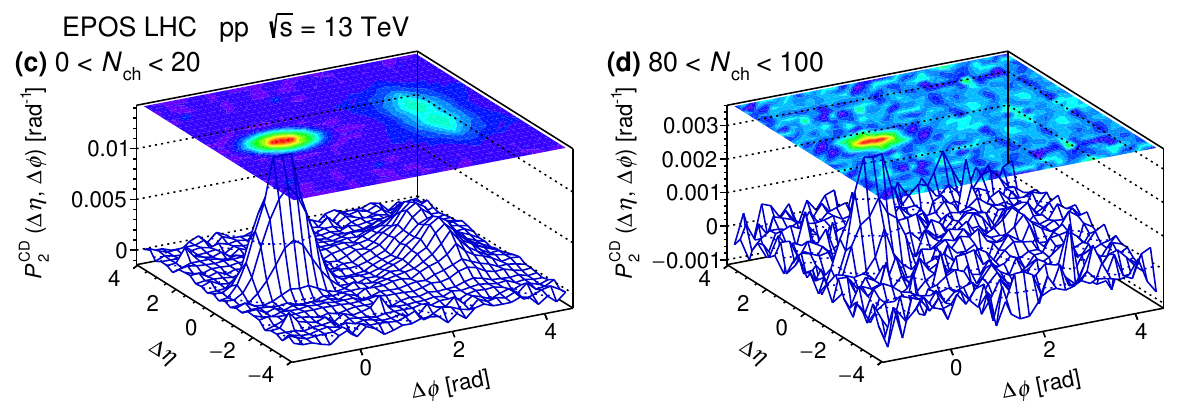}
 \caption{Two-dimensional momentum balance function ($P_{2}^\mathrm{CD}$) for \pythia and \epos model simulation in pp collisions at  $\sqrt{s}$ = 13 TeV. The plots are shown for the integrated spherocity class of the lowest multiplicity $0 < N_\mathrm{ch} < 20$ (left column) and highest multiplicity $80 < N_\mathrm{ch} < 100$ (right column) collisions.}
    \label{fig:p2_epos_pythia}
\end{figure*}

The momentum differential correlation function $P_{2}$ is calculated from the correlator, $\langle \Delta\pt \Delta\pt \rangle$ and it is divided by the square of the mean transverse momentum, $\langle \pt \rangle$, to make it a dimensionless quantity. This can be written as
\begin{center}
\begin{equation}
P_{2} = \frac{\langle  \Delta p_{\rm {T, 1}} \Delta p_{\rm {T, 2}} \rangle (\Delta\eta, \Delta\phi)}{\langle p_{\rm {T}} \rangle^{2}},
\label{equ:p2corrfun}
\end{equation}
\end{center}

where $\Delta p_\mathrm{T}$ = $p_\mathrm{T} - \langle p_\mathrm{T} \rangle$ is calculated for both the trigger (1) and associated (2) particles. $\langle p_\text{T} \rangle$, represents the inclusive mean transverse momentum. 
The differential correlator $\langle \Delta \pt \Delta \pt\rangle$ is written as
\begin{equation}
\begin{aligned}
&\langle \Delta p_{\rm {T}}\Delta p_{\rm {T}}\rangle(\Delta\eta,\Delta\phi)\\
&=\frac{
\int_{p_\mathrm{T,\text{min}}}^{p_\mathrm{T,\text{max}}} \Delta p_\mathrm{T,1}\,\Delta p_\mathrm{T,2}\,
c_{2}'(\mathbf{p}_{1},\mathbf{p}_{2}) \, \mathrm{d}p_\mathrm{T,1}\,\mathrm{d}p_\mathrm{T,2}
}{\int_{p_\mathrm{T,\text{min}}}^{p_\mathrm{T,\text{max}}}c_{2}'(\mathbf{p}_\mathrm{1},\mathbf{p}_{2}) \, \mathrm{d}p_\mathrm{T,1}\,\mathrm{d}p_\mathrm{T,2}
},
\end{aligned}
\end{equation}
here $c_{2}'(\mathbf{p}_{1},\mathbf{p}_{2})$ is the two--particle density distribution expressed as a function of the transverse momenta~\cite{alicep2r2pp}.
The $P_{2}$ observable is particularly sensitive to variations in particle transverse momentum relative to the average \pt~\cite{alicebfpbpbpb,alicep2r2pp, Sahoo:2018uhb, Basu:2020ldt}.  
$P_{2}$ correlations can be calculated using the different charge combinations, $P_{2}^{+,-}, P_{2}^{-,+}, P_{2}^{+,+}$ and $P_{2}^{-,-}$. $+,-$ and $-,+ $ are called unlike-sign correlations and $+, +$ and $-, -$ are called like-sign correlations.
Different particle production processes contribute to unlike-sign and like-sign particle correlations. Unlike-sign pairs are especially sensitive to processes that produce charge-balanced pairs, such as resonance decays or the fragmentation of quark-antiquark pairs, and collective phenomena like flow in the medium. As a result, they can reveal features related to charge conservation and particle-antiparticle production mechanisms. Similarly, like-sign correlations are more sensitive to collective effects, such as quantum statistical correlations (e.g., Bose-Einstein correlations for identical bosons)~\cite{Pratt_ref1, Brown_ref2} and correlations from bulk particle production, collective flow, and Coulomb repulsions~\cite{cmspbpbflow, cmsbf}. The charge-dependent $P_{2}^\mathrm{CD}$ correlations are calculated using the following formula,
\begin{equation} \label{eqn_balfun_p2}
P_{2}^\mathrm{ CD} =\frac{1}{2}[P_{2}^{(+, -)}  + P_{2}^{(-, +)} - P_{2}^{(-, -)} - P_{2}^{(+, +)}].
\end{equation}
The $P_{2}$ observable specifically accounts for how particle momenta deviate from the average transverse momentum and is sensitive to the relative momentum or hardness of the correlations~\cite{alicebfpbpbpb, alicep2r2pp}. This allows it to differentiate between cases where both particles in a pair have transverse momenta below or above the mean $\langle \pt \rangle$, corresponding to soft-soft or hard-hard interactions, respectively. It has additional features for the sensitivity of angular ordering of particle production within jets and the effect from resonance decays as discussed in Ref.~\cite{alicebfpbpbpb, alicep2r2pp}. 

To study these observables as a function of  spherocity  that is classified based on the back-to-back jet topologies to that of isotropic by considering the distributions of final-state particles arising from hadronic and nuclear collisions, as, 
\begin{equation}
    S_{0} = \frac{\pi^2}{4} \min_{\hat{n}} \left( \frac{\sum_i |\vec{p}_{\rm T,i} \times \hat{n}|}{\sum_i \vec{p}_{\rm T,i}} \right)^2.
\end{equation}
Here, the unit vector $\hat{n}$ is selected to minimize the ratio within the brackets. The scaling factor of $\frac{\pi^2}{4}$ guarantees that the $S_{0}$ estimator lies between 0 and 1. To have a meaningful spherocity definition, we have considered events with a minimum of five charged particles at mid rapidity~\cite{arvind_sp, ALICE_SP1, ALICE_SP2, ALICE:2023yuk}. The minimum charged-particle requirement mainly affects the very lowest multiplicity events and therefore has only a mild influence on the  0 $<$ \nch $<$ 20 multiplicity interval considered in this analysis.

The width of the balance function is a measure of the spread of particle-antiparticle correlations in a given variable \( \Omega \), such as $\Delta\eta$ or $\Delta\phi$ difference. Provided that the balance function approaches zero at large $\Delta\eta$, ensuring the absence of a residual background contribution, the width can be calculated using the rms,
\begin{equation}
\sigma = \left[ \frac{\sum_i O(\Omega_i)\, \Omega_i^2}{\sum_i O(\Omega_i)} \right]^{\frac{1}{2}},
\label{eq:bf_width}
\end{equation}
where \( O(\Omega_i) \) denotes the balance function value at the bin centered at \( \Omega_i \), and the summation runs over all bins of the measured distribution. This definition provides a quantitative measure of the width of the balance function, characterizing the separation scale of the correlated particle pairs~\cite{alicebfpbpbpb, alicep2r2pp}. The width of $B$ is estimated in $|\Delta\eta| \leq 3.0$ and for the $|\Delta\phi| \leq \pi/2$ from the one-dimensional projection. For the $P_{2}^\mathrm{CD}$ calculation, the RMS widths are estimated in the region $|\Delta\eta, \Delta\phi| \leq 1.0$, to suppress accumulation of statistical fluctuation.

\section{Model description}
\label{modeldes}
In this study, 500M Monte Carlo (MC) events with \pythia ~\cite{pythia_ref} and \epos~\cite{cmspythia, eposLHC, eposCC} are considered in pp collisions. \textsc{pythia8.306} with CP5 tune is used, which has been optimized to describe a wide range of LHC collision data~\cite{cmspythia}. Multiparton interaction (MPI) and color-reconnection (CR) mechanisms are utilized to mimic the dense partonic environment characteristic of high-multiplicity events~\cite{OrtizVelasquez:2013ofg, Ortiz:2020rwg}. Events are generated at center-of-mass energies of 13 TeV corresponding to the collision energy in Run 2 at the LHC, ensuring a meaningful comparison between simulated and observed distributions~\cite{pythia_ref}.
\begin{figure*}[!ht]  
    \centering  
    \subfigure[]{
        \includegraphics[width=0.3\textwidth]{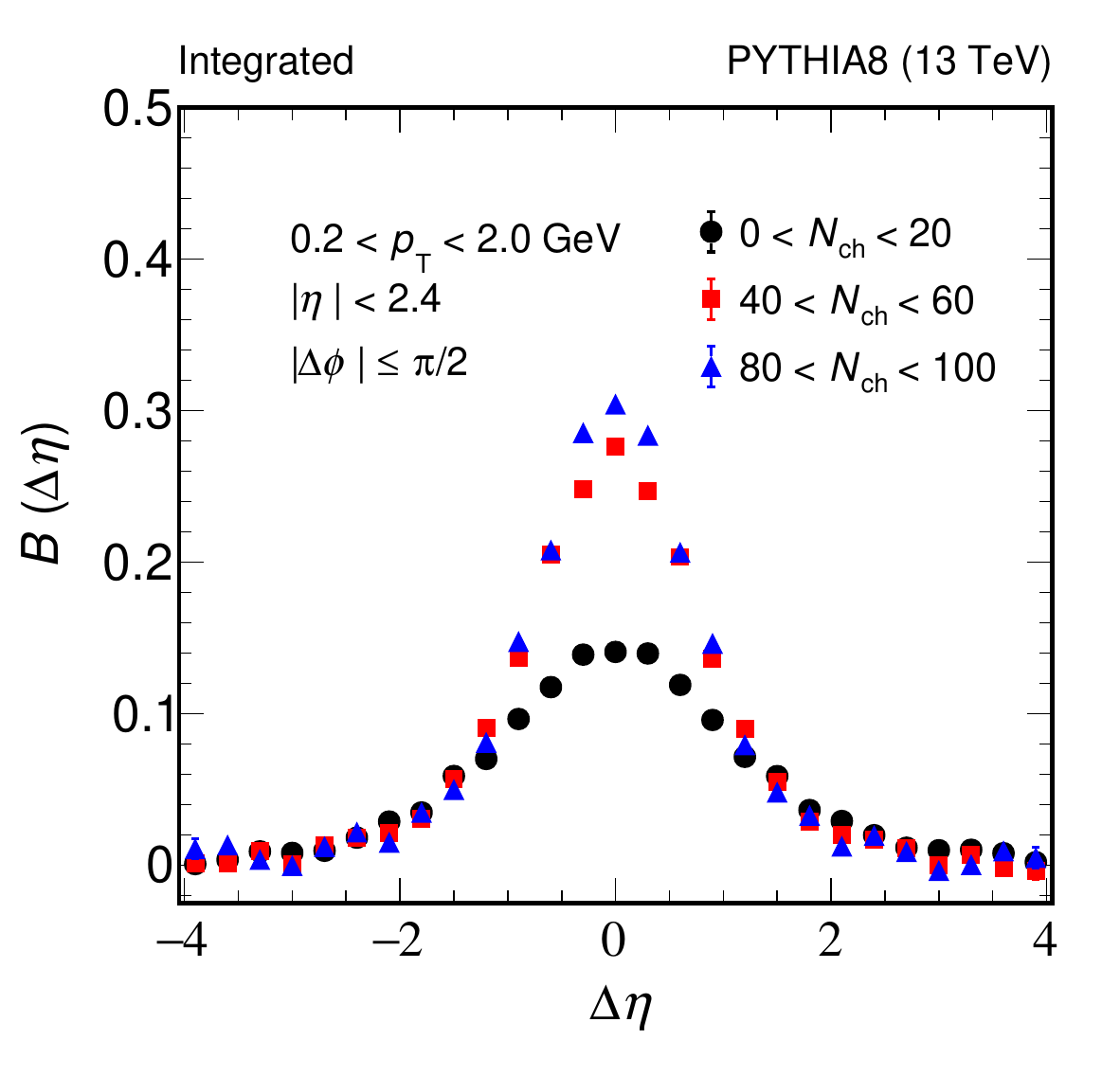}
    }
    \subfigure[]{
        \includegraphics[width=0.3\textwidth]{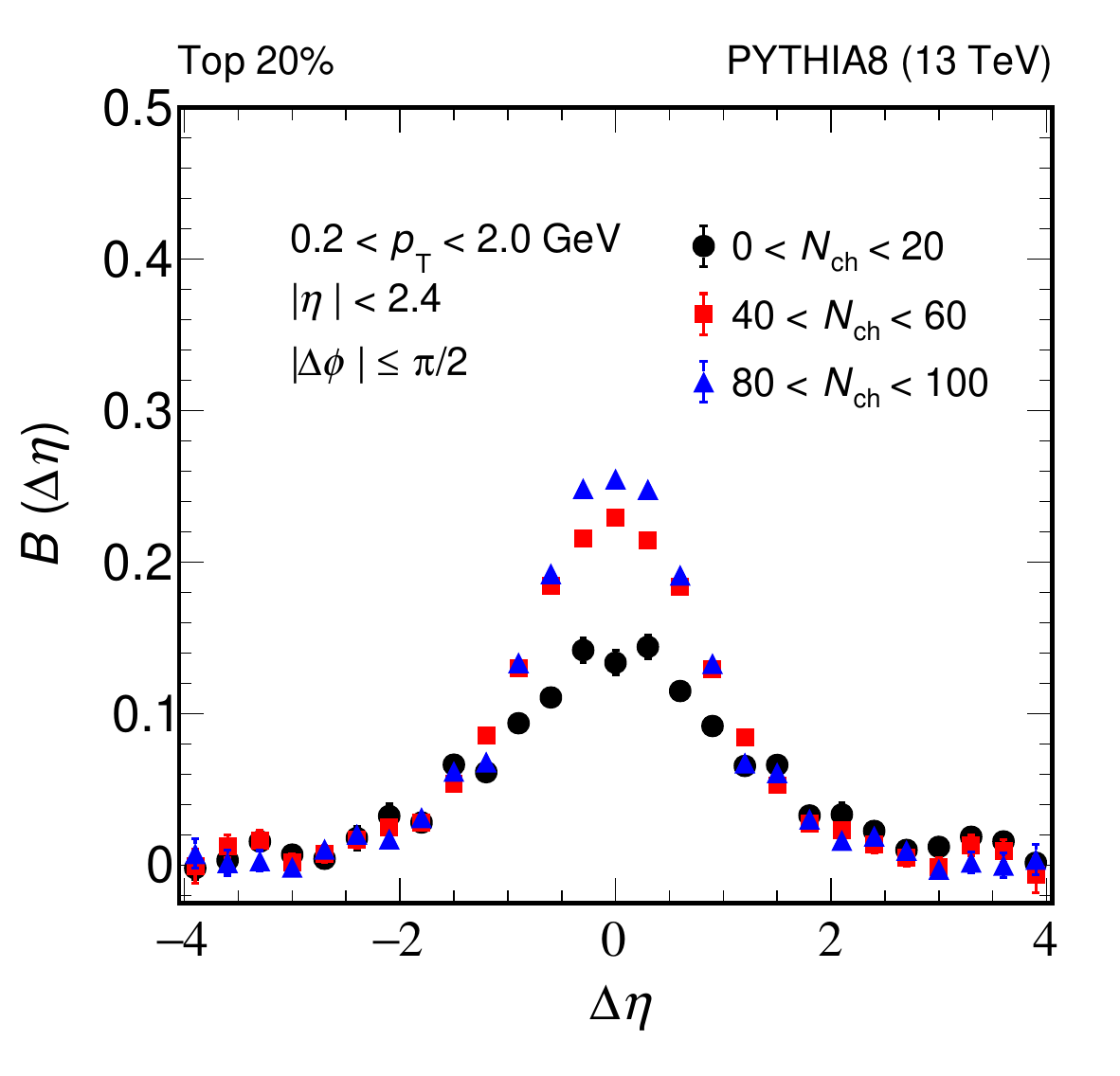}
    }
    \subfigure[]{
        \includegraphics[width=0.3\textwidth]{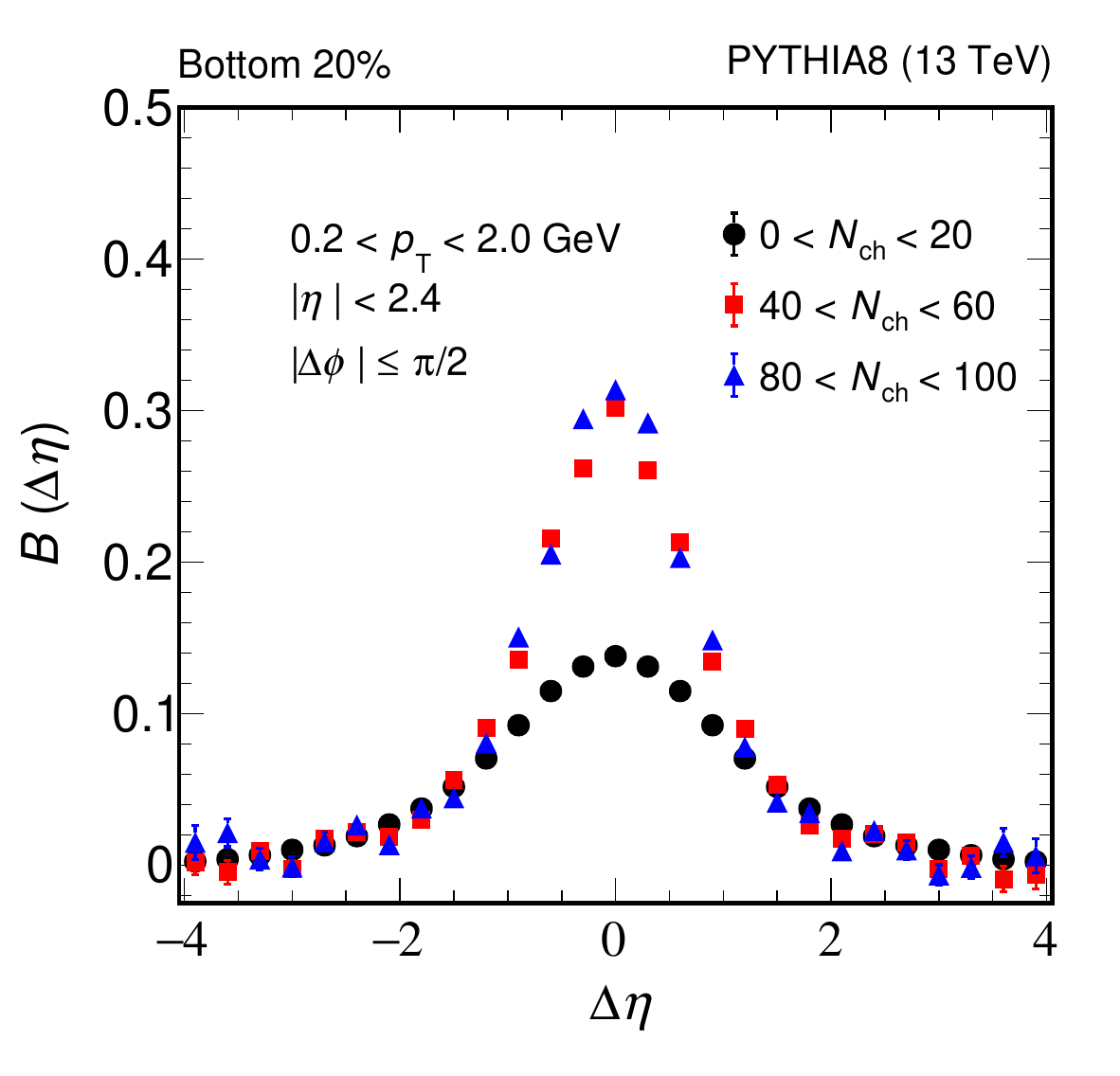}
    }
    \subfigure[]{
        \includegraphics[width=0.3\textwidth]{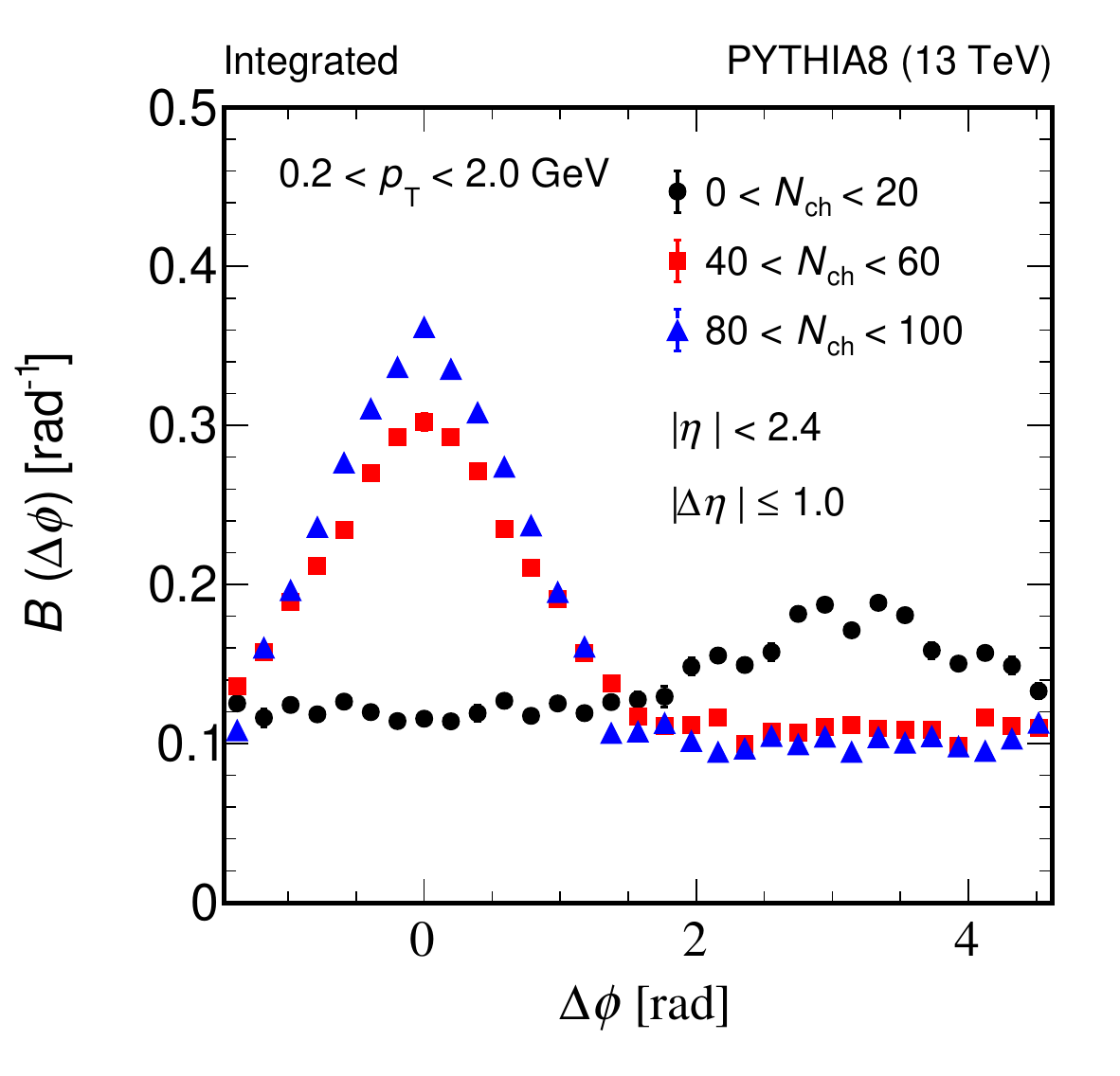}
    }
    \subfigure[]{
        \includegraphics[width=0.3\textwidth]{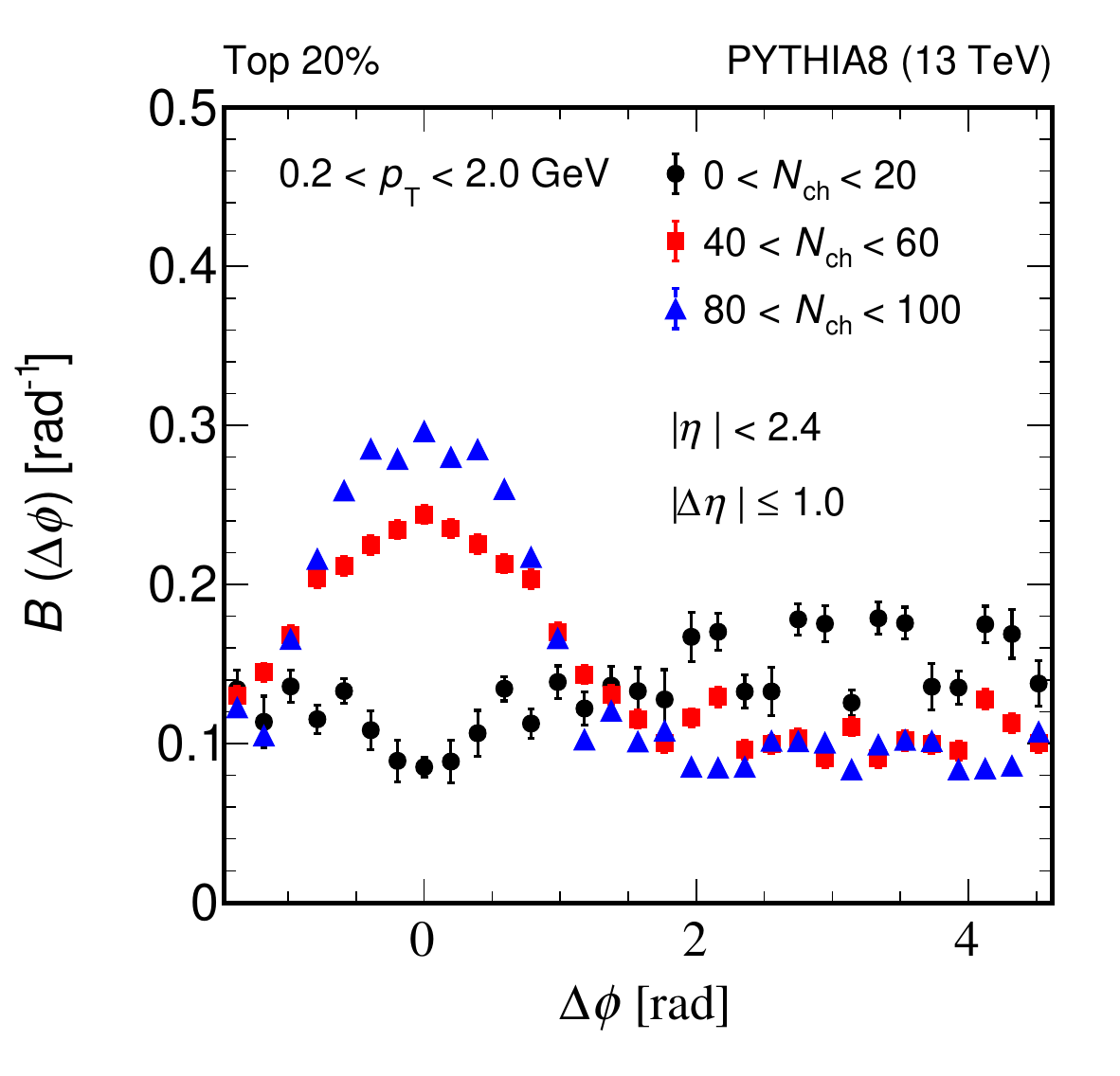}
    }
    \subfigure[]{
        \includegraphics[width=0.3\textwidth]{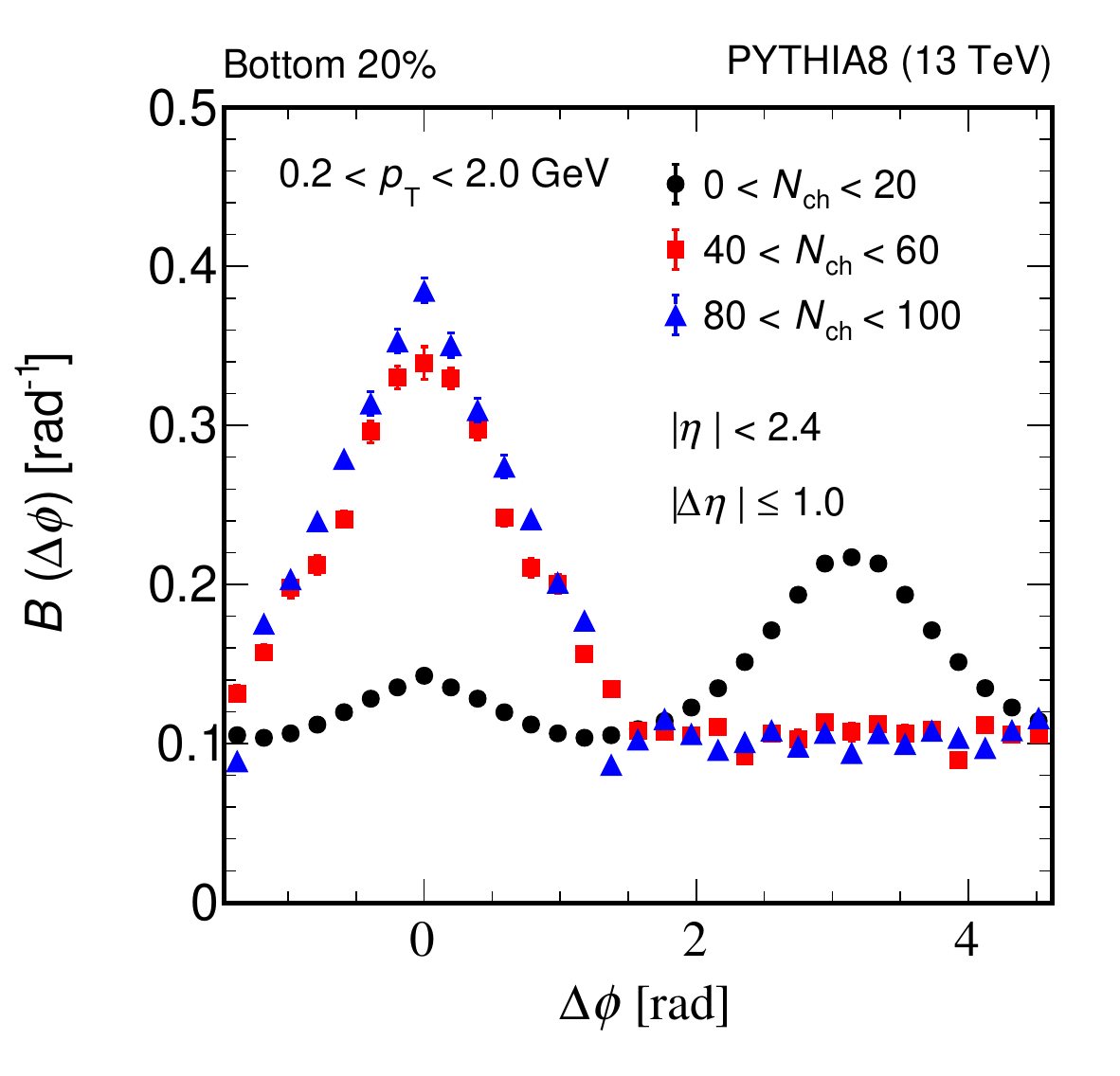}
    }
 \caption{One-dimensional $B$ projections along $\Delta\eta$ and $\Delta\phi$ from \pythia model in pp collisions at $\sqrt{s} =$ 13 TeV. The left column is for the integrated spherocity class, the middle column is for the top 20\%, and the right panel is for the bottom 20\% of the spherocity class. $\Delta\eta$ projections are taken in $|\Delta\phi| \leq \pi/2$ and $\Delta\phi$ projections are taken in $|\Delta\eta| \leq 1$ range.}
    \label{fig:model_plots_1d}
\end{figure*}
MPI accounts for the possibility of several parton-parton interactions occurring within a single pp collision event. This phenomenon contributes significantly to the underlying event activity and particle production. On the other hand, CR models the rearrangement of color strings between partons before hadronization, effectively reducing the total string length and mimicking collective-like behaviour such as collective radial expansion. These two mechanisms are essential for reproducing the structure of two-particle correlations observed in experimental data, particularly in high-multiplicity environments where the partonic overlap is substantial.

In addition to \textsc{pythia}-based calculations, we use the \epos event generator~\cite{eposCC}, a comprehensive MC framework for pp, p–A, and A–A collisions that combines pQCD-inspired parton dynamics with a phenomenological treatment of soft processes via multiple scattering and saturation. A key ingredient is the core-corona separation with subsequent hydrodynamic evolution of the dense core, while the dilute corona hadronizes via strings~\cite{eposCC}. Energy-momentum conservation at the level of individual scatterings, multiple parton interactions, parton-ladder splitting, and string interactions further enables realistic final states. In this study, we employ two \epos configurations: (i) core-corona with hydrodynamic evolution (“core on”), expected to generate finite radial flow, and (ii) a no-core setup (“core off”) that suppresses collective expansion. The difference between these two explicitly isolates radial flow contributions to the shapes and widths of $B$ and $P_{2}^{\mathrm{CD}}$.

\begin{figure*}[!th]  
    \centering
    \subfigure[]{
        \includegraphics[width=0.3\textwidth]{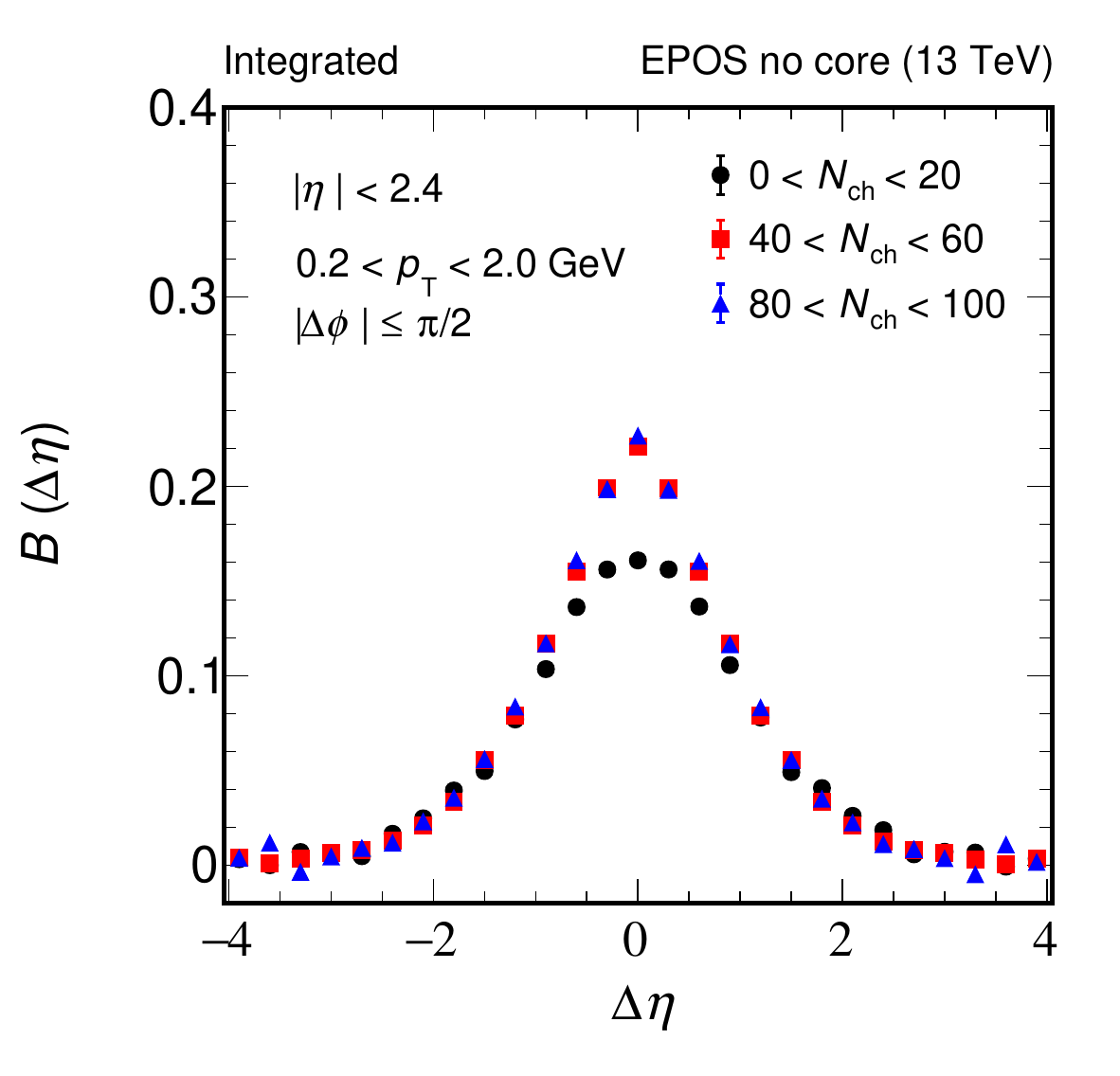}
    }
    \subfigure[]{
        \includegraphics[width=0.3\textwidth]{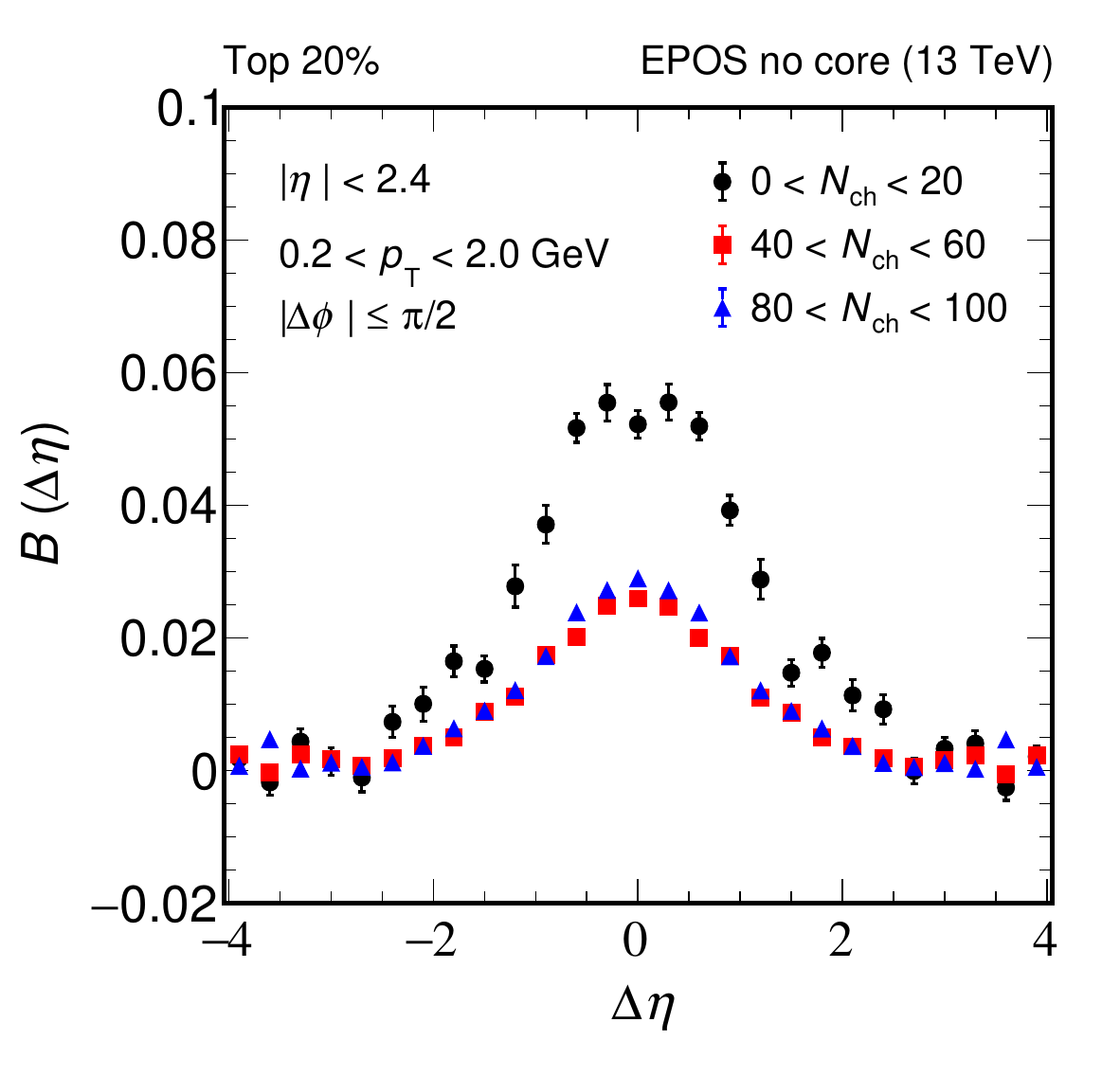}
    }
    \subfigure[]{
        \includegraphics[width=0.3\textwidth]{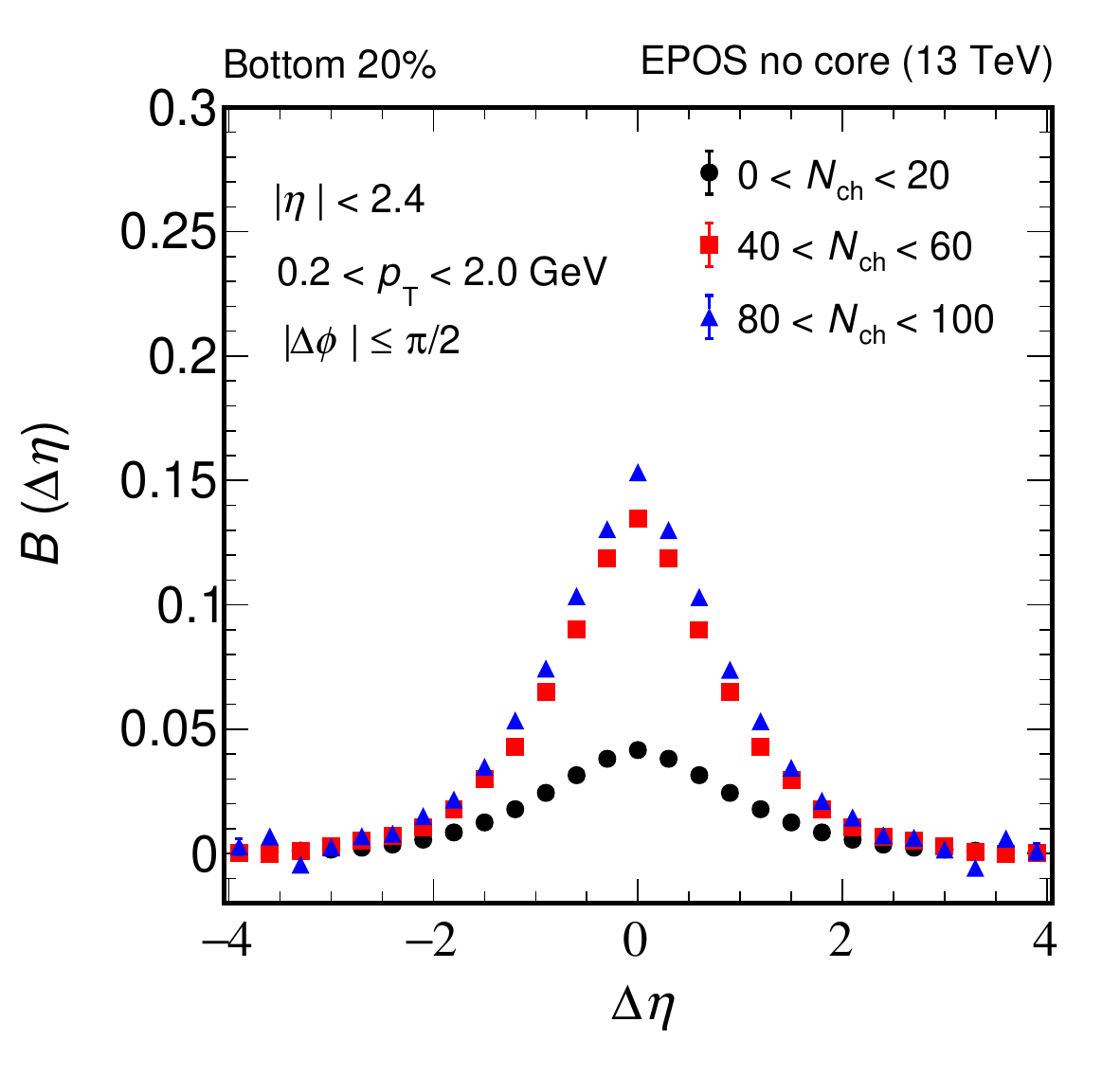}
    }
    \subfigure[]{
        \includegraphics[width=0.3\textwidth]{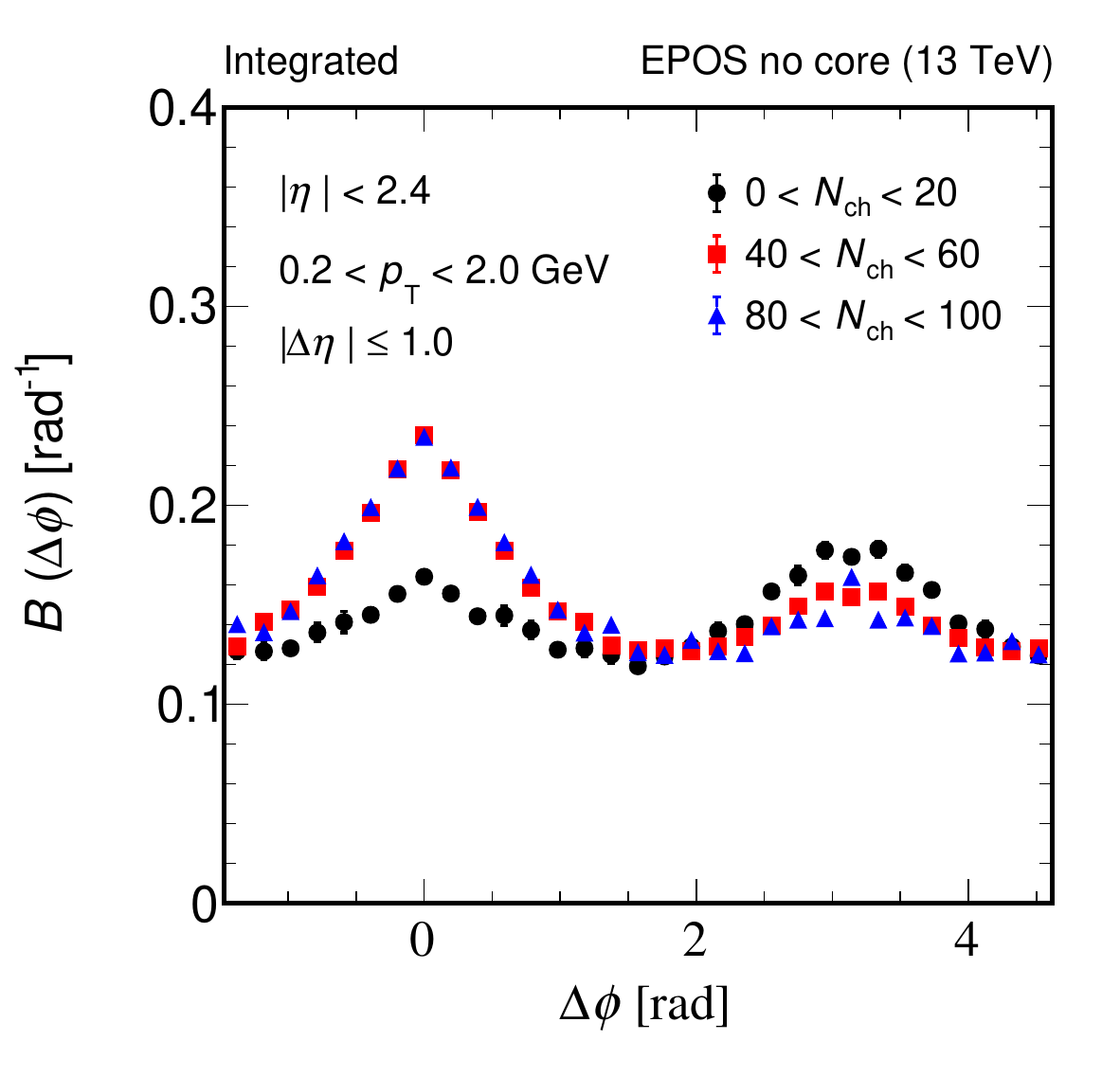}
    }
    \subfigure[]{
        \includegraphics[width=0.3\textwidth]{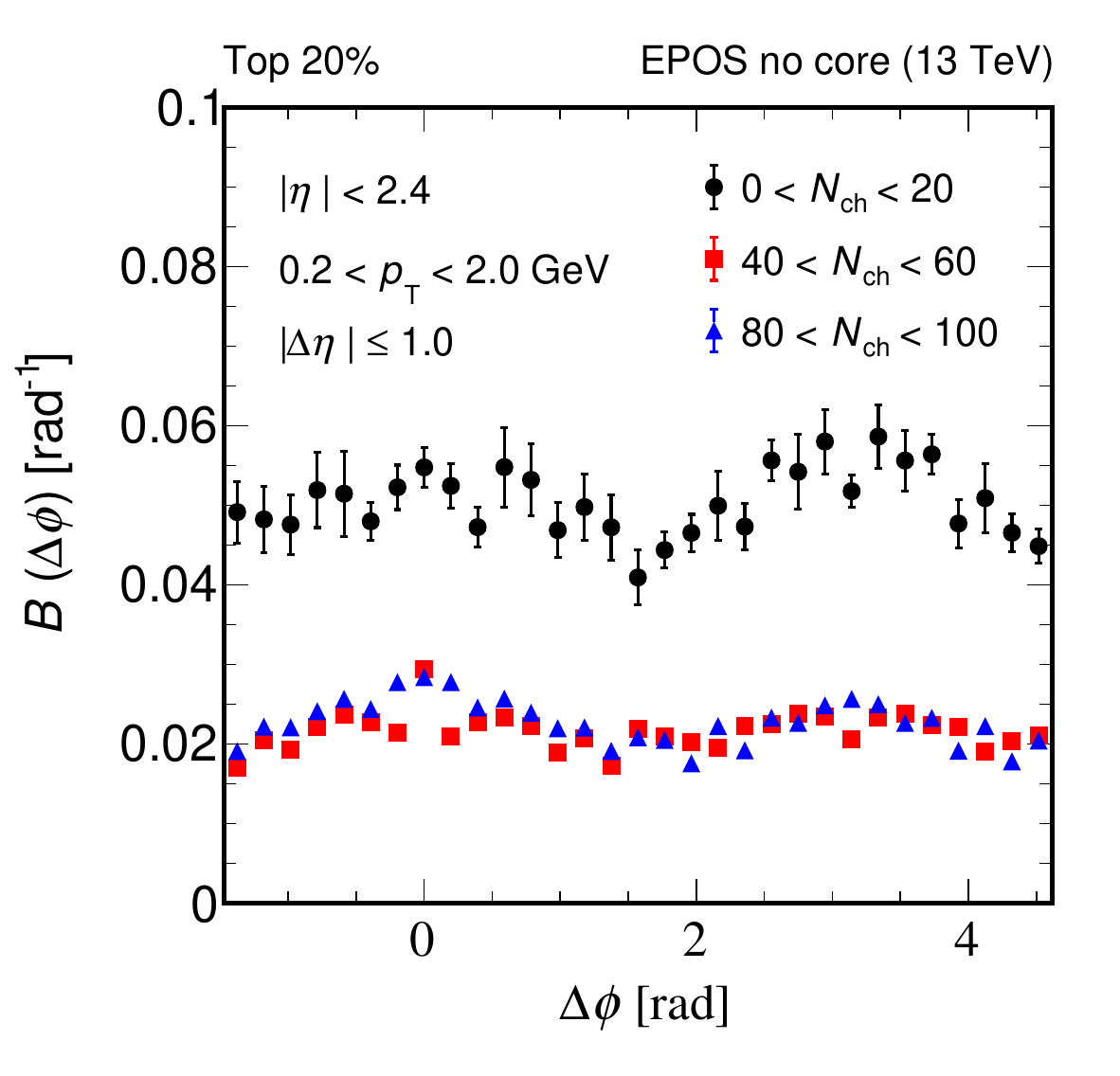}
    }
    \subfigure[]{
        \includegraphics[width=0.3\textwidth]{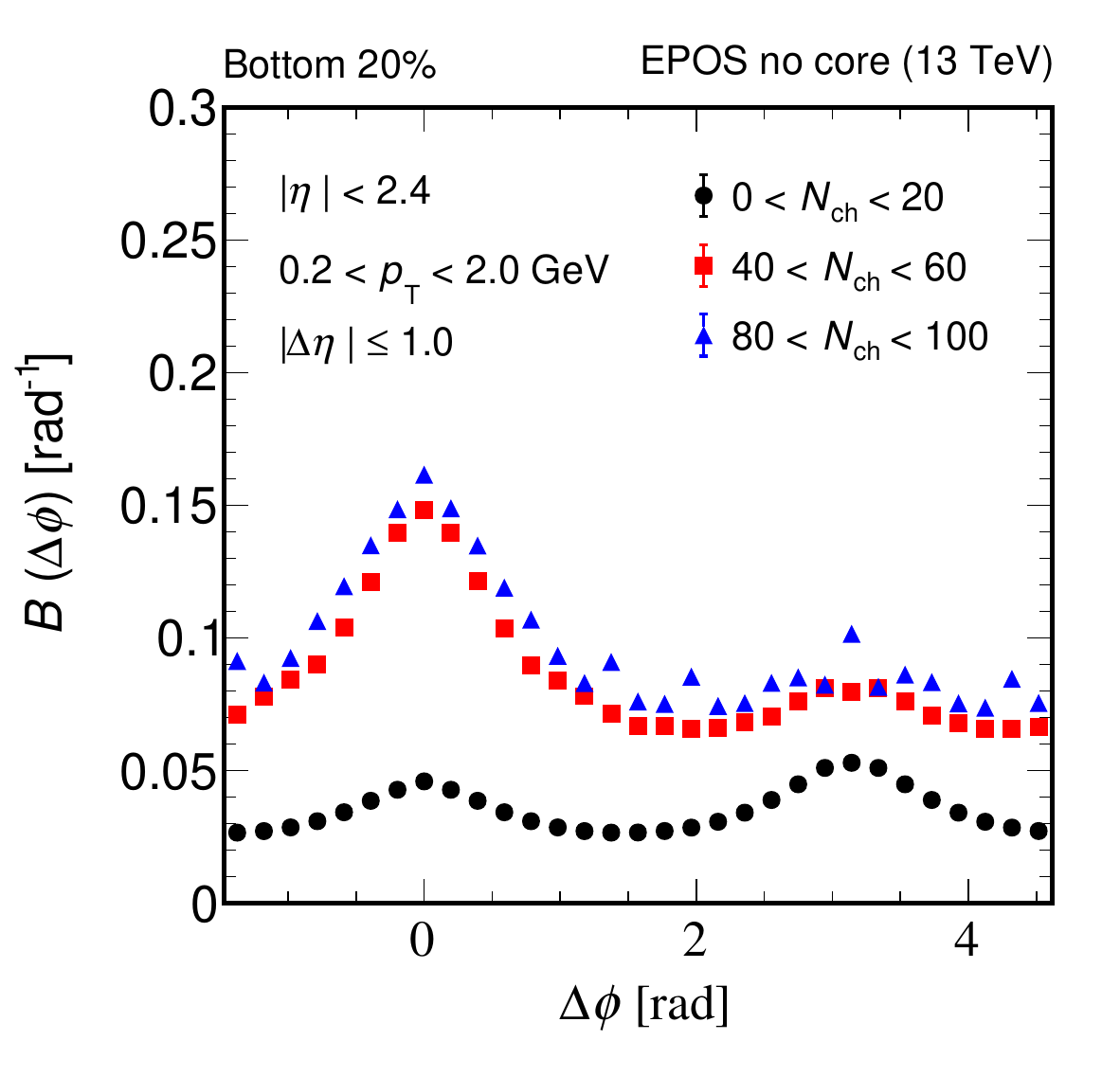}
    }
 \caption{One-dimensional projections of $B$ along $\Delta\eta$ and $\Delta\phi$ from \epos model without core in pp collisions at $\sqrt{s} =$ 13 TeV. The left column is for the integrated spherocity class, the middle column is for the top 20\%, and the right panel is for the bottom 20\% of the spherocity class. $\Delta\eta$ projections are taken in $|\Delta\phi| \leq \pi/2$ and $\Delta\phi$ projections are taken over $|\Delta\eta| \leq 1$ range.}
    \label{fig:model_plots_1d_epos_wcore}
\end{figure*}

\section{Result and discussion}
\label{results}
In this work, for the multiplicity dependence study, the charged particle multiplicity \nch is defined using primary charged particles within $|\eta| < 2.4$ and $p_\mathrm{T} > 0.4$ GeV, following the typical tracking acceptance of the CMS and ATLAS experiments. Figure \ref{fig:sp_epos_pythia} presents the spherocity distributions versus charged-particle multiplicity for (a) \pythia and (b) \epos MC simulations. Both generators exhibit a systematic evolution towards isotropic event configurations with increasing multiplicity. Notably, \epos produces narrower spherocity distributions than \pythia in high-multiplicity regimes, indicating more constrained event topology resulting from the core-corona picture. For this study of balance functions, events are selected for the low (jetty) and high (isotropic) 20\% of spherocity distribution in a given charged-particle multiplicity class.

\subsection{2D correlations}
In this Section, we discuss the two-dimensional number and \pt balance functions in $0 < N_\mathrm{ch}< 20$ and $80 < N_\mathrm{ch}< 100$ intervals from \pythia and \textsc{epos} calculations of pp collisions at $\sqrt{s} = 13$~TeV.   

Figure~\ref{fig:cbf_epos_pythia} presents the plots from the two-dimensional balance functions, $B(\Delta\eta, \Delta\phi)$ obtained using Eq.(\ref{eqn_balfun}). A distinct difference is observed in the correlation structure between the two models. In \epos simulations, the balance functions exhibit a narrow and pronounced peak near $\Delta\eta = 0$ and $\Delta\phi = 0$. This behavior is indicative of strong short-range charge correlations and is consistent with the presence of collectivelike phenomena, such as radial flow, that are present in \epos via hydrodynamic evolution and a core-corona hadronization approach.
Conversely, \pythia shows significantly broader and flatter balance function distributions at the lowest multiplicity. The observed diffused peak structure in both $\Delta\eta$ and $\Delta\phi$ suggests a less constrained charge-balancing mechanism. The collectivelike effects only arise from the interactions of the color strings in \pythia, which seem to have less effect on the width of the balance function as compared to \epos~\cite{Lonnblad:2023kft}.

Similarly, Fig.~\ref{fig:p2_epos_pythia} shows the two-dimensional \pt balance functions, $P_{2}^\mathrm{CD}(\Delta\eta, \Delta\phi)$, obtained using Eq.~(\ref{eqn_balfun_p2}) from \pythia and \epos event generators. A pronounced near-side peak is observed for both models as compared to $B$ for a given \nch bin, indicating that the correlations are dominated by short-range processes.
\begin{figure*}[!ht]  
    \centering
    \subfigure[]{
        \includegraphics[width=0.3\textwidth]{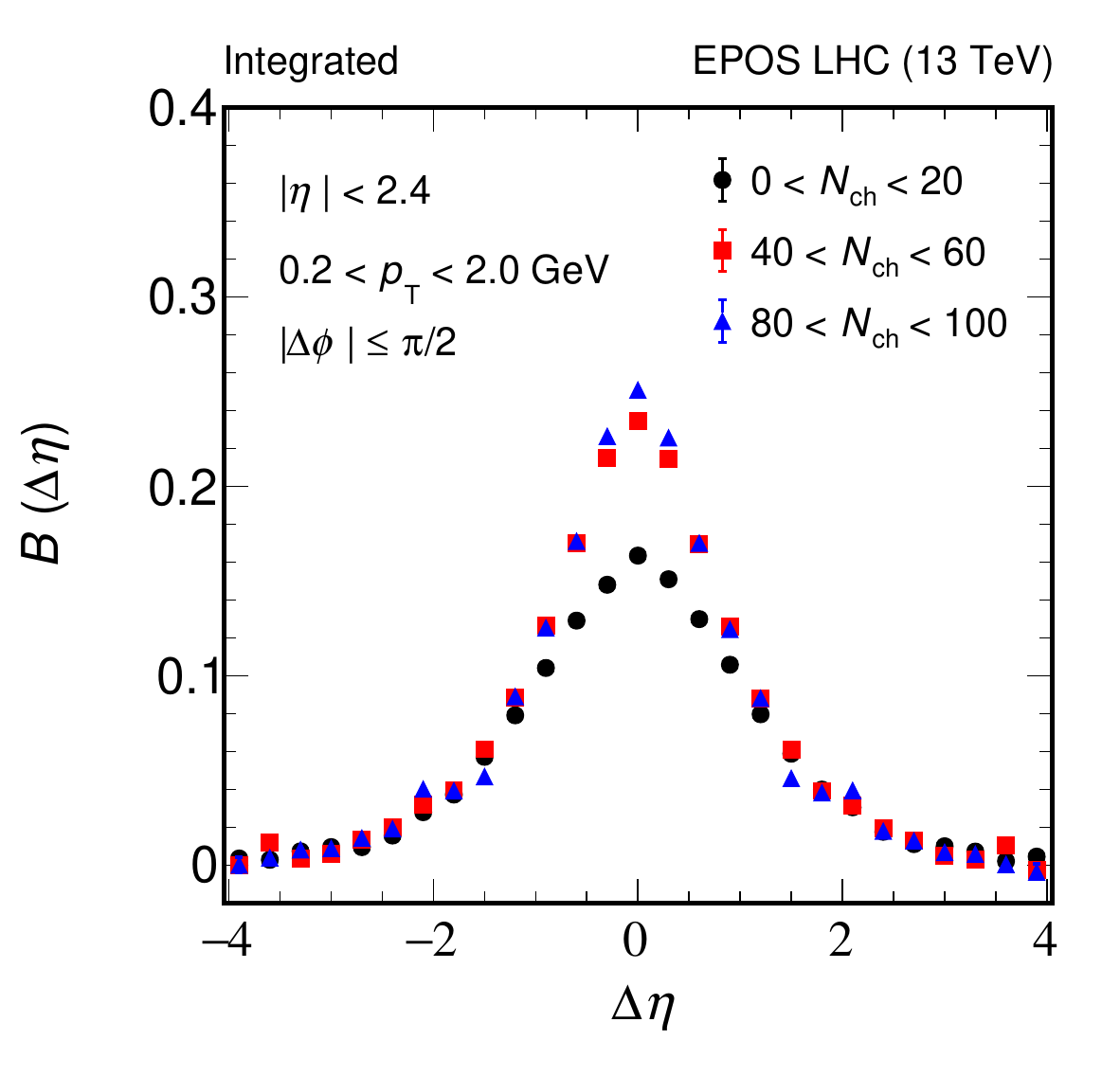}
    }
    \subfigure[]{
        \includegraphics[width=0.3\textwidth]{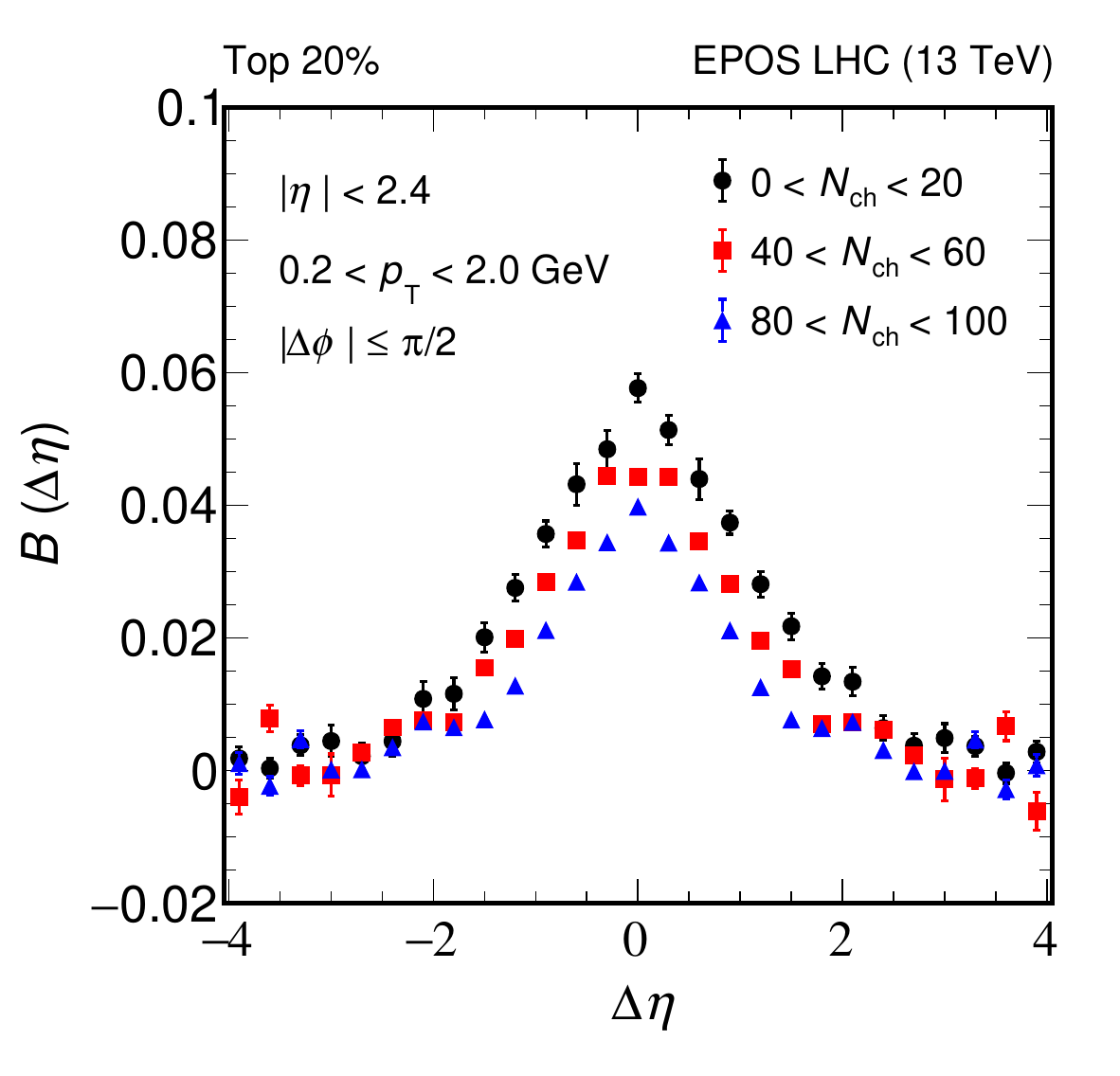}
    }
    \subfigure[]{
        \includegraphics[width=0.3\textwidth]{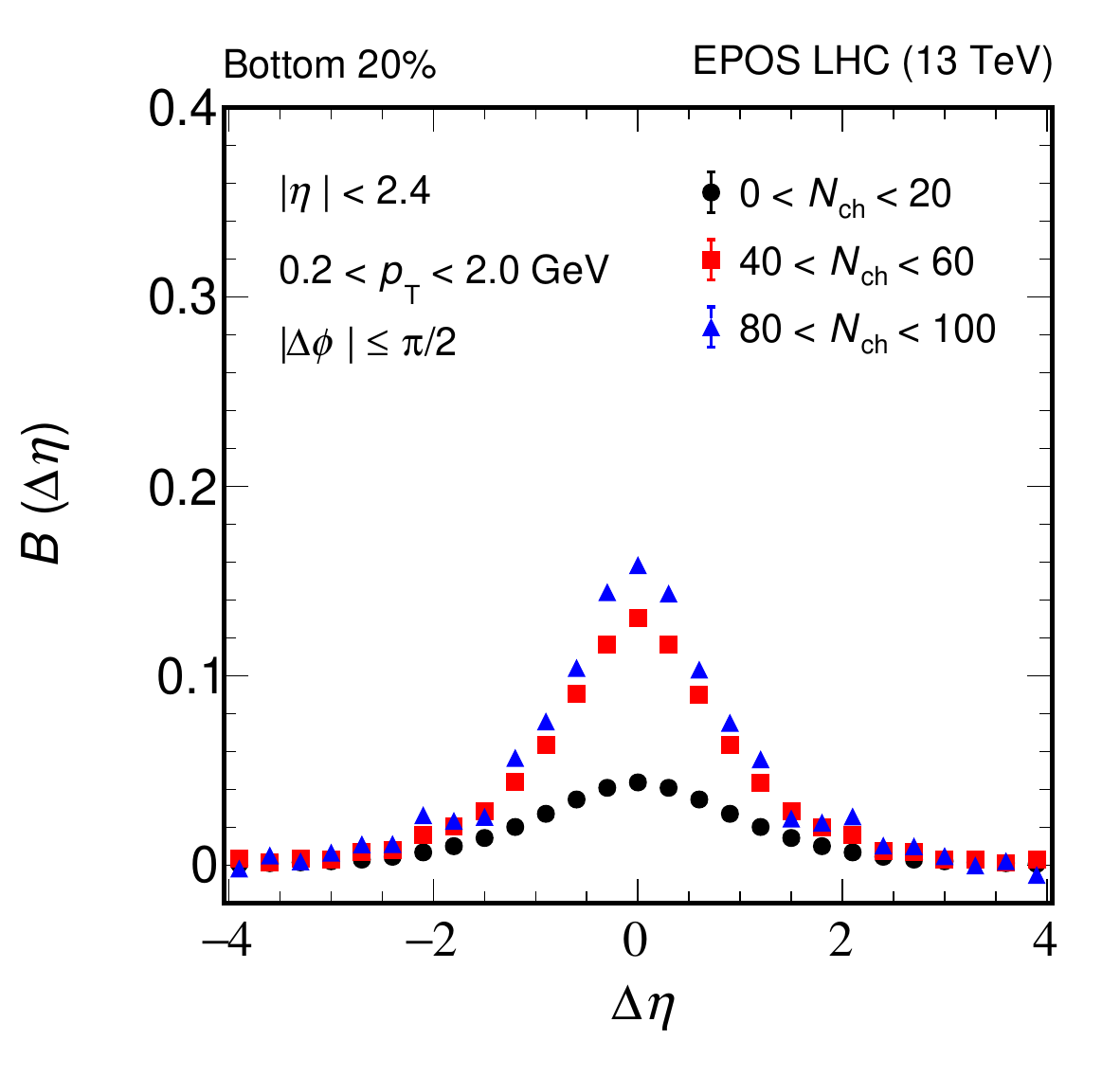}
    }
    \subfigure[]{
        \includegraphics[width=0.3\textwidth]{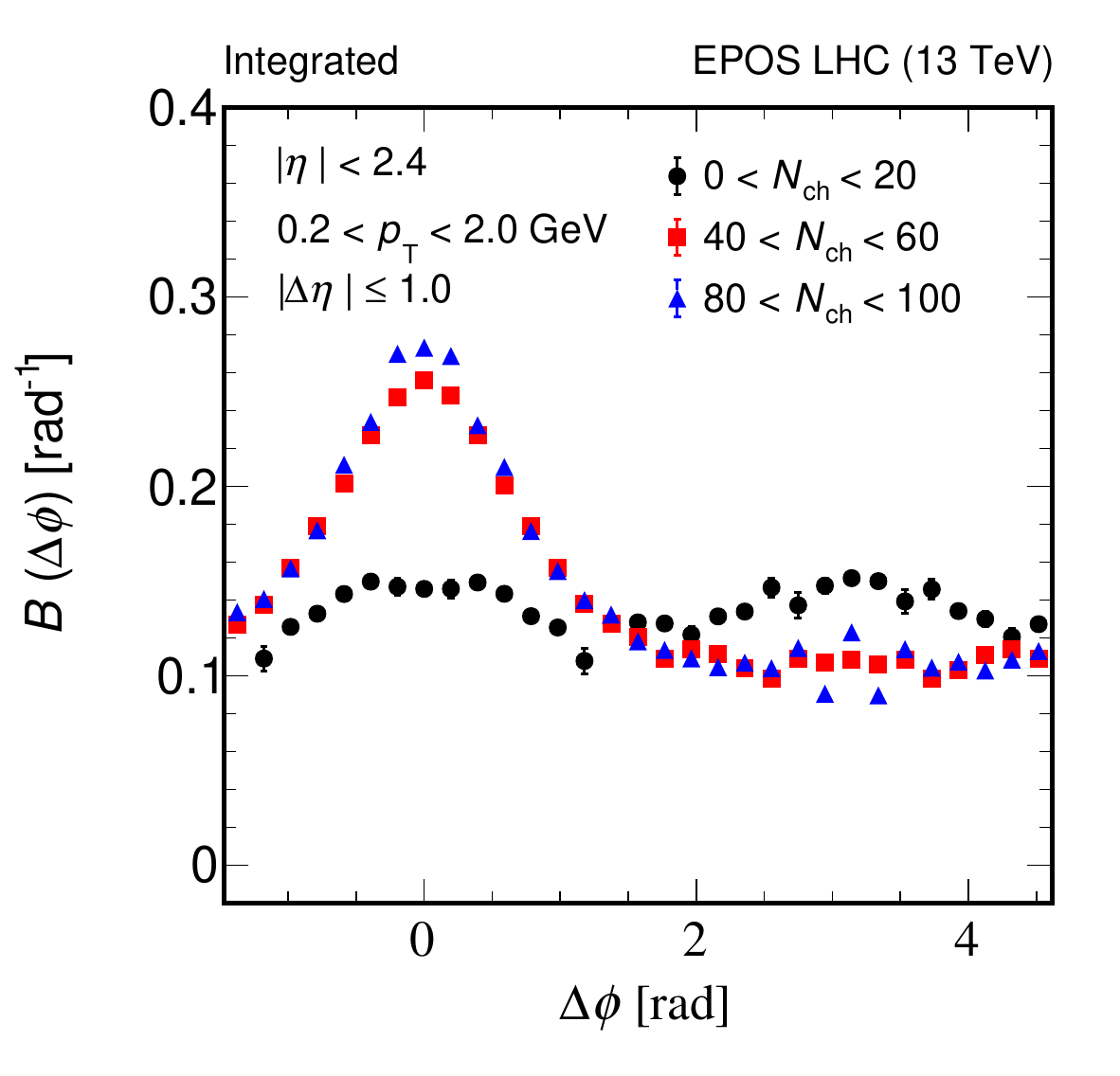}
    }
    \subfigure[]{
        \includegraphics[width=0.3\textwidth]{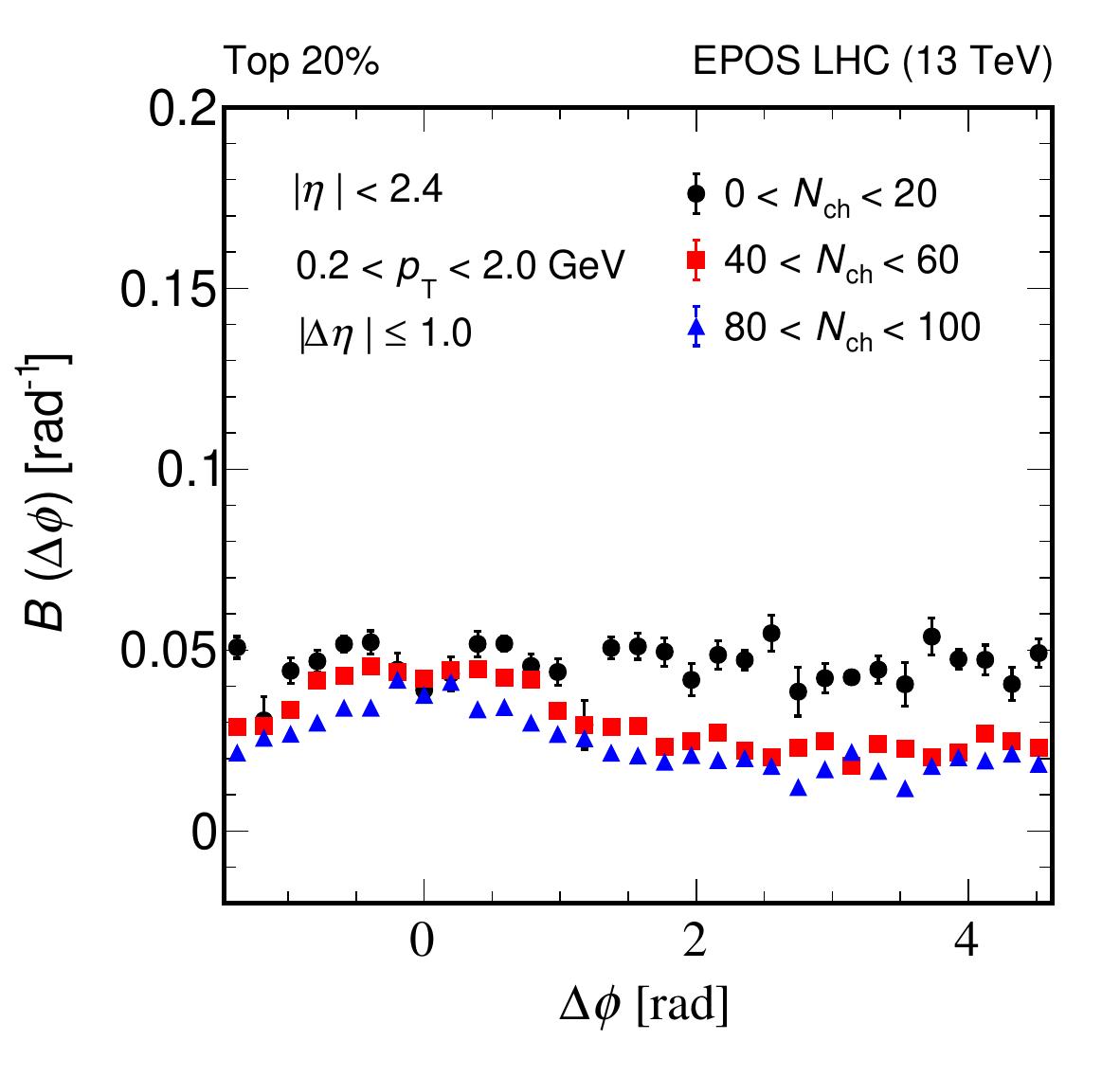}
    }
    \subfigure[]{
        \includegraphics[width=0.3\textwidth]{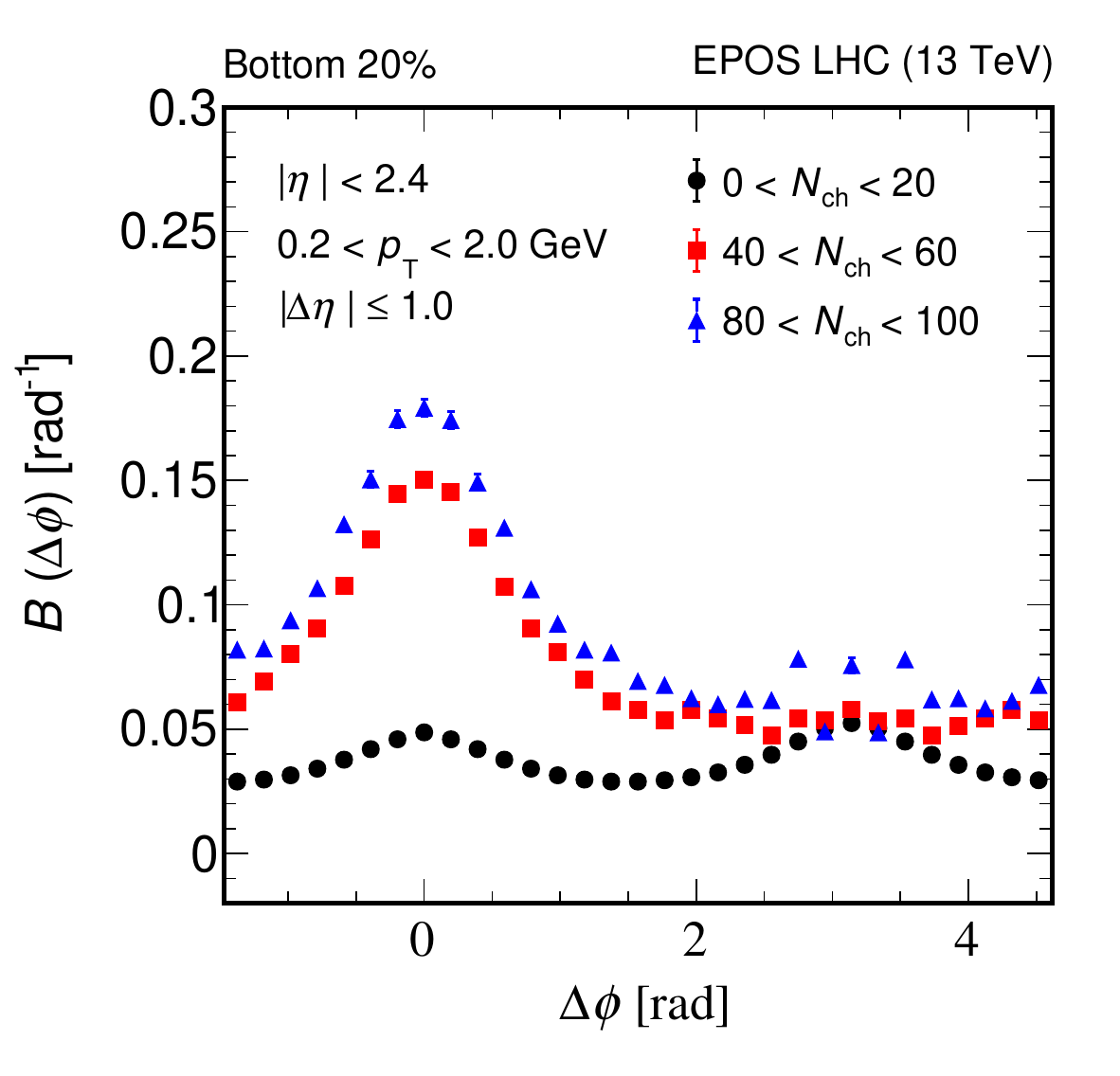}
    }
 \caption{One-dimensional projections of $B$ along $\Delta\eta$ and $\Delta\phi$ from \epos model in pp collisions at $\sqrt{s} =$ 13 TeV. The left column is for the integrated spherocity class, the middle column is for the top 20\%, and the right panel is for the bottom 20\% of the spherocity class. $\Delta\eta$ projections are taken in $|\Delta\phi| \leq \pi/2$ and $\Delta\phi$ projections are taken in $|\Delta\eta| \leq 1$ range.}
    \label{fig:model_plots_1d_epos}
\end{figure*}

\begin{figure*}[!htbp]  
    \centering
    \subfigure[]{
        \includegraphics[width=0.35\textwidth]{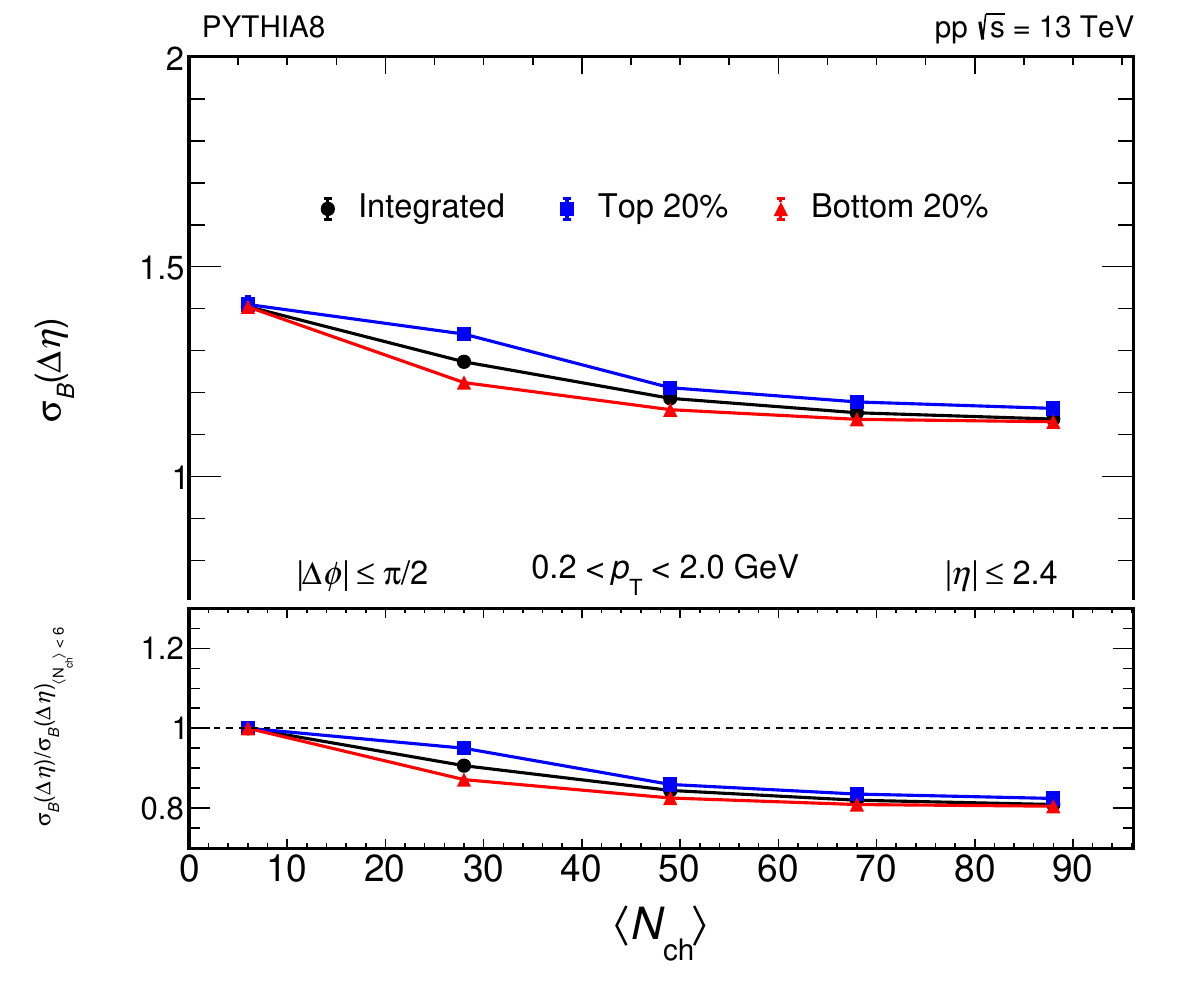}
    }
    \subfigure[]{
        \includegraphics[width=0.35\textwidth]{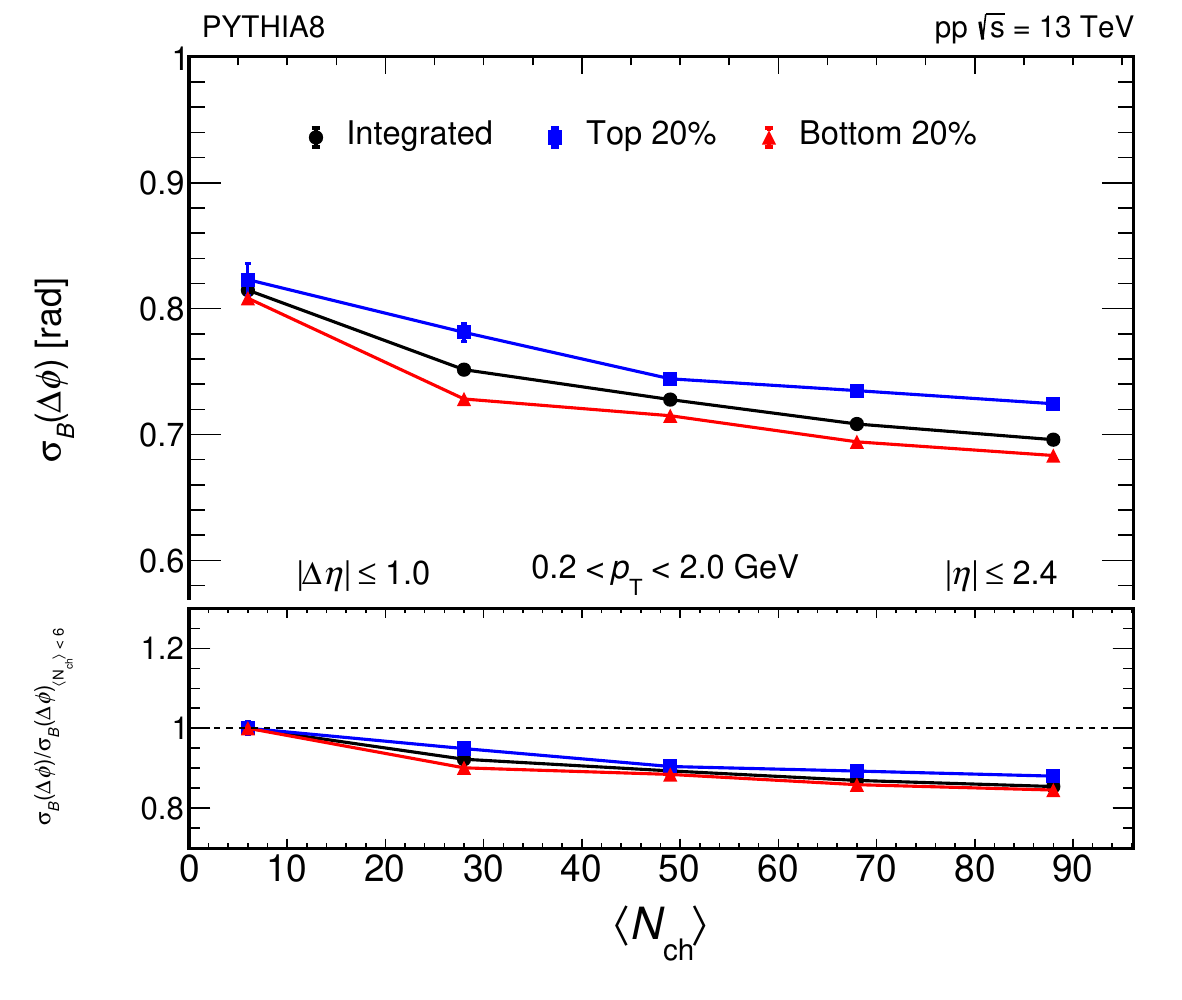}
    }
    \subfigure[]{
        \includegraphics[width=0.35\textwidth]{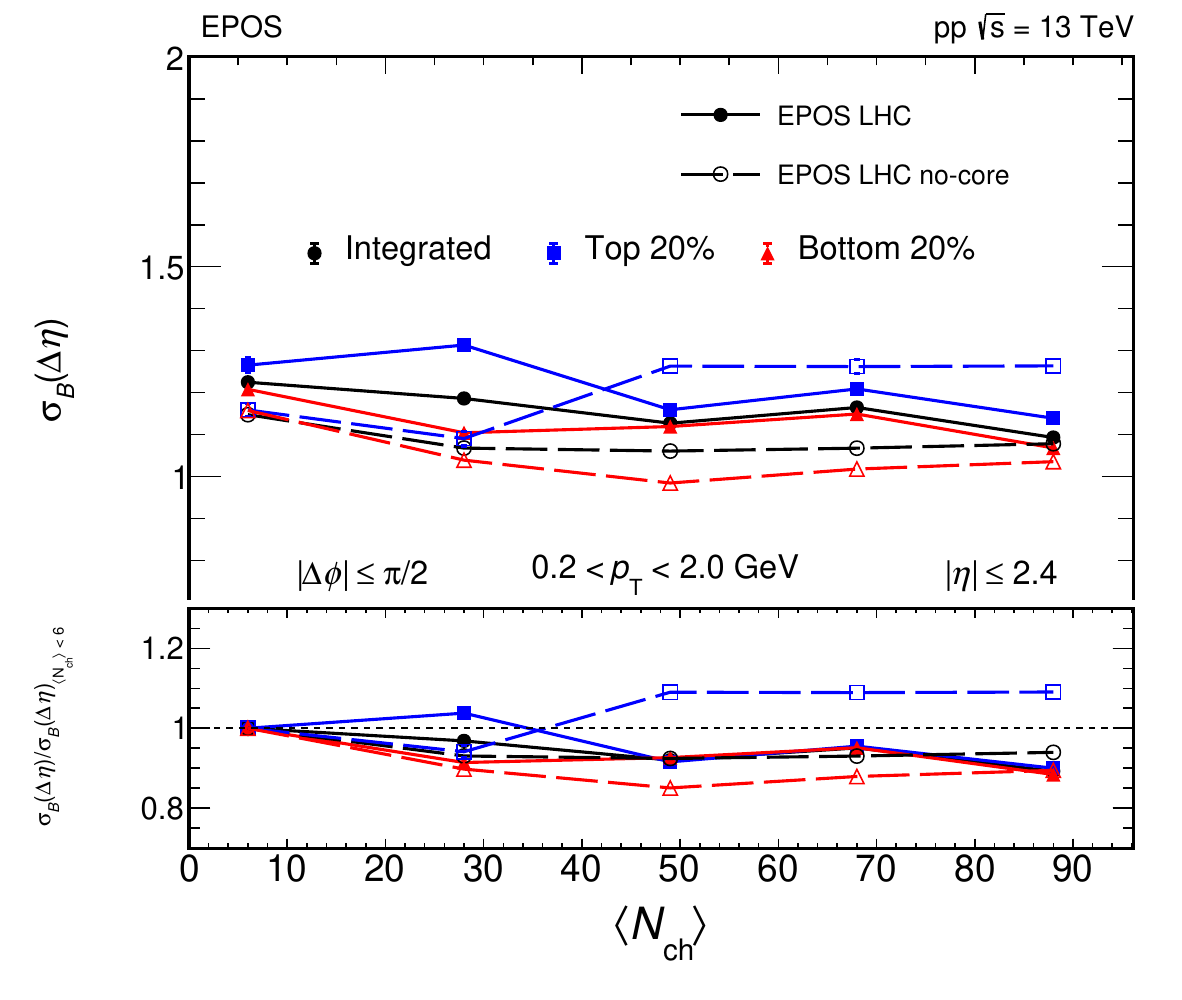} 
    }
    \subfigure[]{
        \includegraphics[width=0.35\textwidth]{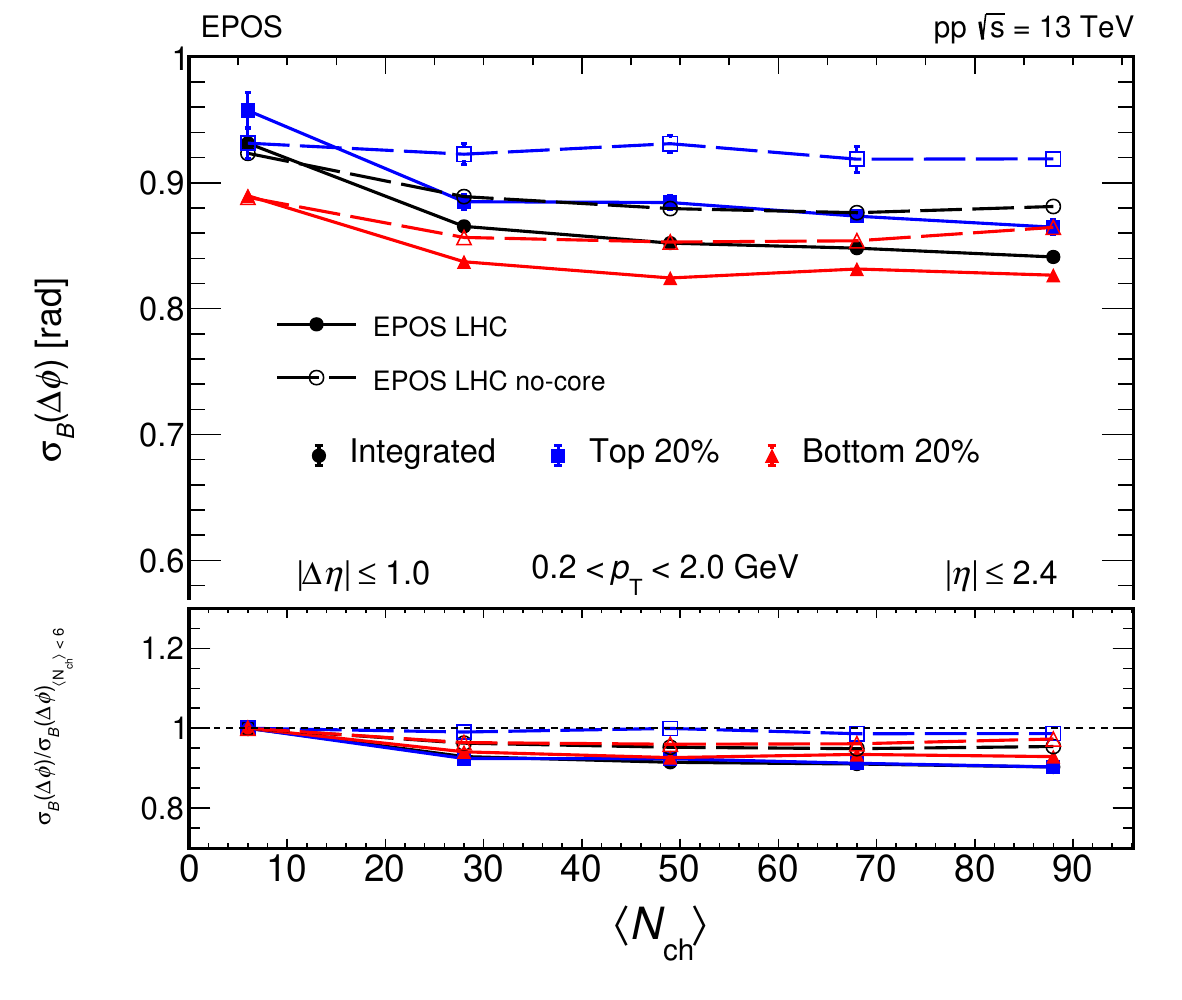}
    }
\caption{The width of $B$ in terms of $\Delta\eta$ (left) and $\Delta\phi$ (right) for \pythia and  \epos in pp collisions at $\sqrt{s} =$ 13 TeV as a function of $\langle N_{\rm ch}\rangle$ and spherocity classes. The top panel shows the rms width and the bottom panel presents the width normalized to its value in the lowest multiplicity interval ($N_{\mathrm{ch}} < 6$). The widths are evaluated within $0.2 < p_{\mathrm{T}} < 2.0$~GeV interval and pseudorapidity acceptance $|\eta| \leq 2.4$.}
    \label{fg:widthModel}
\end{figure*}
\subsection{One-dimensional projections and widths comparison}
Figures~\ref{fig:model_plots_1d},~\ref{fig:model_plots_1d_epos_wcore}, and~\ref{fig:model_plots_1d_epos} present the one-dimensional projections of the charge balance function as a function of pseudorapidity and azimuthal angle, analyzed across different spherocity classes and multiplicity intervals. The figures correspond to different event generator configurations: Fig. ~\ref{fig:model_plots_1d} shows results from \pythia, Figure ~\ref{fig:model_plots_1d_epos_wcore}  displays \epos without the hydrodynamic core, and Fig. ~\ref{fig:model_plots_1d_epos} features \epos calculations with the core enabled. For \pythia, jet-like (bottom 20\%) events consistently produce narrower balance functions than isotropic (top 20\%) events, reflecting more localized charge balancing within jet structures. In contrast, isotropic events, dominated by soft MPI, display broader distributions. These trends are seen in both $\Delta\eta$ and $\Delta\phi$ projections, and are further enhanced at higher multiplicity, emphasizing the role of event topology and particle density in shaping charge-dependent correlations. Figures~\ref{fig:model_plots_1d_epos_wcore} and \ref{fig:model_plots_1d_epos} show the evolution of $B$ without and with the choice of core in the event generation. For $B(\Delta\eta)$ (top panels), when the core is absent, the near-side peak at $\Delta\eta = 0$ appears significantly broader than the corresponding peak in $\Delta\phi$. This broadening is attributed to longitudinal string fluctuations, which cause balancing charges to be more widely separated in pseudorapidity. In contrast, when the core is included, $B(\Delta\phi)$ (bottom panels) exhibits a much sharper near-side peak, indicating that azimuthal correlations between balancing pairs are more localized and less affected by longitudinal diffusion. Notably, the near-side peak in $\Delta\phi$ becomes even narrower and more pronounced in the highest multiplicity class, where the influence of the core is strongest. The suppression of the away-side structure in the top-20\% selection indicates a reduced back-to-back jet contribution relative to the low-multiplicity case, where such correlations remain more pronounced. The distributions are relatively flat at the low-\nch class, indicating a significant dilution of short-range charge correlations in pseudorapidity, which may be related to the longitudinal expansion of the particle-emitting source. Notably, \epos shows a double-peak structure with smaller magnitude at $\Delta\phi \approx 0$ and $\Delta\phi \approx \pi$ than \epos without core and \pythia, particularly in low multiplicity events. This is consistent with the idea that \epos includes core–corona separation, where jetty events can have a hydrodynamic core.


To investigate the interplay between event topology and charge-dependent correlations, the widths of the charge balance function in relative pseudorapidity ($\sigma _{B}(\Delta\eta), \sigma_{P_{2}^\mathrm{CD}}(\Delta\eta)$) and relative azimuthal angle ($\sigma _{B}(\Delta\phi), \sigma_{P_{2}^\mathrm{CD}}(\Delta\phi)$)) are studied as a function of charged-particle multiplicity for three spherocity event classes.

\begin{figure*}[!ht]  
    \centering
    \subfigure[]{
        \includegraphics[width=0.3\textwidth]{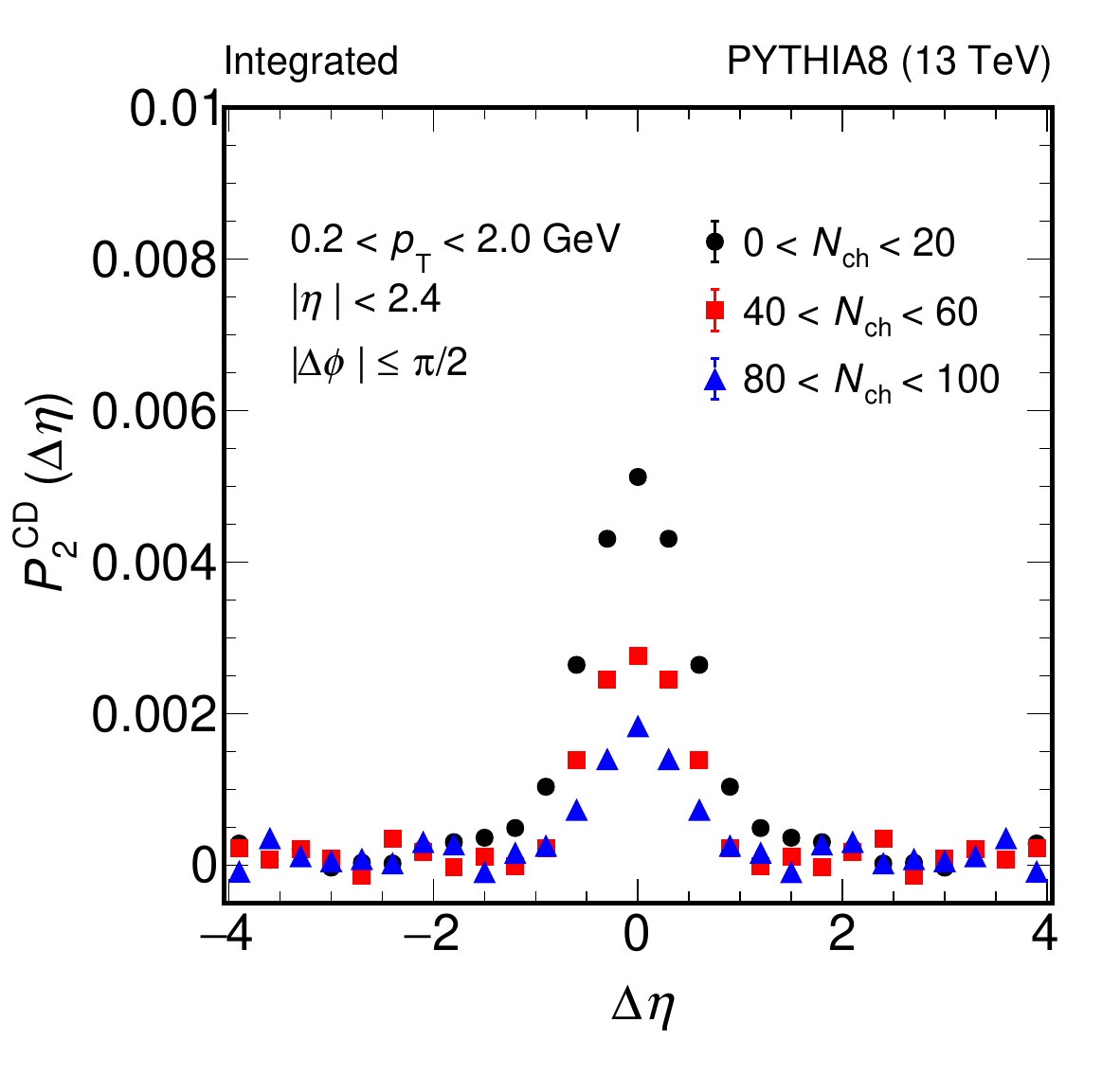}
    }
    \subfigure[]{
        \includegraphics[width=0.3\textwidth]{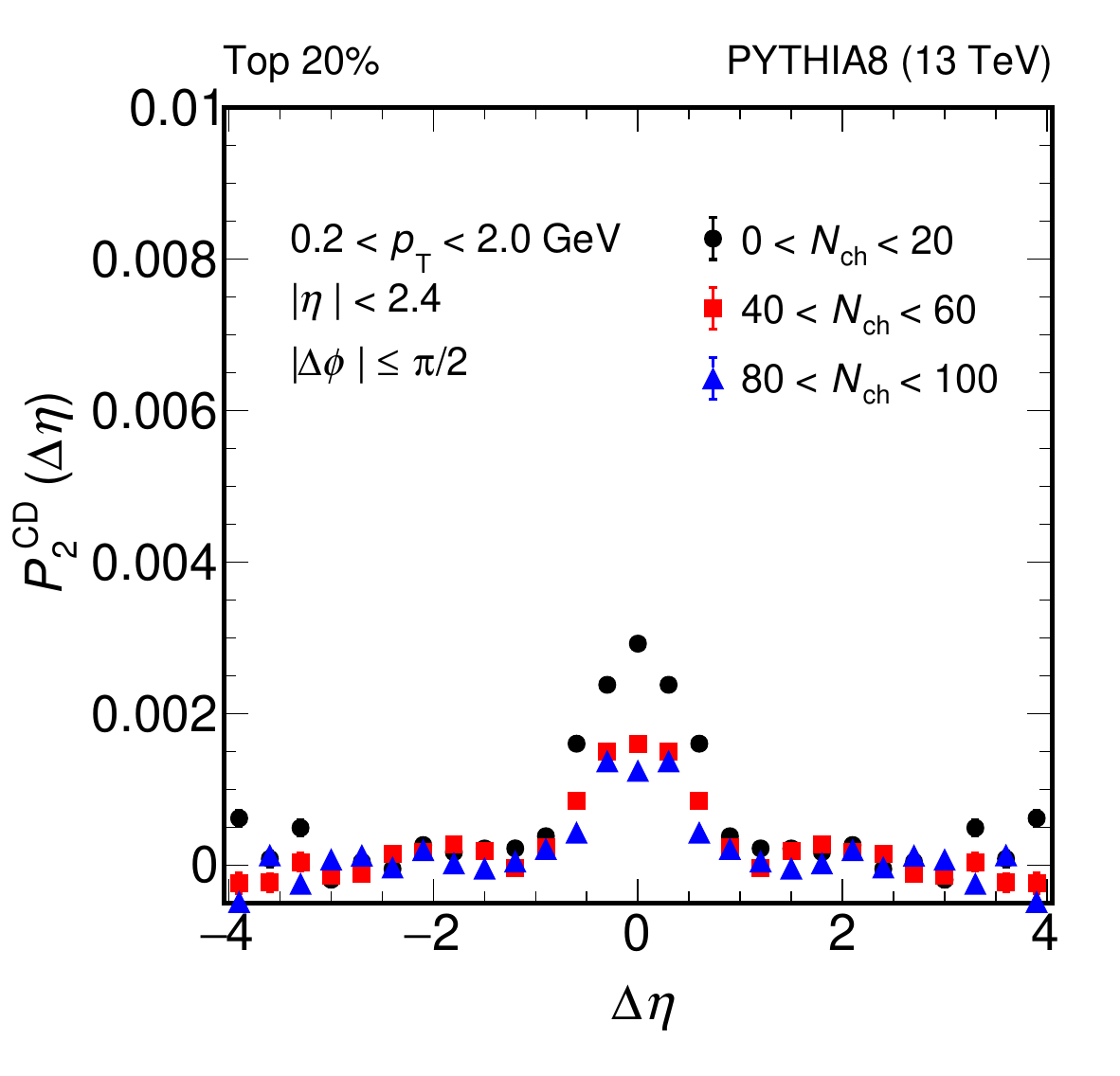}
    }
    \subfigure[]{
        \includegraphics[width=0.3\textwidth]{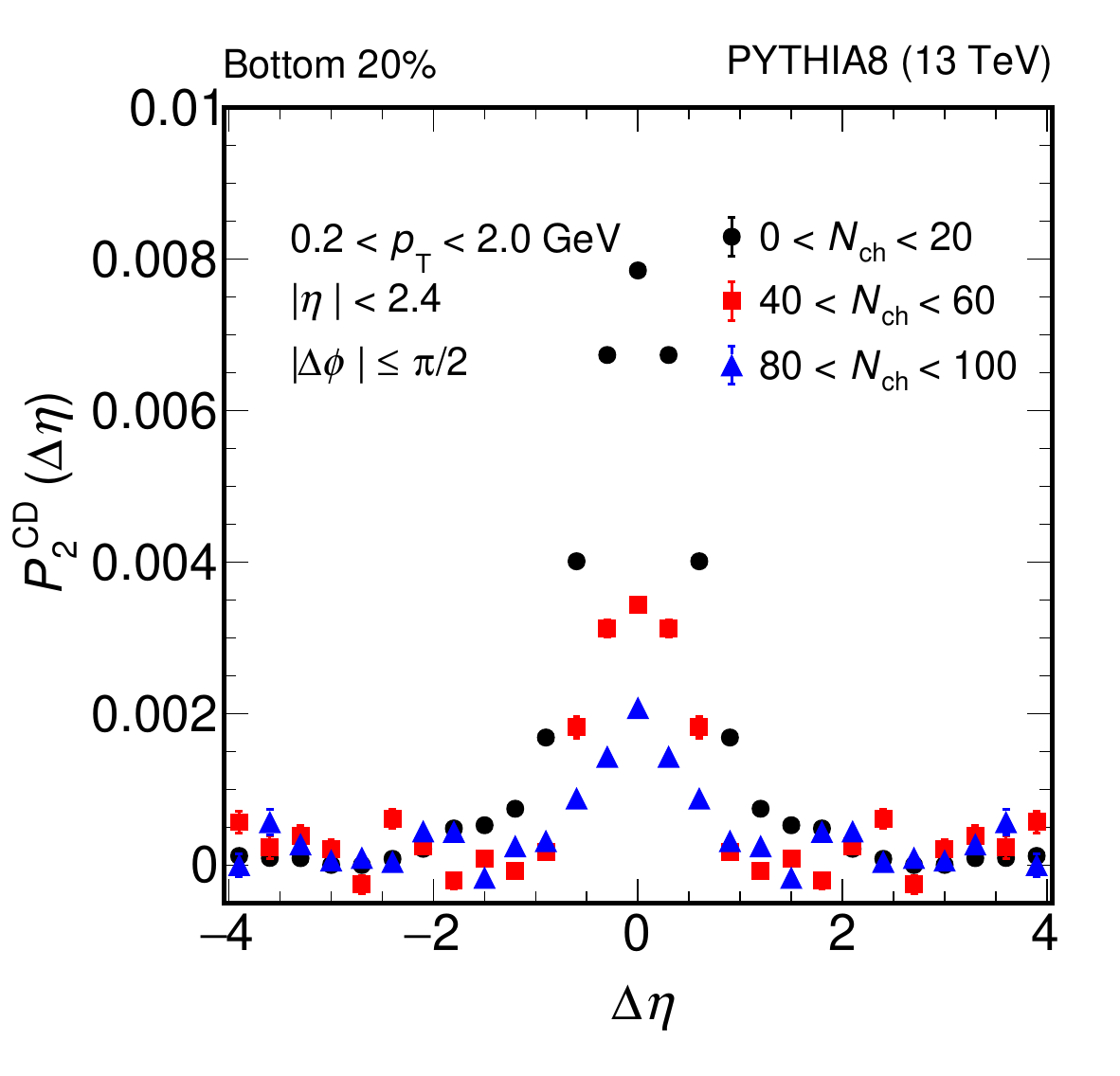}
    }
    \subfigure[]{
        \includegraphics[width=0.3\textwidth]{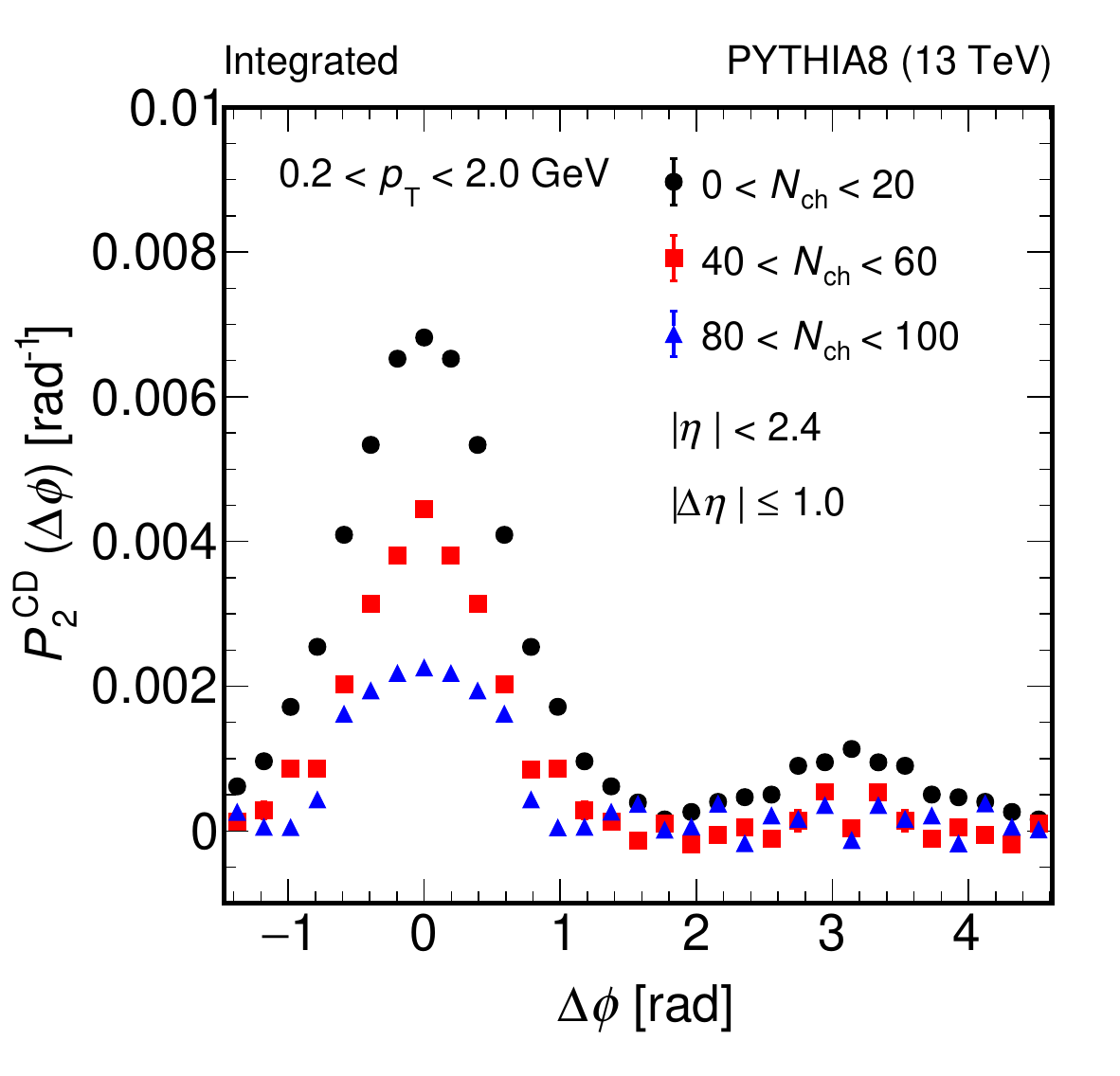}
    }
    \subfigure[]{
        \includegraphics[width=0.3\textwidth]{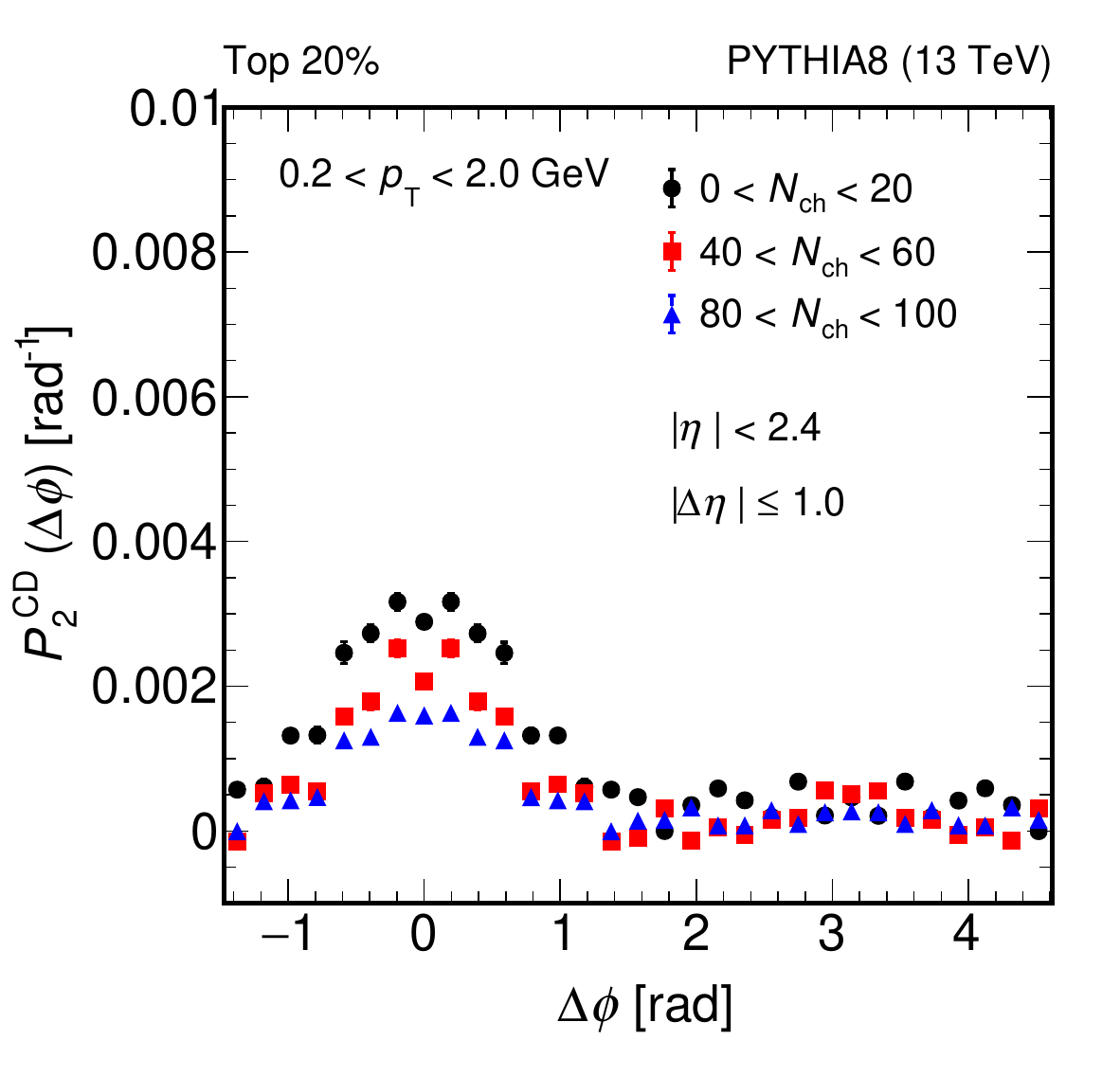}
    }
    \subfigure[]{
        \includegraphics[width=0.3\textwidth]{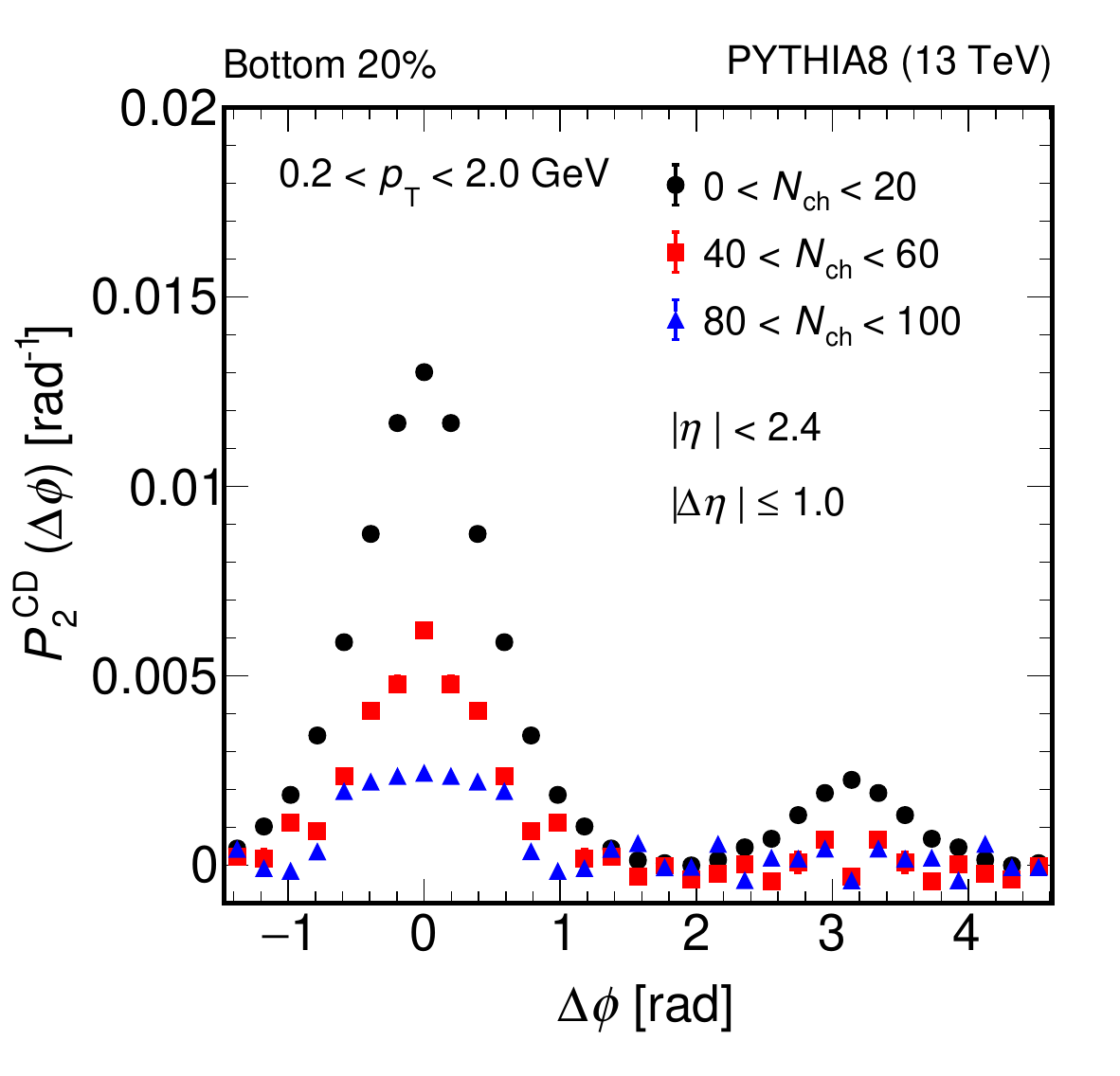}
    }

 \caption{One-dimensional $P_{2}^\mathrm{CD}$ projections along $\Delta\eta$ and $\Delta\phi$ from \pythia model in pp collisions at $\sqrt{s} =$ 13 TeV. The left column is for the integrated spherocity class, the middle column is for the top 20\%, and the right panel is for the bottom 20\% of the spherocity class. $\Delta\eta$ projections are taken in $|\Delta\phi| \leq \pi/2$ and $\Delta\phi$ projections are taken in $|\Delta\eta| \leq 1$ range.}
    \label{fig:1d_pythia_p2cd}
\end{figure*}

Figure~\ref{fg:widthModel} shows the width of number balance functions as a function of $\langle N_{\rm ch}\rangle$  in pp collisions, where $\langle \nch \rangle$ denotes the average charged-particle multiplicity within each multiplicity class. The bottom panels show the width normalized to its value in the lowest multiplicity class, highlighting the relative evolution with $\langle \nch \rangle$. The widths of the charge balance function in both $\Delta\eta$ and $\Delta\phi$ exhibit a strong dependence on charged-particle multiplicity and event topology. In \pythia, both widths decrease steadily with multiplicity, reflecting enhanced correlations among balancing charges in denser events. This narrowing is largely driven by the combined effects of MPI, which increases the density of soft partons, and CR, which reshuffles string topologies to produce more localized hadronization~\cite{OrtizVelasquez:2013ofg}. The reduction in $\Delta\eta$ width indicates that opposite charge pairs are produced more locally in longitudinal phase space, consistent with stronger constraints from local charge conservation and MPI-driven particle production at high multiplicity. By contrast, the narrowing in $\Delta\phi$ reflects azimuthal collimation, which can originate from jet fragmentation in jetty events and CR-induced radial flowlike collectivity in isotropic events~\cite{Ortiz:2020rwg, OrtizVelasquez:2013ofg}. This distinction becomes clearer in \epos: without the hydrodynamic core, the $\Delta\eta$ width remains nearly flat with multiplicity and $\Delta\phi$ shows a little change, while the inclusion of the core leads to a pronounced multiplicity dependence in $\Delta\phi$, highlighting the role of hydrodynamic flow in tightening azimuthal correlations. Event-shape dependence adds further sensitivity: jetty events consistently show narrower widths in both $\Delta\eta$ and $\Delta\phi$ due to collimated jet fragmentation, whereas isotropic events display broader structures, compatible with collectivelike expansion. In $\Delta \phi$, isotropic events exhibit broader balance-function widths than jetlike events, which at high multiplicity accounts for approximately 5\%-7\% larger widths for the isotropic events. 

\begin{figure*}[!hth]  
    \centering
    \subfigure[]{
        \includegraphics[width=0.3\textwidth]{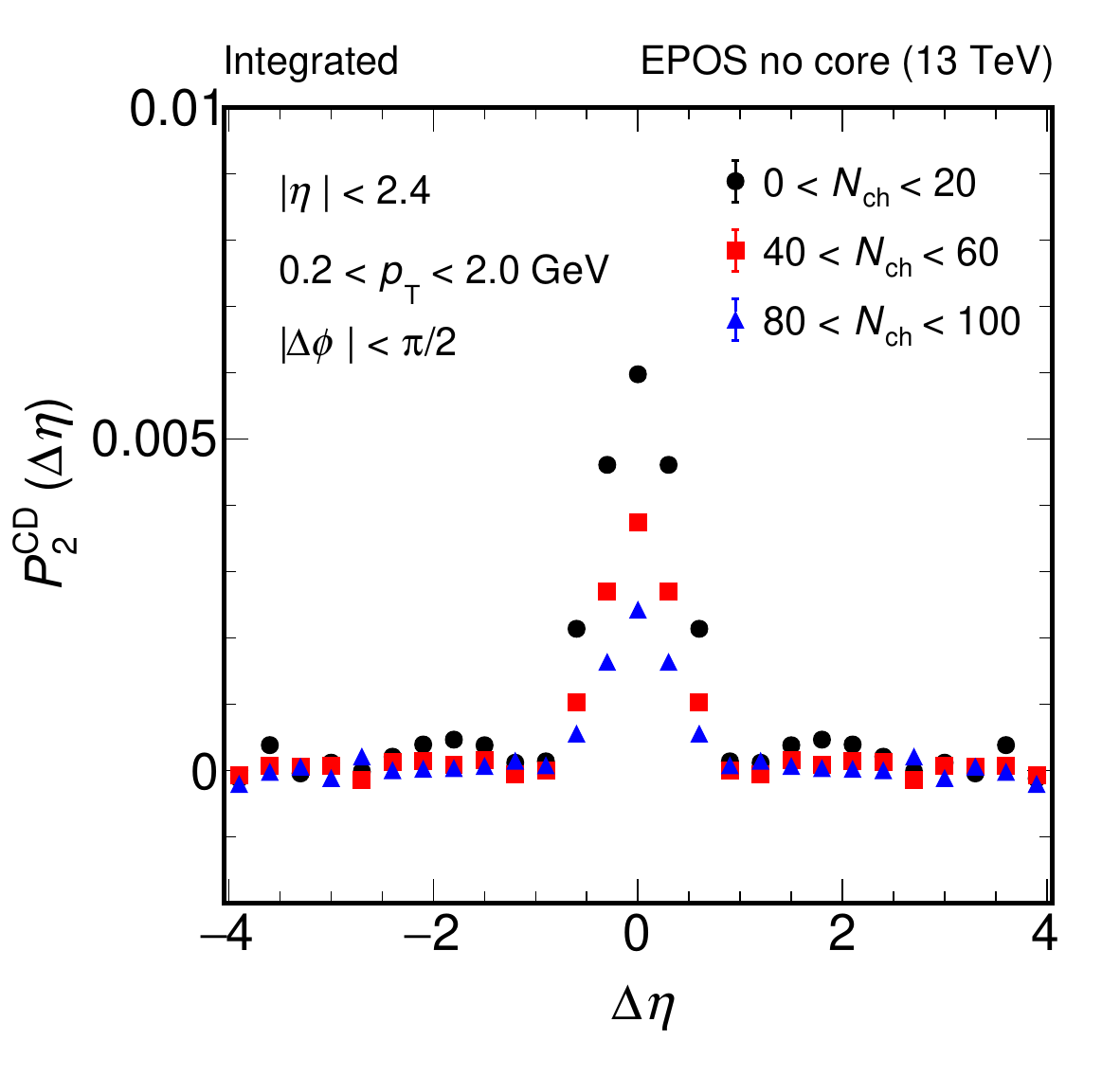}
    }
    \subfigure[]{
        \includegraphics[width=0.3\textwidth]{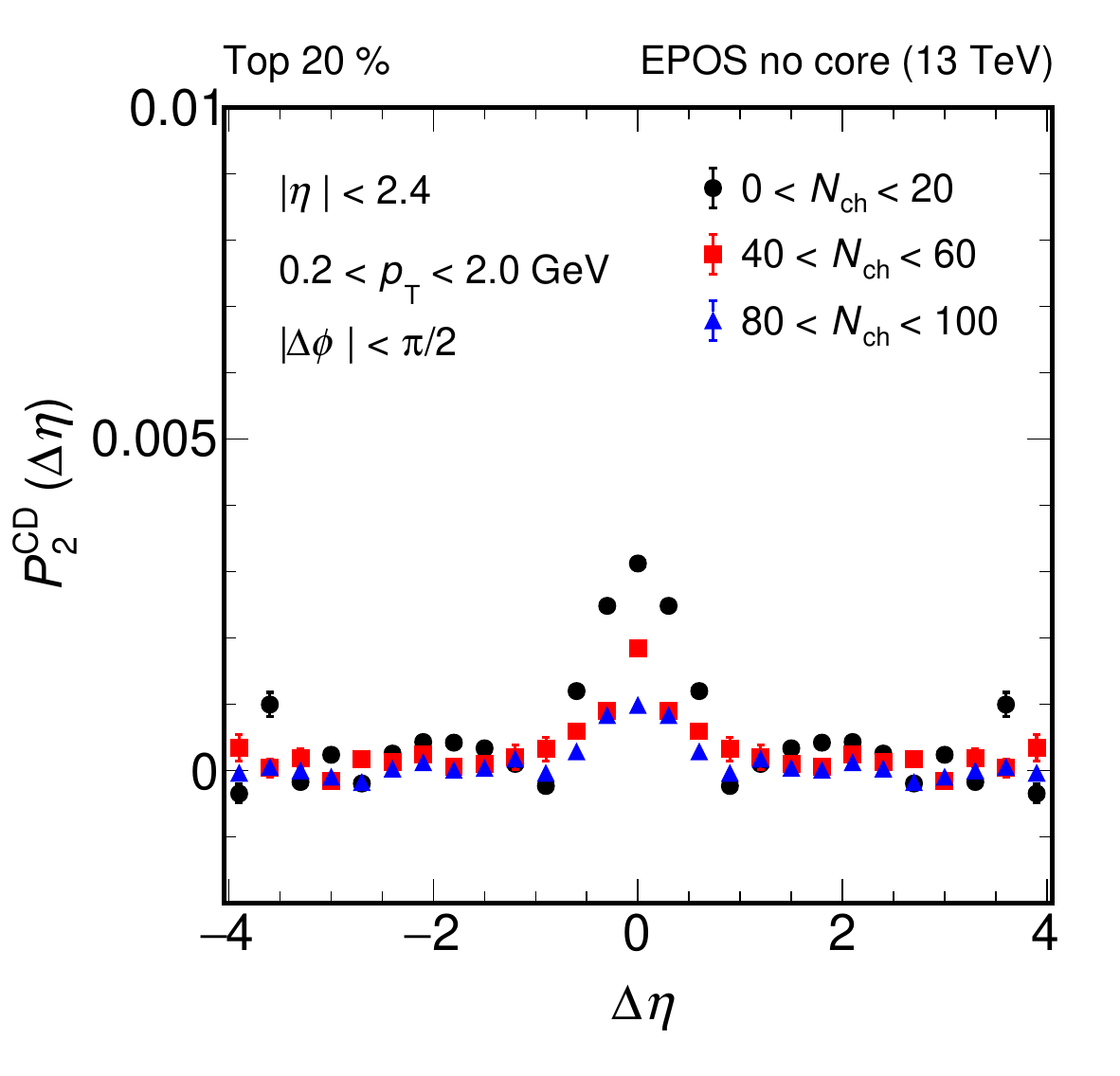}
    }
    \subfigure[]{
        \includegraphics[width=0.3\textwidth]{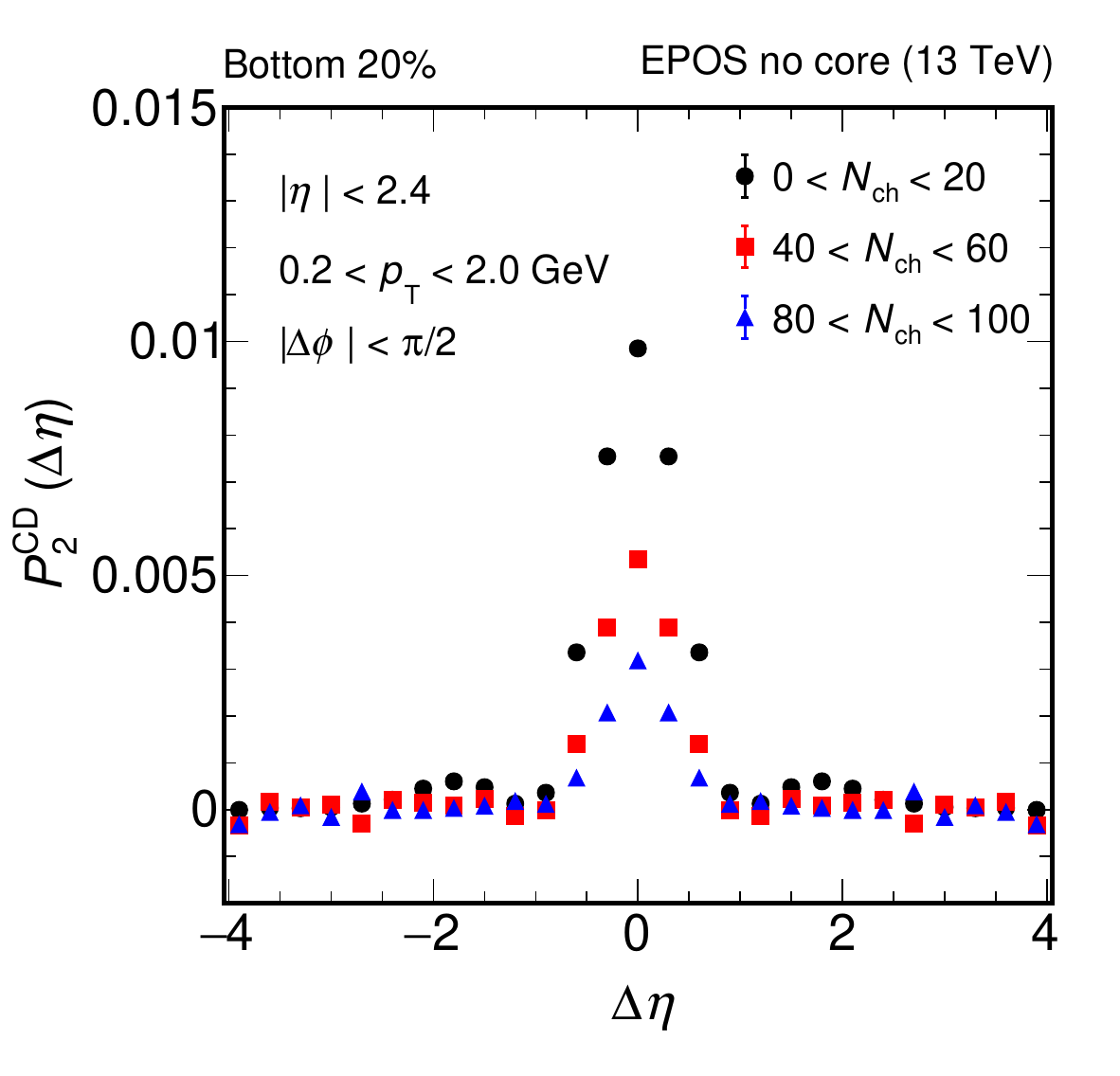}
    }
    \subfigure[]{
        \includegraphics[width=0.3\textwidth]{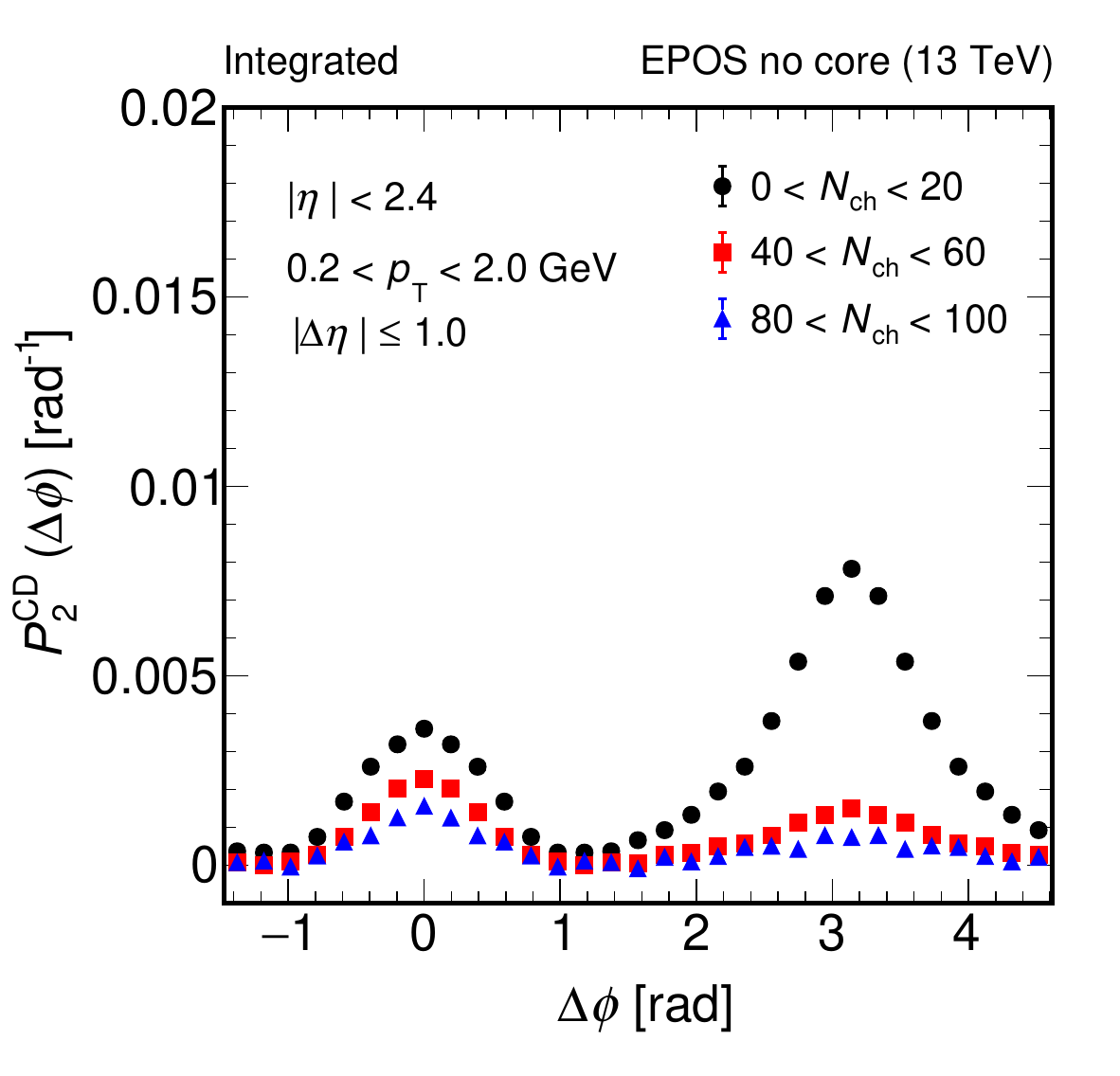}
    }
    \subfigure[]{
        \includegraphics[width=0.3\textwidth]{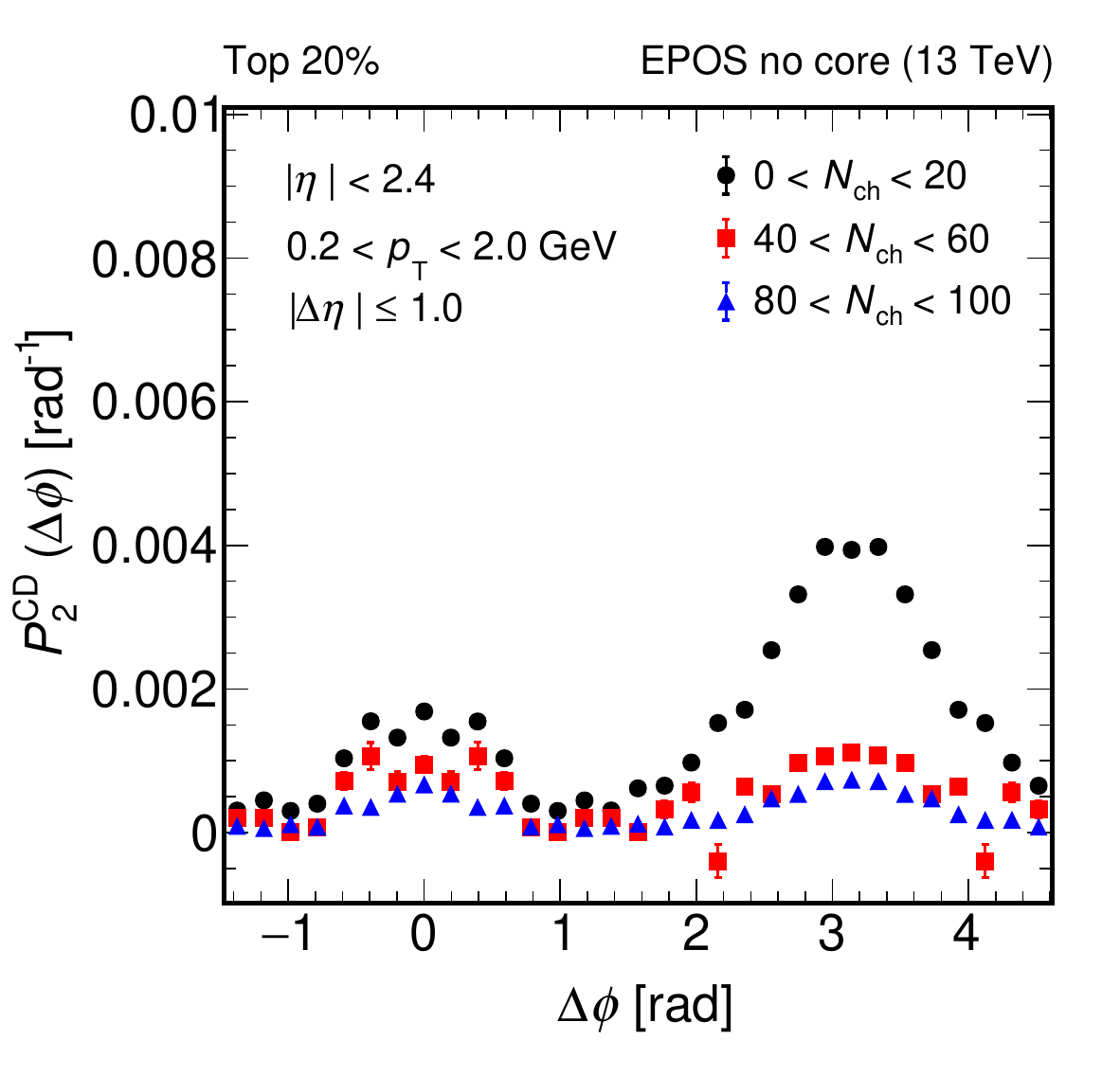}
    }
    \subfigure[]{
        \includegraphics[width=0.3\textwidth]{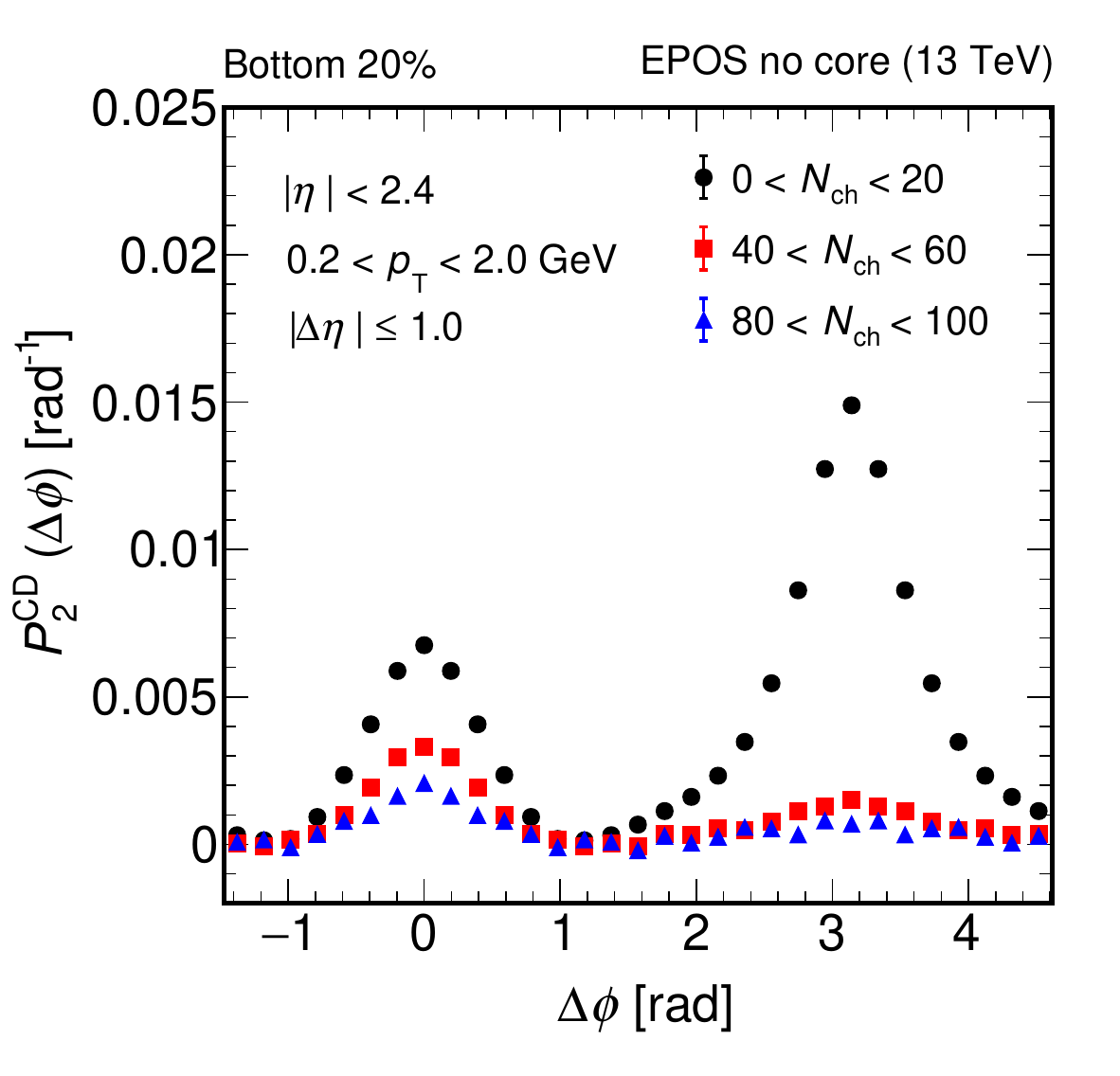}
    }

 \caption{One-dimensional projections of $P_{2}^\mathrm{CD}$ along $\Delta\eta$ and $\Delta\phi$ from \textsc{EPOS} model without core in pp collisions at $\sqrt{s} =$ 13 TeV. The left column is for the integrated spherocity class, the middle column is for the top 20\%, and the right panel is for the bottom 20\% of the spherocity class. $\Delta\eta$ projections are taken in $|\Delta\phi| \leq \pi/2$ and $\Delta\phi$ projections are taken in $|\Delta\eta| \leq 1$ range.}
    \label{fig:1d_epos_p2cd_nocore}
\end{figure*}

Taken together, these results indicate that $\Delta\eta$ narrowing is primarily sensitive to longitudinal charge conservation and MPI, while $\Delta\phi$ narrowing probes radial flowlike dynamics induced by CR or hydrodynamic evolution~\cite{Ortiz:2020rwg, OrtizVelasquez:2013ofg}. This underscores the capability of balance functions, when analyzed in event-shape classes, to disentangle the microscopic mechanisms of hadronization and collective effects in small systems.

\begin{figure*}[!ht]  
    \centering
    \subfigure[]{
        \includegraphics[width=0.3\textwidth]{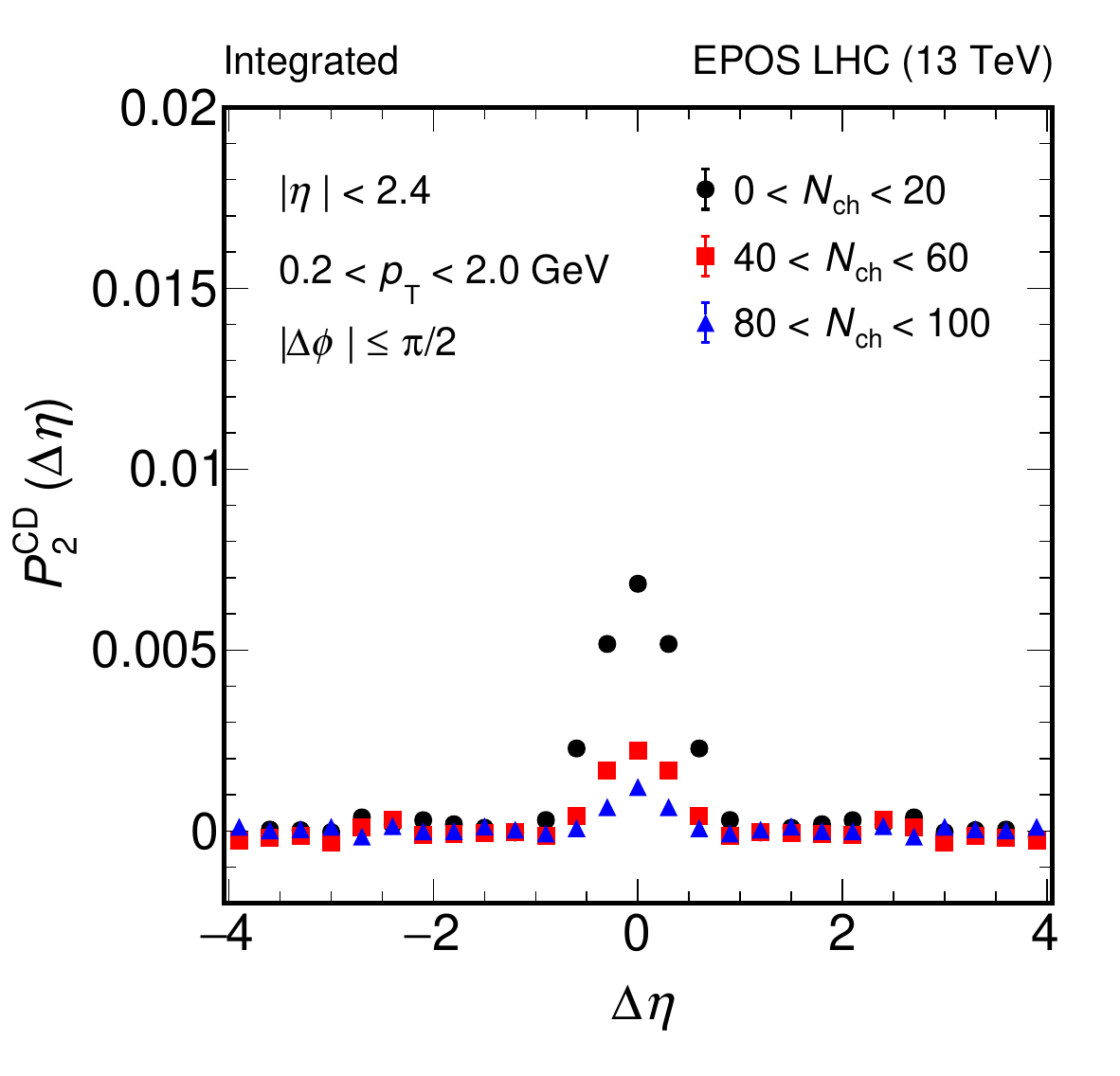}
    }
    \subfigure[]{
        \includegraphics[width=0.3\textwidth]{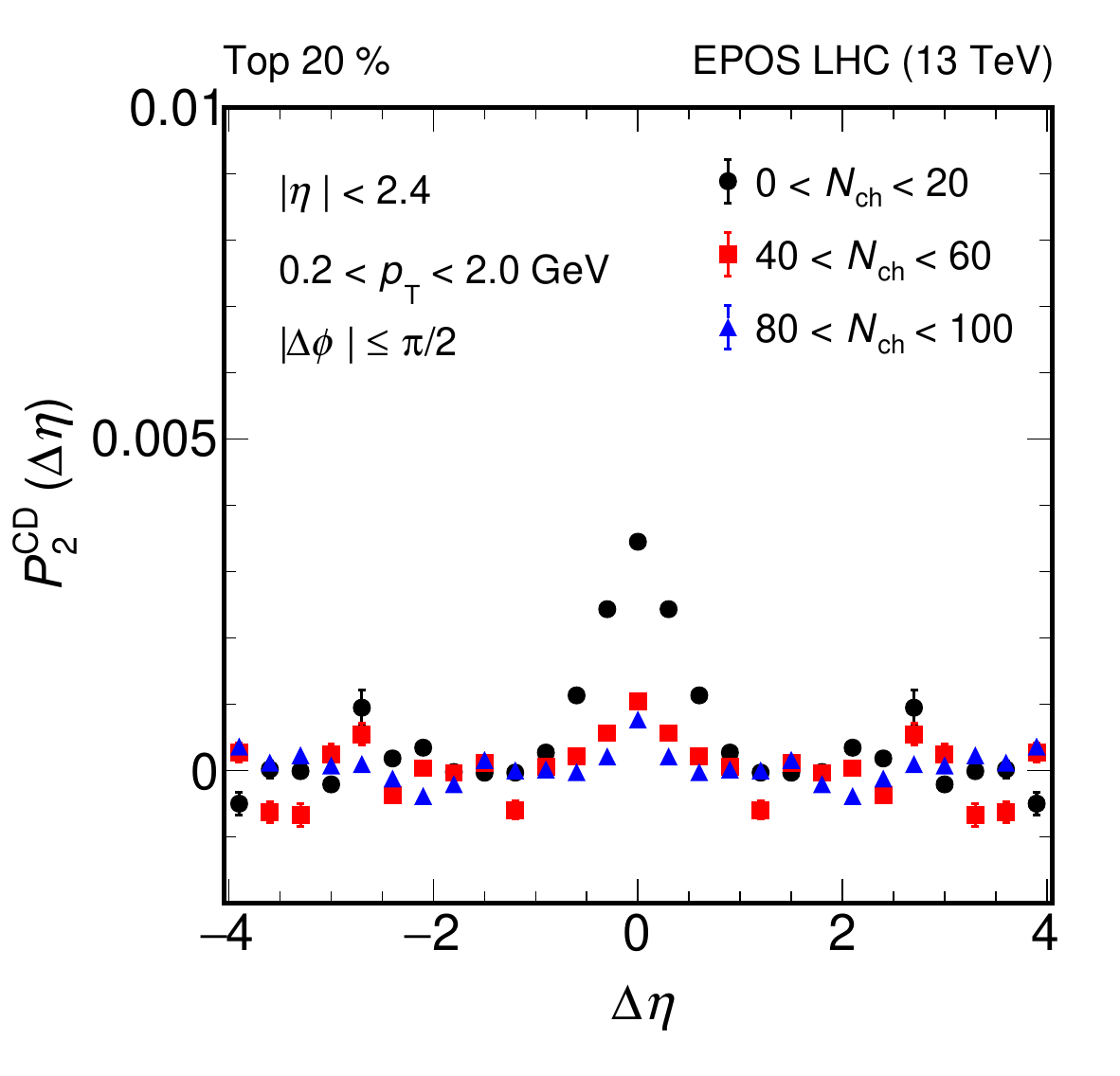}
    }
    \subfigure[]{
        \includegraphics[width=0.3\textwidth]{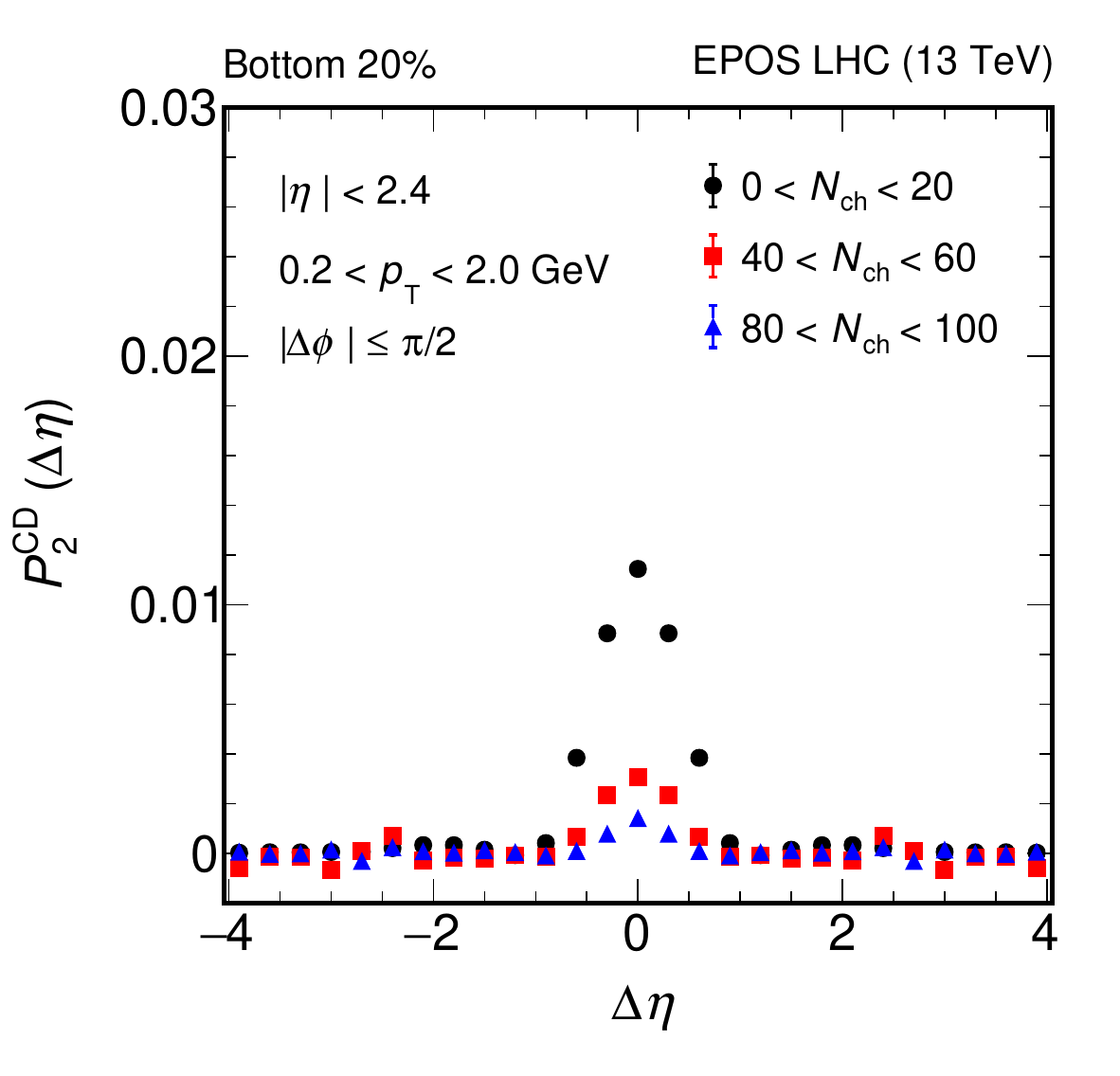}
    }
    \subfigure[]{
        \includegraphics[width=0.3\textwidth]{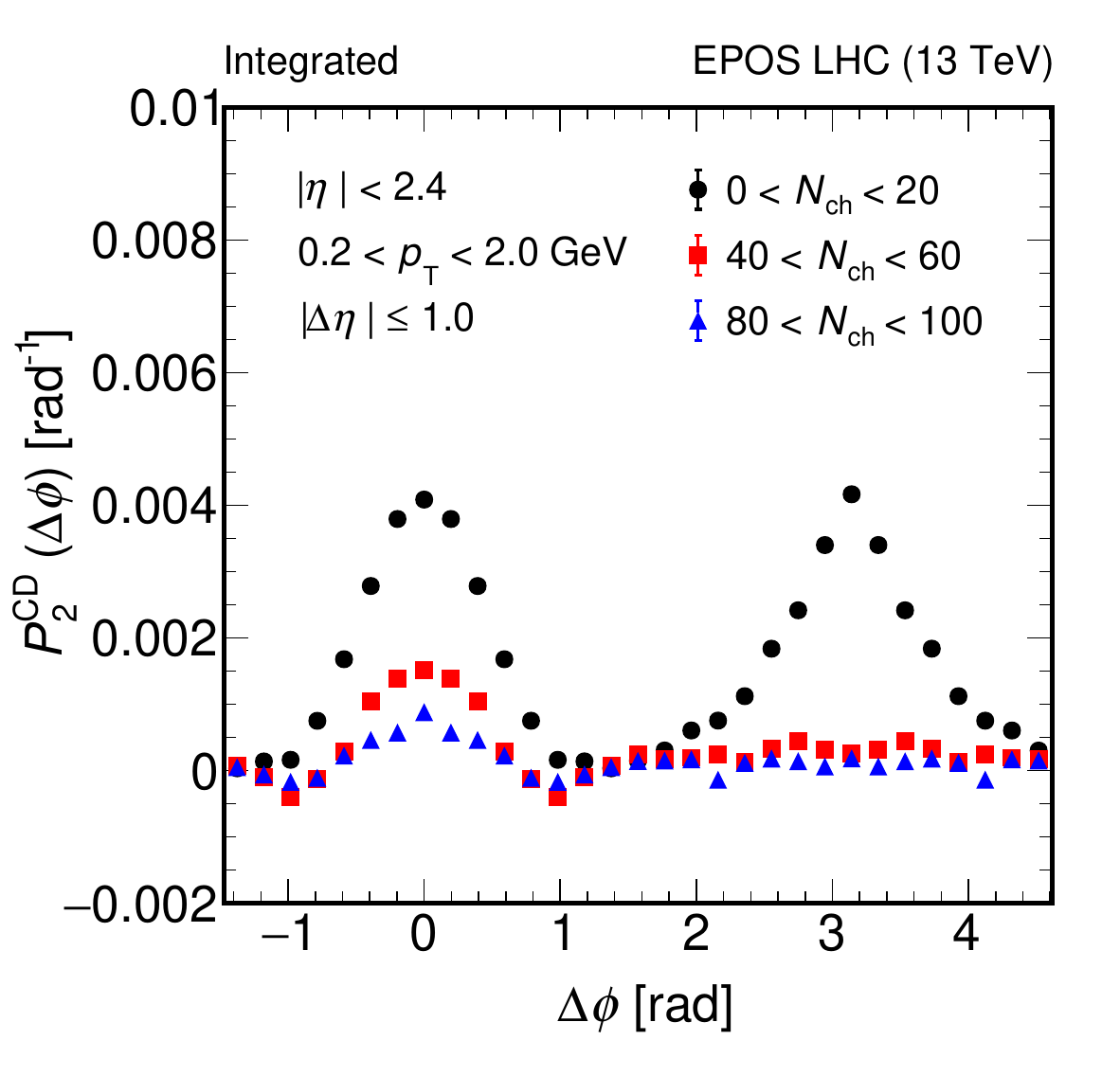}
    }
    \subfigure[]{
        \includegraphics[width=0.3\textwidth]{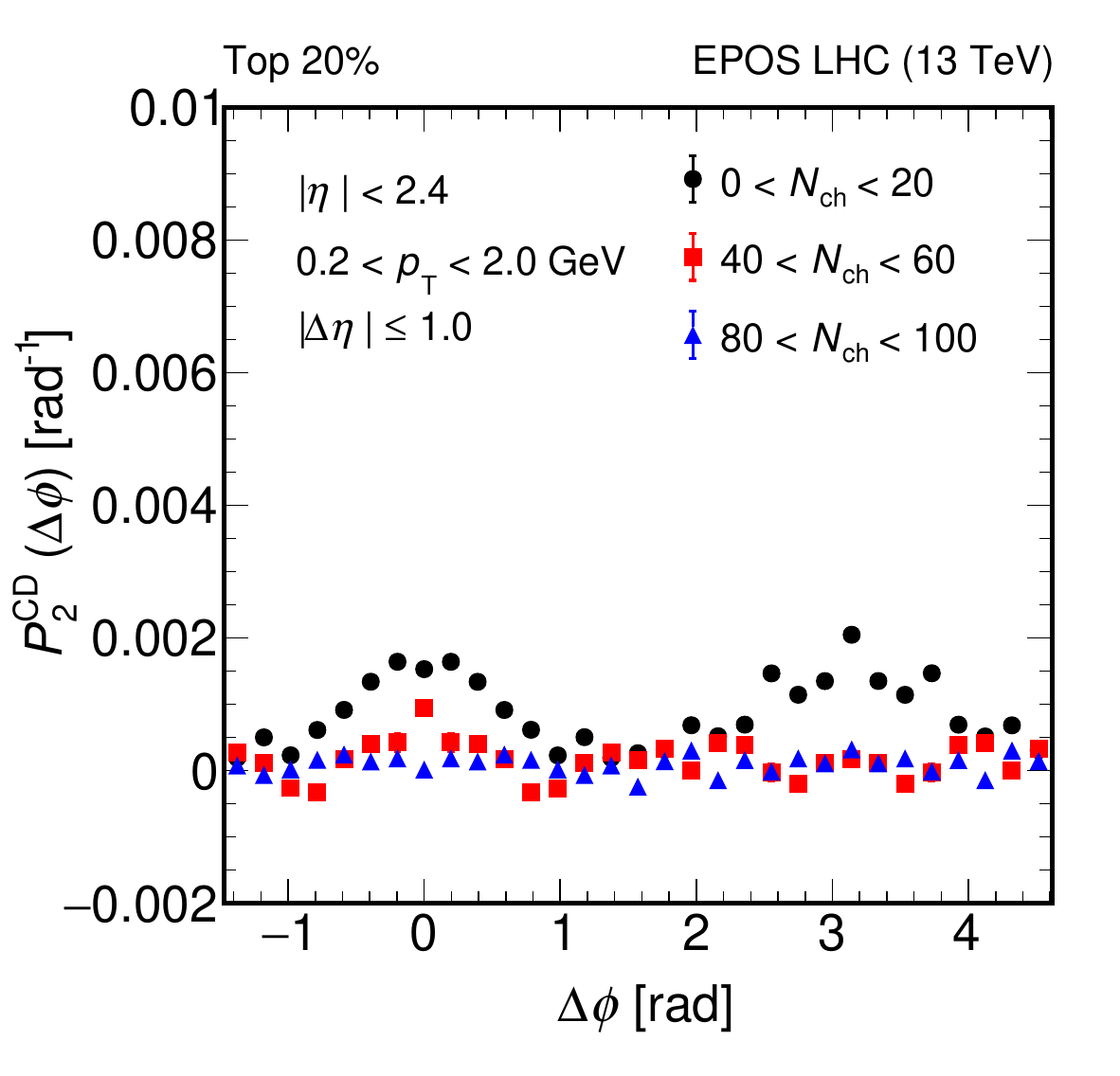}
    }
    \subfigure[]{
        \includegraphics[width=0.3\textwidth]{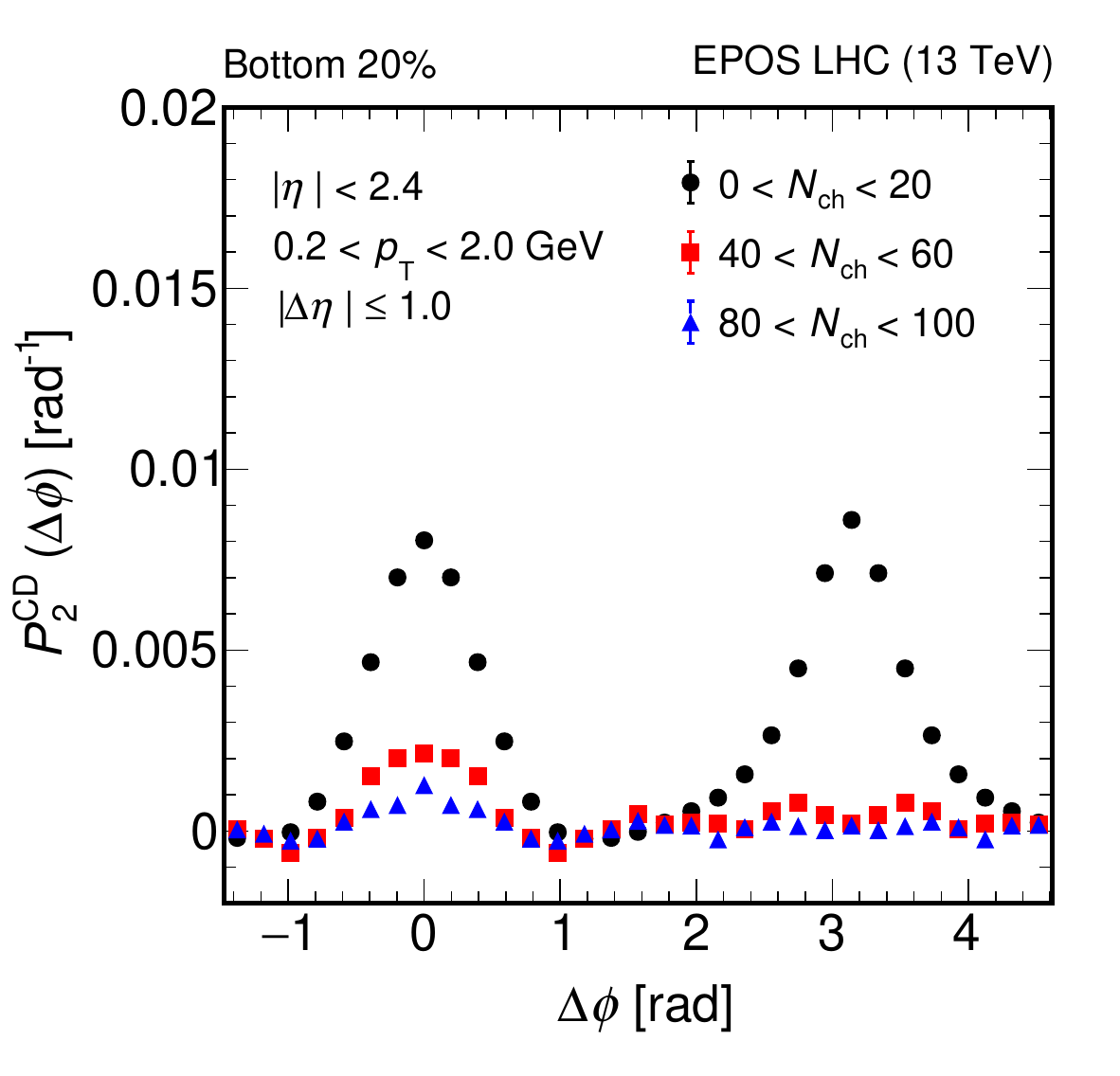}
    }   
 \caption{One-dimensional projections of $P_{2}^\mathrm{CD}$ along $\Delta\eta$ and $\Delta\phi$ from \textsc{epos} model in pp collisions at $\sqrt{s} =$ 13 TeV. The left column is for the integrated spherocity class, the middle column is for the top 20\%, and the right panel is for the bottom 20\% of the spherocity class. $\Delta\eta$ projections are taken in $|\Delta\phi| \leq \pi/2$ and $\Delta\phi$ projections are taken in $|\Delta\eta| \leq 1$ range.}
\label{fig:1d_epos_p2cd}
\end{figure*}

Figures~\ref{fig:1d_pythia_p2cd},~\ref{fig:1d_epos_p2cd_nocore} and~\ref{fig:1d_epos_p2cd}  present the one-dimensional projection of $P_{2}^\mathrm{CD}$ along $\Delta\eta$ and $\Delta\phi$ for three multiplicity classes  in \pythia and \epos without and with core model. Each column corresponds to a different spherocity class: the left column shows the case for spherocity integrated, the middle column corresponds to the top 20\% spherocity (isotropic events), and the right column represents the bottom 20\% (jetlike events). The correlator $P_{2}^\mathrm{CD}$, which emphasizes momentum correlations between unlike-sign particle pairs, shows a prominent peak at small $\Delta\eta$ and $\Delta\phi$, especially at low multiplicities. This indicates strong short-range correlations typical of jets and resonance decays, where particle pairs are produced in closer angles and tend to carry correlated transverse momenta. As the event multiplicity increases, the magnitude of $P_{2}^\mathrm{CD}$ decreases in both $\Delta\eta$ and $\Delta\phi$ projections. This trend reflects the dilution of momentum correlations in more populated events, consistent with the dominance of softer particle production mechanisms.

The $\Delta\phi$ distributions [panels (d)--(f)] exhibit away-side ($\Delta\phi \approx \pi$) structures. The near-side enhancement is associated with localized momentum correlations, while the away-side peak is indicative of back-to-back particle emission, primarily from dijet fragmentation. In all spherocity selection, the away-side structure is most pronounced, reflecting the stronger role of jet fragmentation in low-multiplicity events in \epos, while the away-side structure is suppressed in \pythia calculations, suggesting reduced dijet dominance and possibly enhanced isotropic emission patterns. The overall correlation strength decreases with multiplicity, which may result from an increased contribution of soft particle production mechanisms in high-multiplicity events. Conversely, isotropic events yield flatter and broader distributions, suggesting reduced momentum correlation strength arising from mechanisms such as MPIs. With increasing multiplicity, the width becomes narrower in both model simulations~\cite{Lonnblad:2023kft}.

$P_{2}^\mathrm{CD}$ correlator is affected by both the angular correlations and the momentum-correlated pairs like those from the same jet or resonance, and significantly affects the distribution structure. It has a strong influence from the jet fragmentation where particles are produced close in angle and share similar momenta, resulting in a sharper near-side distribution compared to the $B$. Collective phenomena like flow or charge balancing across larger $\Delta\phi$ are visible in $B$, hence it has a wider distribution. However, $P_{2}^\mathrm{CD}$ largely filters out such contributions because these long-range pairs often do not exhibit strong \pt covariance.
\begin{figure*} [!htb] 
    \centering
    \subfigure[]{
        \includegraphics[width=0.35\textwidth]{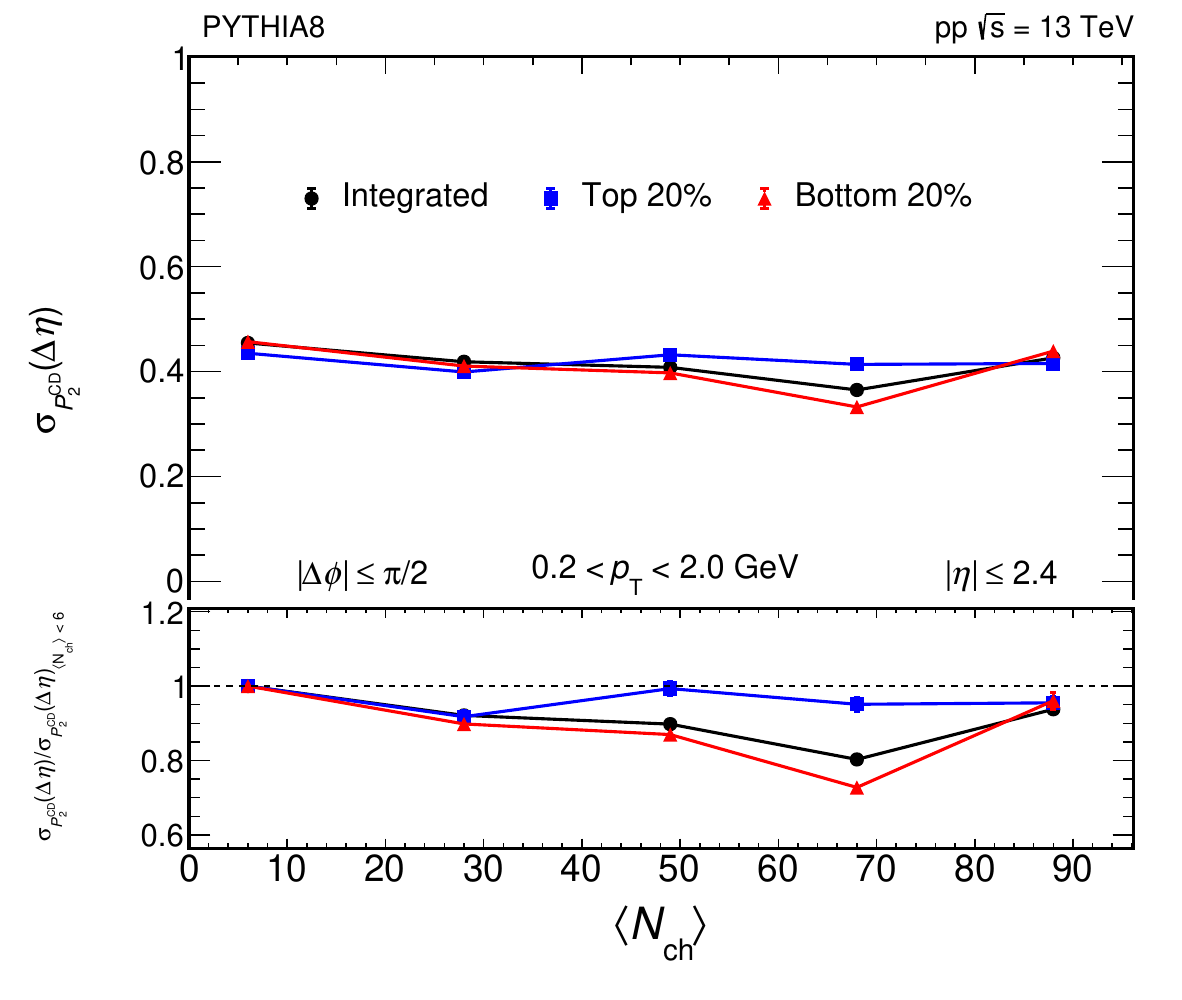}
    }
    \subfigure[]{
        \includegraphics[width=0.35\textwidth]{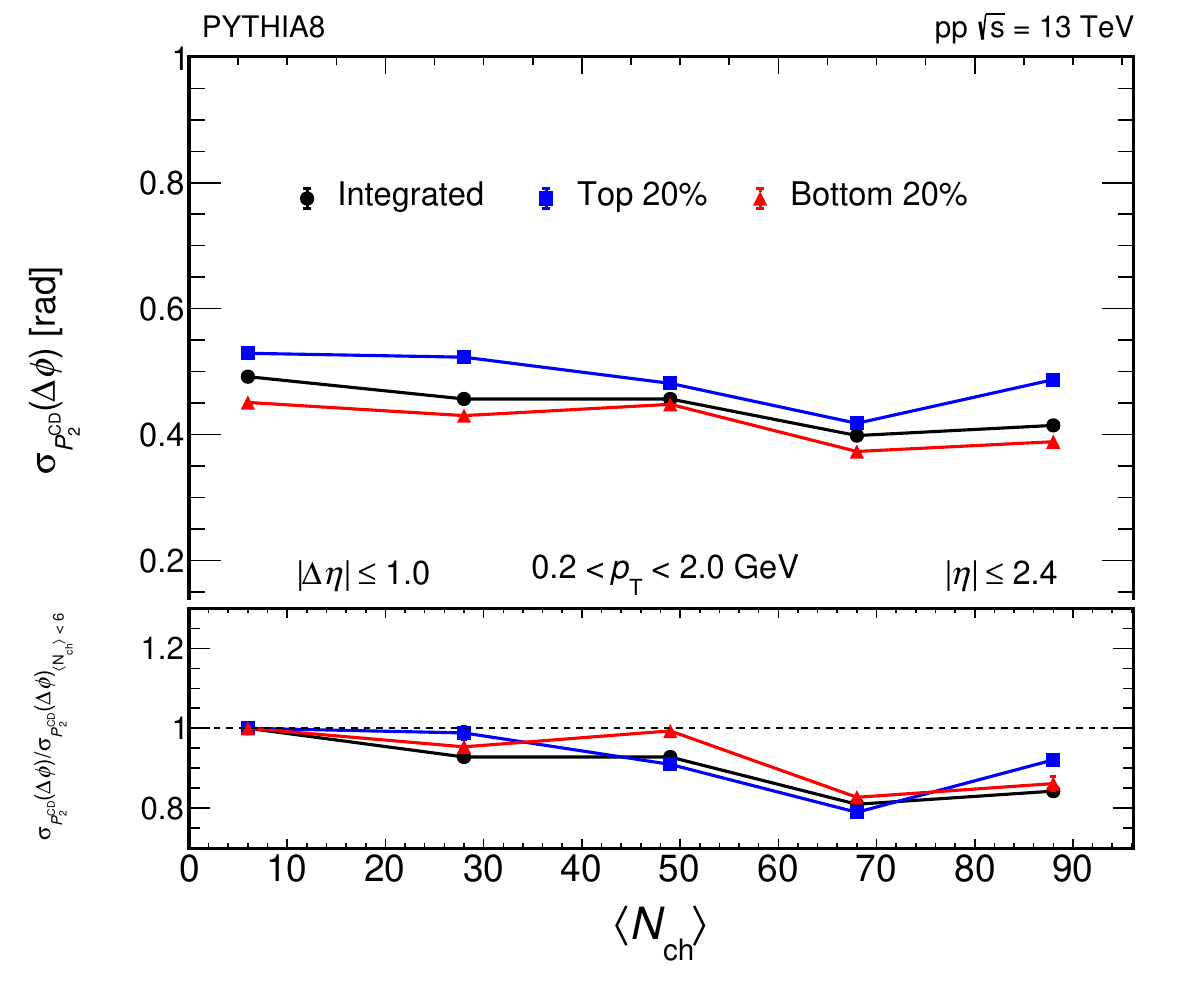}
    }
    \subfigure[]{
        \includegraphics[width=0.35\textwidth]{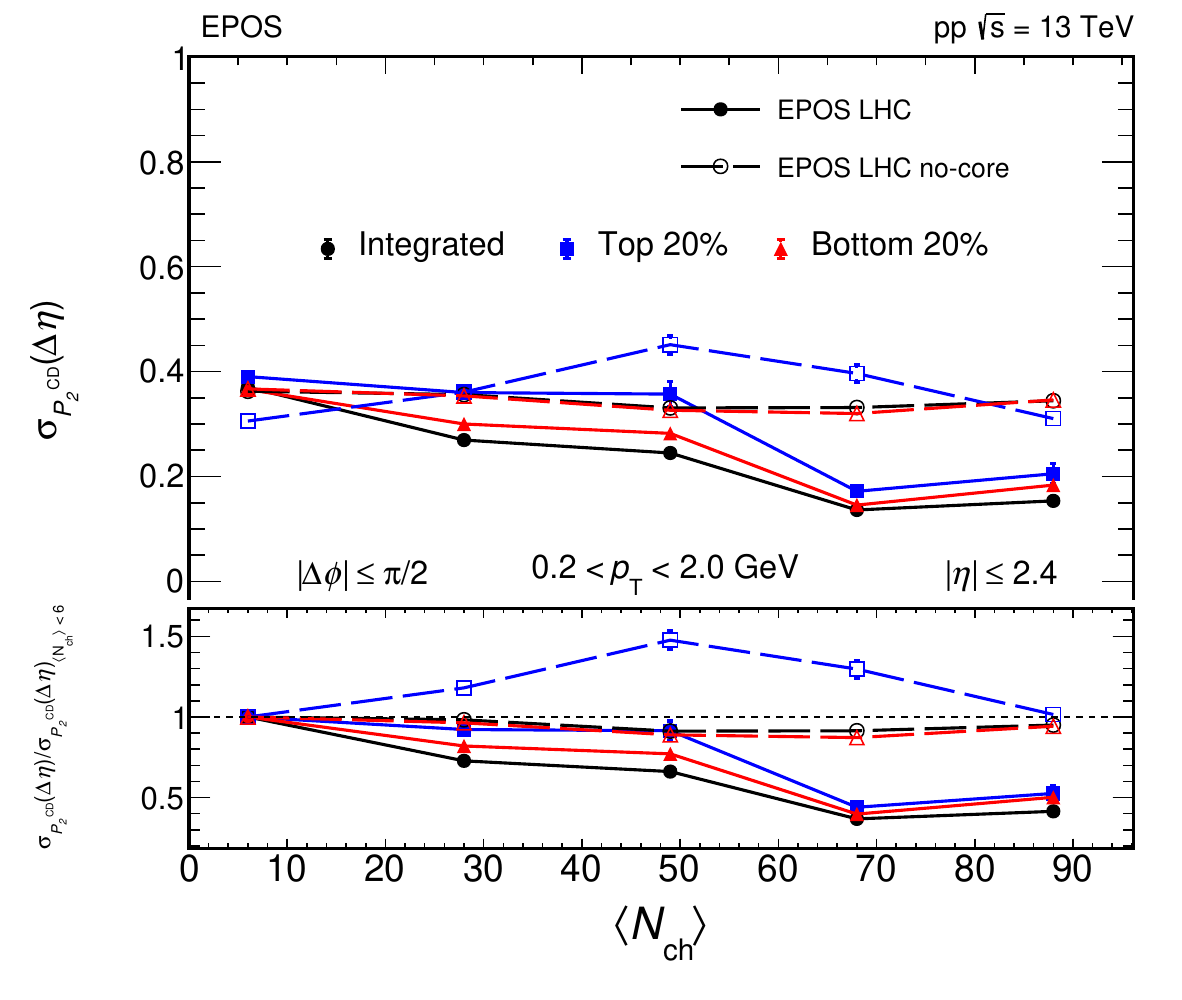}
    }
    \subfigure[]{
        \includegraphics[width=0.35\textwidth]{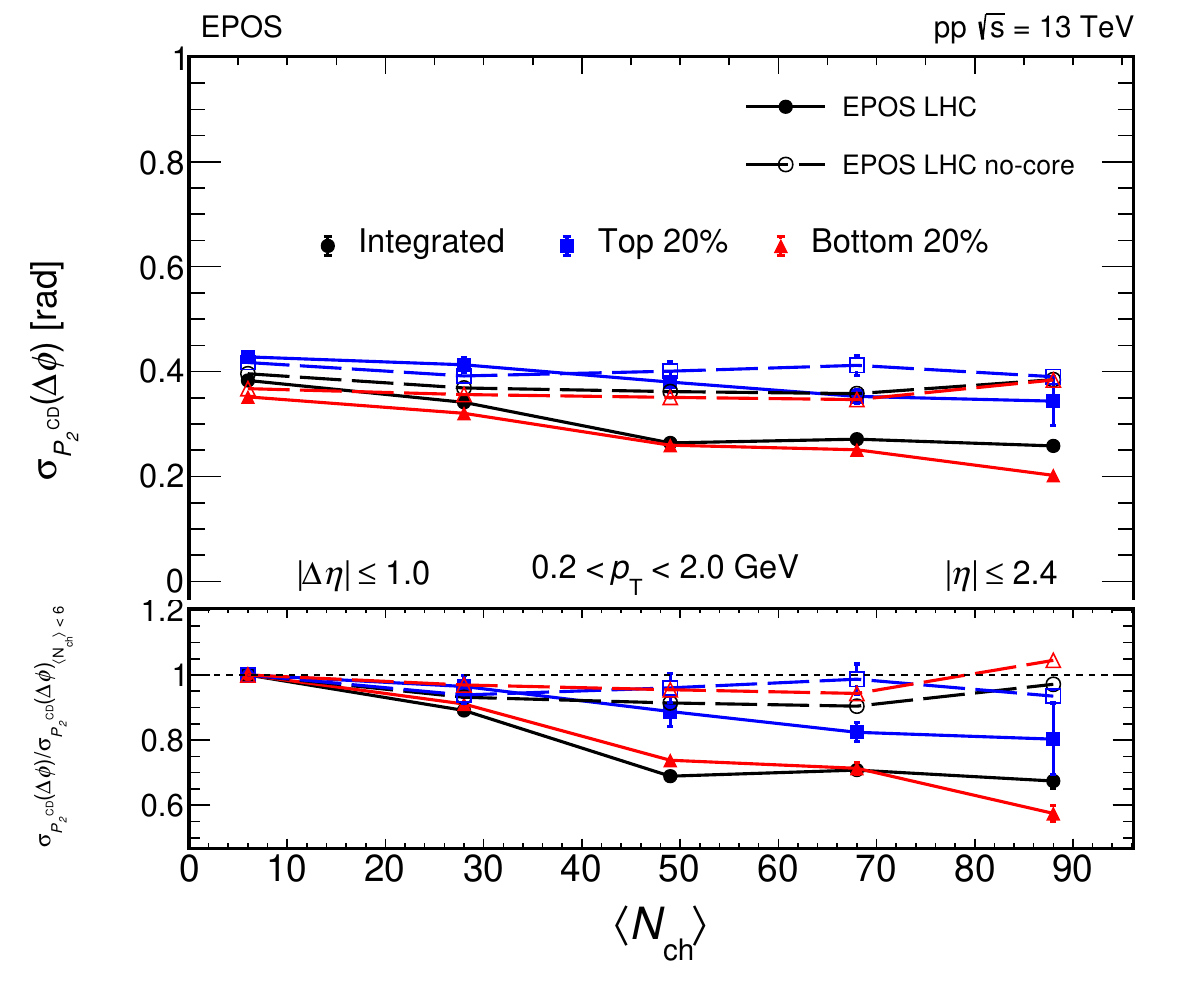}
    }
\caption{Width of $P_{2}^\mathrm{CD}$ in terms of $\Delta\eta$ (left) and $\Delta\phi$ (right) for \pythia and \epos  model simulations in pp collisions at $\sqrt{s} =$ 13 TeV as a function of $\langle\nch\rangle$ and spherocity classes. The top panel shows the rms width and the bottom panel presents the width normalized to its value in the lowest multiplicity interval ($N_{\mathrm{ch}} < 6$). The widths are evaluated within $0.2 < p_{\mathrm{T}} < 2.0$~GeV interval and pseudorapidity acceptance $|\eta| \leq 2.4$.}
\label{fg:p2widthModel}
\end{figure*}
Figure~\ref{fg:p2widthModel} presents the multiplicity dependence of the charge-dependent $P_{2}$ correlator widths in $\Delta\eta$ and $\Delta\phi$ for three spherocity classes in pp collisions at $\sqrt{s}=13$ TeV. In \pythia, the widths in both $\Delta\eta$ and $\Delta\phi$ exhibit only a mild decrease with increasing $\langle N_{\rm ch}\rangle$, reflecting the dominant role of MPI and CR in shaping transverse momentum correlations. Jetty events show slightly narrower widths than isotropic events, particularly in $\Delta\phi$, but the overall separation remains modest, consistent with the absence of collective effects~\cite{OrtizVelasquez:2013ofg}. In contrast, \epos exhibits a stronger sensitivity to event topology, especially when the hydrodynamic core is enabled. For $\Delta\eta$ (panel c), the no-core configuration shows nearly flat widths with $\langle\nch\rangle$, whereas the core-enabled scenario displays clear narrowing at high multiplicities, indicating the localization of momentum balancing by collective expansion. In $\Delta\phi$ (panel d), the topology separation becomes more pronounced in the core case, where jetty events yield narrow $\Delta\phi$ widths, while isotropic events remain broad, highlighting the influence of collective flow. Thus, while \pythia captures jet-induced collimation through MPI+CR, the enhanced topology dependence in \epos with core reflects the presence of mediumlike effects. Together, these results demonstrate that $P_{2}^\mathrm{CD}$ widths provide complementary sensitivity to both jet fragmentation dynamics and possible collective behavior in small collision systems.

\section{Summary}
\label{summary}
This study explores charge-dependent number and momentum correlations through balance functions in proton-proton collisions at $\sqrt{s} = 13$ TeV, using transverse spherocity and charged-particle multiplicity as key event classifiers. Balance functions are evaluated in terms of relative pseudorapidity ($\Delta\eta$) and azimuthal angle ($\Delta\phi$) for particles in the transverse momentum range $0.2 < p_{\rm T} < 2.0$ GeV. Two event generators, \pythia and \epos, are employed to probe the underlying particle production mechanisms. While \epos incorporates hydrodynamic evolution and core-corona separation, \pythia relies on parton showering, MPI, and CR.

Our results show that the widths of balance functions in both $\Delta\eta$ and $\Delta\phi$ decrease systematically with increasing charged-particle multiplicity, reflecting that charge-balancing pairs are produced more locally in high-multiplicity events. This narrowing is most pronounced in jetlike (low-spherocity) events, highlighting the localization of charge conservation within hard-QCD dominated processes, such as jet fragmentation. Isotropic (high-spherocity) events, on the other hand, exhibit broader balance functions, pointing toward a greater influence from soft particle production, multiparton interactions, and possible collective flow effects. Isotropic events exhibit widths approximately 5\%larger than 7\%  jetlike events in $\Delta \phi$, reflecting broader charge correlations in soft-dominated topologies.
The comparison between \pythia\ and \epos reveals a clear separation as a function of $\Delta\eta$ and $\Delta\phi$. For the $\Delta\eta$ balance functions, \epos with the hydrodynamic core enabled produces narrower widths than \pythia, indicating stronger localization of charge-balancing pairs in pseudorapidity. In contrast, the $\Delta\phi$ widths obtained with \epos are generally larger than those predicted by \pythia. However, within the \epos model, enabling the hydrodynamic core leads to a stronger multiplicity dependence of the $\Delta\phi$ width compared to the no-core configuration, highlighting the role of hydrodynamic evolution in shaping the azimuthal evolution of charge correlations. In contrast, in the no-core scenario (where hydrodynamic flow is absent), the broad widths and weaker multiplicity dependence suggest a picture dominated by independent fragmentation and less collective behavior.

This differential study, incorporating event topology, provides new insight into the competing mechanisms at play in small collision systems. By analyzing the balance function as a function of spherocity and multiplicity, we directly probe the interplay between localized charge conservation in jets and medium-driven collective effects. The contrasting results from \epos with core versus no-core configurations, alongside \pythia, highlight the sensitivity of the balance function to different stages of the collision evolution and their respective hadronization mechanisms. Looking forward, the high-precision measurements achievable with LHC Run 3 data will enable robust experimental tests of these theoretical scenarios, offering essential constraints to improve the modeling of hadronization, color reconnection, and the possible emergence of collectivity in small systems. This work thus sets the stage for disentangling the origins of collective phenomena and refining our understanding of QCD dynamics in high-energy pp collisions.
\begin{acknowledgements}
S. C. B and A. K acknowledge support under the INFN postdoctoral fellowship.
\end{acknowledgements}

\bibliographystyle{utphys}   
\bibliography{reference}

\providecommand{\href}[2]{#2}\begingroup\raggedright\begin{thebibliography}{10}

\bibitem{cmswhite}
{\bfseries CMS} Collaboration, A.~Hayrapetyan {\em et~al.}, ``{Overview of
  high-density QCD studies with the CMS experiment at the LHC}'',
  \href{https://doi.org/10.1016/j.physrep.2024.11.007}{{\em Phys. Rept.}
  {\bfseries 1115} (2025) 219},
  \href{https://arxiv.org/abs/2405.10785}{{\ttfamily arXiv:2405.10785
  [nucl-ex]}}.

\bibitem{ALICE:2022wpn}
{\bfseries ALICE} Collaboration, S.~Acharya {\em et~al.}, ``{The ALICE
  experiment: a journey through QCD}'',
  \href{https://doi.org/10.1140/epjc/s10052-024-12935-y}{{\em Eur. Phys. J. C}
  {\bfseries 84} (2024) 813},
  \href{https://arxiv.org/abs/2211.04384}{{\ttfamily arXiv:2211.04384
  [nucl-ex]}}.

\bibitem{PHENIX:2004vcz}
{\bfseries PHENIX} Collaboration, K.~Adcox {\em et~al.}, ``{Formation of dense
  partonic matter in relativistic nucleus-nucleus collisions at RHIC:
  Experimental evaluation by the PHENIX collaboration}'',
  \href{https://doi.org/10.1016/j.nuclphysa.2005.03.086}{{\em Nucl. Phys. A}
  {\bfseries 757} (2005) 184},
  \href{https://arxiv.org/abs/nucl-ex/0410003}{{\ttfamily
  arXiv:nucl-ex/0410003}}.

\bibitem{starqgp}
{\bfseries STAR} Collaboration, J.~Adams {\em et~al.}, ``{Experimental and
  theoretical challenges in the search for the quark gluon plasma: The STAR
  Collaboration's critical assessment of the evidence from RHIC collisions}'',
  \href{https://doi.org/10.1016/j.nuclphysa.2005.03.085}{{\em Nucl. Phys. A}
  {\bfseries 757} (2005) 102},
  \href{https://arxiv.org/abs/nucl-ex/0501009}{{\ttfamily
  arXiv:nucl-ex/0501009}}.

\bibitem{qgpmed2}
{\bfseries PHOBOS} Collaboration, B.~B. Back {\em et~al.}, ``{The PHOBOS
  perspective on discoveries at RHIC}'',
  \href{https://doi.org/10.1016/j.nuclphysa.2005.03.084}{{\em Nucl. Phys. A}
  {\bfseries 757} (2005) 28},
  \href{https://arxiv.org/abs/nucl-ex/0410022}{{\ttfamily
  arXiv:nucl-ex/0410022}}.

\bibitem{QGPmedium1}
{\bfseries BRAHMS} Collaboration, I.~Arsene {\em et~al.}, ``{Quark gluon plasma
  and color glass condensate at RHIC? The Perspective from the BRAHMS
  experiment}'', \href{https://doi.org/10.1016/j.nuclphysa.2005.02.130}{{\em
  Nucl. Phys. A} {\bfseries 757} (2005) 1},
  \href{https://arxiv.org/abs/nucl-ex/0410020}{{\ttfamily
  arXiv:nucl-ex/0410020}}.

\bibitem{ALICE:pruneau}
{\bfseries ALICE} Collaboration, S.~Acharya {\em et~al.}, ``{General balance
  functions of identified charged hadron pairs of (\ensuremath{\pi},K,p) in
  Pb\textendash{}Pb collisions at $\sqrt{s_{_\mathrm{NN}}}$= 2.76 TeV}'',
  \href{https://doi.org/10.1016/j.physletb.2022.137338}{{\em Phys. Lett. B}
  {\bfseries 833} (2022) 137338},
  \href{https://arxiv.org/abs/2110.06566}{{\ttfamily arXiv:2110.06566
  [nucl-ex]}}.

\bibitem{cmsbf}
{\bfseries CMS} Collaboration, A.~Tumasyan {\em et~al.}, ``{Multiplicity and
  transverse momentum dependence of charge-balance functions in pPb and PbPb
  collisions at LHC energies}'',
  \href{https://doi.org/10.1007/JHEP08(2024)148}{{\em JHEP} {\bfseries 08}
  (2024) 148}, \href{https://arxiv.org/abs/2307.11185}{{\ttfamily
  arXiv:2307.11185 [nucl-ex]}}.

\bibitem{jet1}
{\bfseries CMS} Collaboration, S.~Chatrchyan {\em et~al.}, ``{Observation and
  studies of jet quenching in PbPb collisions at nucleon-nucleon center-of-mass
  energy = 2.76 TeV}'', \href{https://doi.org/10.1103/PhysRevC.84.024906}{{\em
  Phys. Rev. C} {\bfseries 84} (2011) 024906},
  \href{https://arxiv.org/abs/1102.1957}{{\ttfamily arXiv:1102.1957
  [nucl-ex]}}.

\bibitem{jet2}
{\bfseries HADES} Collaboration, S.~Harabasz, ``{Multi-differential pattern of
  low-mass $e^+ e^-$ excess from $\sqrt{s_{_\mathrm{NN}}}$=2.4 GeV Au$+$Au
  collisions with HADES}'',
  \href{https://doi.org/10.1016/j.nuclphysa.2018.09.052}{{\em Nucl. Phys. A}
  {\bfseries 982} (2019) 771}.

\bibitem{jet3}
{\bfseries STAR} Collaboration, L.~Adamczyk {\em et~al.}, ``{Measurements of
  jet quenching with semi-inclusive hadron+jet distributions in Au+Au
  collisions at $\sqrt{s_{_\mathrm{NN}}}$ = 200 GeV}'',
  \href{https://doi.org/10.1103/PhysRevC.96.024905}{{\em Phys. Rev. C}
  {\bfseries 96} (2017) 024905},
  \href{https://arxiv.org/abs/1702.01108}{{\ttfamily arXiv:1702.01108
  [nucl-ex]}}.

\bibitem{Shuryak}
E.~V. Shuryak, ``{On the origin of the 'Ridge' phenomenon induced by jets in
  heavy ion collisions}'',
  \href{https://doi.org/10.1103/PhysRevC.76.047901}{{\em Phys. Rev. C}
  {\bfseries 76} (2007) 047901},
  \href{https://arxiv.org/abs/0706.3531}{{\ttfamily arXiv:0706.3531
  [nucl-th]}}.

\bibitem{ALICE:2023ulm}
{\bfseries ALICE} Collaboration, S.~Acharya {\em et~al.}, ``{Emergence of
  Long-Range Angular Correlations in Low-Multiplicity Proton-Proton
  Collisions}'', \href{https://doi.org/10.1103/PhysRevLett.132.172302}{{\em
  Phys. Rev. Lett.} {\bfseries 132} (2024) 172302},
  \href{https://arxiv.org/abs/2311.14357}{{\ttfamily arXiv:2311.14357
  [nucl-ex]}}.

\bibitem{ALICE:2016fzo}
{\bfseries ALICE} Collaboration, J.~Adam {\em et~al.}, ``{Enhanced production
  of multi-strange hadrons in high-multiplicity proton-proton collisions}'',
  \href{https://doi.org/10.1038/nphys4111}{{\em Nature Phys.} {\bfseries 13}
  (2017) 535}, \href{https://arxiv.org/abs/1606.07424}{{\ttfamily
  arXiv:1606.07424 [nucl-ex]}}.

\bibitem{Baty:2021ugw}
A.~Baty, P.~Gardner, and W.~Li, ``{Novel observables for exploring QCD
  collective evolution and quantum entanglement within individual jets}'',
  \href{https://doi.org/10.1103/PhysRevC.107.064908}{{\em Phys. Rev. C}
  {\bfseries 107} (2023) 064908},
  \href{https://arxiv.org/abs/2104.11735}{{\ttfamily arXiv:2104.11735
  [hep-ph]}}.

\bibitem{cmsppflow}
{\bfseries CMS} Collaboration, V.~Khachatryan {\em et~al.}, ``{Measurement of
  long-range near-side two-particle angular correlations in pp collisions at
  $\sqrt{s}=$13 TeV}'',
  \href{https://doi.org/10.1103/PhysRevLett.116.172302}{{\em Phys. Rev. Lett.}
  {\bfseries 116} (2016) 172302},
  \href{https://arxiv.org/abs/1510.03068}{{\ttfamily arXiv:1510.03068
  [nucl-ex]}}.

\bibitem{CMS:2023iam}
{\bfseries CMS} Collaboration, A.~Hayrapetyan {\em et~al.}, ``{Observation of
  Enhanced Long-Range Elliptic Anisotropies Inside High-Multiplicity Jets in pp
  Collisions at $\sqrt{s}$ = 13\,\,TeV}'',
  \href{https://doi.org/10.1103/PhysRevLett.133.142301}{{\em Phys. Rev. Lett.}
  {\bfseries 133} (2024) 142301},
  \href{https://arxiv.org/abs/2312.17103}{{\ttfamily arXiv:2312.17103
  [hep-ex]}}.

\bibitem{Manea:2024qgd}
A.~Manea, C.~Pruneau, D.~C. Brandibur, A.~Danu, A.~F. Dobrin, V.~Gonzalez, and
  S.~Basu, ``{Investigating late-stage particle production in pp collisions
  with~balance functions}'',
  \href{https://doi.org/10.1140/epjc/s10052-025-14049-5}{{\em Eur. Phys. J. C}
  {\bfseries 85} (2025) 323},
  \href{https://arxiv.org/abs/2411.11207}{{\ttfamily arXiv:2411.11207
  [hep-ph]}}.

\bibitem{scottpratt}
S.~Pratt, ``{General Charge Balance Functions, A Tool for Studying the Chemical
  Evolution of the Quark-Gluon Plasma}'',
  \href{https://doi.org/10.1103/PhysRevC.85.014904}{{\em Phys. Rev. C}
  {\bfseries 85} (2012) 014904},
  \href{https://arxiv.org/abs/1109.3647}{{\ttfamily arXiv:1109.3647
  [nucl-th]}}.

\bibitem{bfscott}
W.-H. Zhou, H.~Liu, F.~Li, Y.-F. Sun, J.~Xu, and C.~M. Ko, ``{Elliptic flow
  splittings in the Polyakov\textendash{}Nambu\textendash{}Jona-Lasinio
  transport model}'', \href{https://doi.org/10.1103/PhysRevC.104.044901}{{\em
  Phys. Rev. C} {\bfseries 104} (2021) 044901},
  \href{https://arxiv.org/abs/2105.09518}{{\ttfamily arXiv:2105.09518
  [nucl-th]}}.

\bibitem{alicebfpbpbpb}
{\bfseries ALICE} Collaboration, S.~Acharya {\em et~al.}, ``{Two particle
  differential transverse momentum and number density correlations in p-Pb and
  Pb-Pb at the LHC}'', \href{https://doi.org/10.1103/PhysRevC.100.044903}{{\em
  Phys. Rev. C} {\bfseries 100} (2019) 044903},
  \href{https://arxiv.org/abs/1805.04422}{{\ttfamily arXiv:1805.04422
  [nucl-ex]}}.

\bibitem{alicep2r2pp}
{\bfseries ALICE} Collaboration, S.~Acharya {\em et~al.}, ``{Measurements of
  differential two-particle number and transverse momentum correlation
  functions in pp collisions at $\sqrt{\textit{s}}$ = 13 TeV}'',
  \href{https://doi.org/10.1140/epjc/s10052-025-14531-0}{{\em Eur. Phys. J. C}
  {\bfseries 85} (2025) 866},
  \href{https://arxiv.org/abs/2411.07059}{{\ttfamily arXiv:2411.07059
  [nucl-ex]}}.

\bibitem{Parida}
T.~Parida, P.~Bozek, and S.~Chatterjee, ``{Charm balance function in
  relativistic heavy-ion collisions}'',
  \href{https://doi.org/10.1103/PhysRevC.109.014903}{{\em Phys. Rev. C}
  {\bfseries 109} (2024) 014903},
  \href{https://arxiv.org/abs/2308.14446}{{\ttfamily arXiv:2308.14446
  [nucl-th]}}.

\bibitem{STARBF1}
{\bfseries STAR} Collaboration, B.~I. Abelev {\em et~al.}, ``{Longitudinal
  scaling property of the charge balance function in Au + Au collisions at 200
  GeV}'', \href{https://doi.org/10.1016/j.physletb.2010.05.028}{{\em Phys.
  Lett. B} {\bfseries 690} (2010) 239},
  \href{https://arxiv.org/abs/1002.1641}{{\ttfamily arXiv:1002.1641
  [nucl-ex]}}.

\bibitem{STARBF2}
{\bfseries EHS, NA22} Collaboration, M.~R. Atayan {\em et~al.}, ``{Boost
  invariance and multiplicity dependence of the charge balance functionin
  $\pi^{+} p$ and $K^{+} p$ collisions at $\sqrt{s}$ = 22-GeV/c}'',
  \href{https://doi.org/10.1016/j.physletb.2006.04.027}{{\em Phys. Lett. B}
  {\bfseries 637} (2006) 39},
  \href{https://arxiv.org/abs/hep-ex/0506027}{{\ttfamily
  arXiv:hep-ex/0506027}}.

\bibitem{icpaqgp}
S.~A. Voloshin, ``Heavy ion collisions: {Correlations and Fluctuations} in
  particle production'', \href{https://doi.org/10.1088/1742-6596/50/1/013}{{\em
  J. Phys.: Conf. Ser.} {\bfseries 50} (2006) 111},
\href{https://arxiv.org/abs/nucl-ex/0505003}{{\ttfamily arXiv:nucl-ex/0505003
  [nucl-ex]}}.

\bibitem{Bialas:1}
A.~Bialas and J.~Rafelski, ``{Balance of baryon number in the quark coalescence
  model}'', \href{https://doi.org/10.1016/j.physletb.2005.11.084}{{\em Phys.
  Lett. B} {\bfseries 633} (2006) 488--491},
  \href{https://arxiv.org/abs/hep-ph/0508084}{{\ttfamily
  arXiv:hep-ph/0508084}}.

\bibitem{Bialas:2}
A.~Bialas, ``{Balance functions in coalescence model}'',
  \href{https://doi.org/10.1016/j.physletb.2003.10.106}{{\em Phys. Lett. B}
  {\bfseries 579} (2004) 31--38},
  \href{https://arxiv.org/abs/hep-ph/0308245}{{\ttfamily
  arXiv:hep-ph/0308245}}.

\bibitem{originalsp}
A.~Banfi, G.~P. Salam, and G.~Zanderighi, ``{Phenomenology of event shapes at
  hadron colliders}'', \href{https://doi.org/10.1007/JHEP06(2010)038}{{\em
  JHEP} {\bfseries 06} (2010) 038},
  \href{https://arxiv.org/abs/1001.4082}{{\ttfamily arXiv:1001.4082 [hep-ph]}}.

\bibitem{arvind_sp}
A.~Khuntia, S.~Tripathy, A.~Bisht, and R.~Sahoo, ``{Event shape engineering and
  multiplicity dependent study of identified particle production in proton +
  proton collisions at $\sqrt{s}$ = 13 TeV using PYTHIA8}'',
  \href{https://doi.org/10.1088/1361-6471/abb1f8}{{\em J. Phys. G} {\bfseries
  48} (2021) 035102}, \href{https://arxiv.org/abs/1811.04213}{{\ttfamily
  arXiv:1811.04213 [hep-ph]}}.

\bibitem{ALICE_SP1}
{\bfseries ALICE} Collaboration, S.~Acharya {\em et~al.}, ``{Charged-particle
  production as a function of multiplicity and transverse spherocity in pp
  collisions at $\sqrt{s} =5.02$ and 13 TeV}'',
  \href{https://doi.org/10.1140/epjc/s10052-019-7350-y}{{\em Eur. Phys. J. C}
  {\bfseries 79} (2019) 857},
  \href{https://arxiv.org/abs/1905.07208}{{\ttfamily arXiv:1905.07208
  [nucl-ex]}}.

\bibitem{ALICE_SP2}
{\bfseries ALICE} Collaboration, S.~Acharya {\em et~al.}, ``{Light-flavor
  particle production in high-multiplicity pp collisions at $ \sqrt{\textrm{s}}
  $= 13 TeV as a function of transverse spherocity}'',
  \href{https://doi.org/10.1007/JHEP05(2024)184}{{\em JHEP} {\bfseries 05}
  (2024) 184}, \href{https://arxiv.org/abs/2310.10236}{{\ttfamily
  arXiv:2310.10236 [hep-ex]}}.

\bibitem{pythia_ref}
C.~Bierlich {\em et~al.}, ``{A comprehensive guide to the physics and usage of
  PYTHIA 8.3}'', \href{https://doi.org/10.21468/SciPostPhysCodeb.8}{{\em
  SciPost Phys. Codeb.} {\bfseries 2022} (2022) 8},
  \href{https://arxiv.org/abs/2203.11601}{{\ttfamily arXiv:2203.11601
  [hep-ph]}}.

\bibitem{eposLHC}
T.~Pierog, I.~Karpenko, J.~M. Katzy, E.~Yatsenko, and K.~Werner, ``{EPOS LHC:
  Test of collective hadronization with data measured at the CERN Large Hadron
  Collider}'', \href{https://doi.org/10.1103/PhysRevC.92.034906}{{\em Phys.
  Rev. C} {\bfseries 92} (2015) 034906},
  \href{https://arxiv.org/abs/1306.0121}{{\ttfamily arXiv:1306.0121 [hep-ph]}}.

\bibitem{eposCC}
K.~Werner, ``{Core-corona procedure and microcanonical hadronization to
  understand strangeness enhancement in proton-proton and heavy ion collisions
  in the EPOS4 framework}'',
  \href{https://doi.org/10.1103/PhysRevC.109.014910}{{\em Phys. Rev. C}
  {\bfseries 109} (2024) 014910},
  \href{https://arxiv.org/abs/2306.10277}{{\ttfamily arXiv:2306.10277
  [hep-ph]}}.

\bibitem{Pruneau:2019baa}
C.~A. Pruneau, ``Role of baryon number conservation in measurements of
  fluctuations'', \href{https://doi.org/10.1103/PhysRevC.100.034905}{{\em Phys.
  Rev. C} {\bfseries 100} (2019) 034905},
  \href{https://arxiv.org/abs/1903.04591}{{\ttfamily arXiv:1903.04591
  [nucl-th]}}.

\bibitem{mult_def}
{\bfseries CMS} Collaboration, A.~Tumasyan {\em et~al.}, ``{Correlations
  between azimuthal anisotropy and mean transverse momentum in pp, pPb, and
  peripheral PbPb collisions}'',
  \href{https://arxiv.org/abs/2410.04578}{{\ttfamily arXiv:2410.04578
  [nucl-ex]}}.

\bibitem{STAR:bf1}
{\bfseries STAR} Collaboration, M.~M. Aggarwal {\em et~al.}, ``{Balance
  Functions from Au$+$Au, $d+$Au, and $p+p$ Collisions at
  $\sqrt{s_\mathrm{NN}}$ = 200 GeV}'',
  \href{https://doi.org/10.1103/PhysRevC.82.024905}{{\em Phys. Rev. C}
  {\bfseries 82} (2010) 024905},
  \href{https://arxiv.org/abs/1005.2307}{{\ttfamily arXiv:1005.2307
  [nucl-ex]}}.

\bibitem{cmsppridge}
{\bfseries CMS} Collaboration, V.~Khachatryan {\em et~al.}, ``Observation of
  long-range, near-side angular correlations in proton-proton collisions at the
  {LHC}'', \href{https://doi.org/10.1007/jhep09(2010)091}{{\em JHEP} {\bfseries
  09} (2010) 091},
\href{https://arxiv.org/abs/1009.4122}{{\ttfamily arXiv:1009.4122 [hep-ex]}}.

\bibitem{cmspbpbflow}
{\bfseries CMS} Collaboration, S.~Chatrchyan {\em et~al.}, ``{Long-range and
  short-range dihadron angular correlations in central PbPb collisions at a
  nucleon-nucleon center of mass energy of $\sqrt{s_{_\mathrm{NN}}}$ = 2.76
  TeV}'', \href{https://doi.org/10.1007/JHEP07(2011)076}{{\em JHEP} {\bfseries
  07} (2011) 076}, \href{https://arxiv.org/abs/1105.2438}{{\ttfamily
  arXiv:1105.2438 [nucl-ex]}}.

\bibitem{hin18008}
{\bfseries CMS} Collaboration, A.~Tumasyan {\em et~al.}, ``{Two-particle
  azimuthal correlations in \ensuremath{\gamma}p interactions using pPb
  collisions at $\sqrt{s_{_\mathrm{NN}}}$ = 8.16TeV}'',
  \href{https://doi.org/10.1016/j.physletb.2023.137905}{{\em Phys. Lett. B}
  {\bfseries 844} (2023) 137905},
  \href{https://arxiv.org/abs/2204.13486}{{\ttfamily arXiv:2204.13486
  [nucl-ex]}}.

\bibitem{CMSPP}
{\bfseries CMS} Collaboration, A.~M. Sirunyan {\em et~al.}, ``{Elliptic flow of
  charm and strange hadrons in high-multiplicity pPb collisions at
  $\sqrt{s_{_\mathrm{NN}}} =$ 8.16 TeV}'',
  \href{https://doi.org/10.1103/PhysRevLett.121.082301}{{\em Phys. Rev. Lett.}
  {\bfseries 121} (2018) 082301},
  \href{https://arxiv.org/abs/1804.09767}{{\ttfamily arXiv:1804.09767
  [hep-ex]}}.

\bibitem{Sahoo:2018uhb}
B.~Sahoo, B.~K. Nandi, P.~Pujahari, S.~Basu, and C.~Pruneau, ``{Simulation
  studies of
  $R_{2}$({\ensuremath{\Delta}}{\ensuremath{\eta}},{\ensuremath{\Delta}}{\ensuremath{\varphi}})
  and
  $P_{2}$({\ensuremath{\Delta}}{\ensuremath{\eta}},{\ensuremath{\Delta}}{\ensuremath{\varphi}})
  correlation functions in pp collisions with the PYTHIA and HERWIG models}'',
  \href{https://doi.org/10.1103/PhysRevC.100.024909}{{\em Phys. Rev. C}
  {\bfseries 100} (2019) 024909},
  \href{https://arxiv.org/abs/1810.09747}{{\ttfamily arXiv:1810.09747
  [nucl-ex]}}.

\bibitem{Basu:2020ldt}
S.~Basu, V.~Gonzalez, J.~Pan, A.~Knospe, A.~Marin, C.~Markert, and C.~Pruneau,
  ``{Differential two-particle number and momentum correlations with the AMPT,
  UrQMD, and EPOS models in Pb-Pb collisions at $\sqrt{s_{_\mathrm{NN}}}$ =
  2.76 TeV}'', \href{https://doi.org/10.1103/PhysRevC.104.064902}{{\em Phys.
  Rev. C} {\bfseries 104} (2021) 064902},
  \href{https://arxiv.org/abs/2001.07167}{{\ttfamily arXiv:2001.07167
  [nucl-ex]}}.

\bibitem{Pratt_ref1}
S.~Pratt and J.~Vredevoogd, ``{Femtoscopy in Relativistic Heavy Ion Collisions
  and its Relation to Bulk Properties of QCD Matter}'',
  \href{https://doi.org/10.1103/PhysRevC.79.069901}{{\em Phys. Rev. C}
  {\bfseries 78} (2008) 054906},
  \href{https://arxiv.org/abs/0809.0516}{{\ttfamily arXiv:0809.0516
  [nucl-th]}}. [Erratum: Phys.Rev.C 79, 069901 (2009)].

\bibitem{Brown_ref2}
R.~H. Brown and R.~Q. Twiss, ``{Correlation between Photons in two Coherent
  Beams of Light}'', \href{https://doi.org/10.1038/177027a0}{{\em Nature}
  {\bfseries 177} (1956) 27--29}.

\bibitem{ALICE:2023yuk}
{\bfseries ALICE} Collaboration, S.~Acharya {\em et~al.}, ``{Production of
  pions, kaons, and protons as a function of the relative transverse activity
  classifier in pp collisions at $ \sqrt{s} $ = 13 TeV}'',
  \href{https://doi.org/10.1007/JHEP06(2023)027}{{\em JHEP} {\bfseries 06}
  (2023) 027}, \href{https://arxiv.org/abs/2301.10120}{{\ttfamily
  arXiv:2301.10120 [nucl-ex]}}.

\bibitem{cmspythia}
{\bfseries CMS} Collaboration, A.~Tumasyan {\em et~al.}, ``{CMS PYTHIA~8 colour
  reconnection tunes based on underlying-event data}'',
  \href{https://doi.org/10.1140/epjc/s10052-023-11630-8}{{\em Eur. Phys. J. C}
  {\bfseries 83} (2023) 587},
  \href{https://arxiv.org/abs/2205.02905}{{\ttfamily arXiv:2205.02905
  [hep-ex]}}.

\bibitem{OrtizVelasquez:2013ofg}
A.~Ortiz~Velasquez, P.~Christiansen, E.~Cuautle~Flores, I.~Maldonado~Cervantes,
  and G.~Pai{\'c}, ``{Color Reconnection and Flowlike Patterns in $pp$
  Collisions}'', \href{https://doi.org/10.1103/PhysRevLett.111.042001}{{\em
  Phys. Rev. Lett.} {\bfseries 111} (2013) 042001},
  \href{https://arxiv.org/abs/1303.6326}{{\ttfamily arXiv:1303.6326 [hep-ph]}}.

\bibitem{Ortiz:2020rwg}
A.~Ortiz, A.~Paz, J.~D. Romo, S.~Tripathy, E.~A. Zepeda, and I.~Bautista,
  ``{Multiparton interactions in $pp$ collisions from machine learning-based
  regression}'', \href{https://doi.org/10.1103/PhysRevD.102.076014}{{\em Phys.
  Rev. D} {\bfseries 102} (2020) 076014},
  \href{https://arxiv.org/abs/2004.03800}{{\ttfamily arXiv:2004.03800
  [hep-ph]}}.

\bibitem{Lonnblad:2023kft}
L.~L{\"o}nnblad and H.~Shah, ``{Baryon correlations in Pythia}'',
  \href{https://doi.org/10.1140/epjc/s10052-023-12271-7}{{\em Eur. Phys. J. C}
  {\bfseries 83} (2023) 1105},
  \href{https://arxiv.org/abs/2309.01557}{{\ttfamily arXiv:2309.01557
  [hep-ph]}}.

\end{thebibliography}\endgroup
\end{document}